\documentclass[12pt]{article}

\usepackage{fancyhdr}

\usepackage{newtxtext,newtxmath}
\usepackage{graphicx}
\graphicspath{{../}}  %% for local pdflatex'ing

    \usepackage[dvipsnames]{xcolor} % colors in comment macros
\usepackage[letterpaper,margin=1in]{geometry}

\renewenvironment{abstract}
	{\quotation}
	{\endquotation}

\date{}

\makeatletter
\renewcommand{\fnum@figure}{\textbf{Figure \thefigure}}
\renewcommand{\fnum@table}{\textbf{Table \thetable}}
\makeatother

\usepackage{scicite}

\usepackage{url}

    \definecolor{ao}{rgb}{0.0, 0.4, 0.0}
    \definecolor{jb}{rgb}{0.4,0.0, 0.0}
    
    \newcommand{\TA}[1]{}
    \newcommand{\JH}[1]{}
    \newcommand{\RXA}[1]{}
    \newcommand{\BDT}[1]{}
    \newcommand{\JB}[1]{}
    
    \newcommand{\chardp}{\ensuremath{\theta_{CHP}}}

    \newcommand{\argmax}{\mathop{\mathrm{arg\,max}}}
    
    \newcommand{\unit}[1]{\ensuremath{\, \mathrm{#1}}}
\def\scititle{Improving cosmological reach of a gravitational wave observatory using Deep Loop Shaping}

\title{\bfseries \boldmath \scititle}

\author{
Jonas~Buchli$^{1\dagger\ast}$,
Brendan~Tracey$^{1\dagger}$,
Tomislav~Andric$^{2, 3\dagger}$,
Christopher~Wipf$^{4\dagger}$,\and
Yu~Him~Justin~Chiu$^{1\dagger}$,
Matthias~Lochbrunner$^{1\dagger}$,
Craig~Donner$^{1\dagger}$,
Rana~X~Adhikari$^{4\dagger\ast}$,\and
Jan~Harms$^{2,3\dagger\ast}$,
Iain~Barr$^{1}$,
Roland~Hafner$^{1}$,
Andrea~Huber$^{1}$,\and
Abbas~Abdolmaleki$^{1}$,
Charlie~Beattie$^{1}$,
Joseph~Betzwieser$^{4}$,
Serkan~Cabi$^{1}$,\and
Jonas~Degrave$^{1}$,
Yuzhu~Dong$^{1}$,
Leslie~Fritz$^{1}$,
Anchal~Gupta$^{4}$,
Oliver~Groth$^{1}$,\and
Sandy~Huang$^{1}$,
Tamara~Norman$^{1}$,
Hannah~Openshaw$^{1}$,
Jameson~Rollins$^{4}$,\and
Greg~Thornton$^{1}$,
George~van~den~Driessche$^{1}$,
Markus~Wulfmeier$^{1}$,\and
Pushmeet~Kohli$^{1\ast}$,
Martin~Riedmiller$^{1}$, \and
The LIGO Instrument Team$^{5}$\and
\small$^{1}$Google DeepMind, London, UK\and
\small$^{2}$Gran Sasso Science Institute (GSSI), L'Aquila, Italy\and
\small$^{3}$Laboratori Nazionali del Gran Sasso, Assergi (INFN), Italy\and
\small$^{4}$LIGO Laboratory, Division of Physics, Math, and Astronomy, California Institute of Technology, Pasadena, USA\and
\small$^\dagger$These authors contributed equally to this work\and
\small$^\ast$Corresponding author. Email: buchli@google.com, rana@caltech.edu, jan.harms@gssi.it, pushmeet@google.com\\
\small$^{5}$LIGO Instrument Team authors and affiliations are listed in
the supplementary materials.}

\begin{document}

\maketitle

\begin{abstract} \bfseries \boldmath
Improved low-frequency sensitivity of gravitational wave observatories would unlock study of intermediate-mass black hole mergers, binary black hole eccentricity, and provide early warnings for multi-messenger observations of binary neutron star mergers. Today’s mirror stabilization control injects harmful noise, constituting a major obstacle to sensitivity improvements. We eliminated this noise through Deep Loop Shaping, a reinforcement learning method using frequency domain rewards. 
We proved our methodology on the LIGO Livingston Observatory (LLO). 
Our controller reduced control noise in the 10--30\,Hz band by over 30x, and up to 100x in sub-bands surpassing the design goal motivated by the quantum limit. 
These results highlight the potential of Deep Loop Shaping to improve current and future GW observatories, and more broadly instrumentation and control systems.
\end{abstract}

\pagestyle{fancy}
\fancyhead[RO,LE]{Preprint - Published in Science 10.1126/science.adw1291}

\noindent

\begin{figure}
    \centering
 \includegraphics[width=0.5\columnwidth]{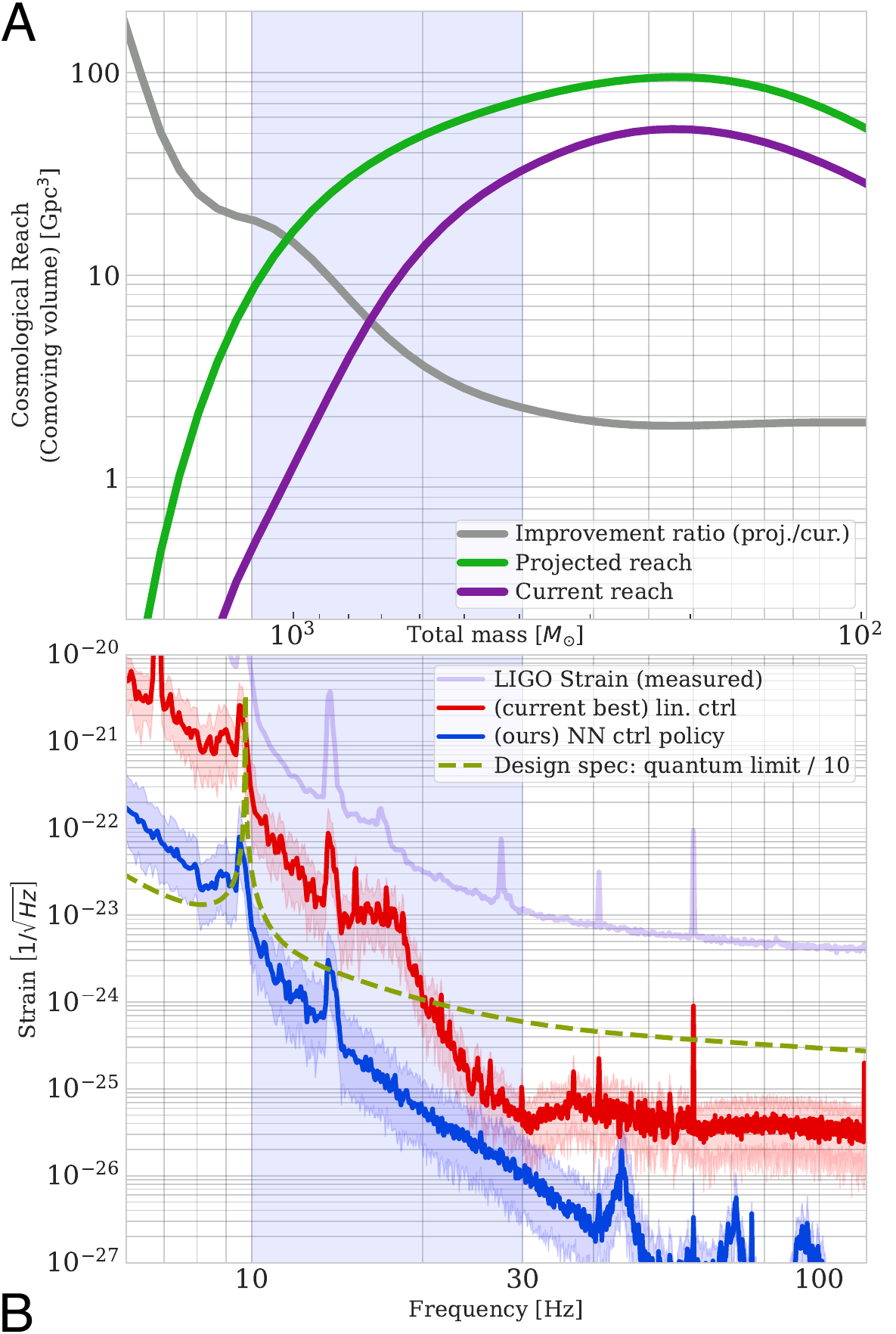}
 \caption{
 \textbf{Cosmological reach and strain noise from control.}
 (A) The plot shows the volume in space explored with binary black hole merger waveforms~\cite{khanFrequencydomainGravitationalWaves2016} for different cases of technical noise. The x-axis in (A) is the total mass of the equal-mass binary pair. This corresponds to the x-axis in (B), the frequency of the first quasi-normal mode of a Schwarzschild black hole with such a mass, as measured in the source frame.
 The purple trace shows the reach of LIGO as of March 2024. 
 The green trace shows the volumetric improvement in the case where the technical noise is removed entirely. 
 Many of the known technical noise sources are linked to controls. 
 (B) LIGO's noise budget and controller performance. 
 Purple: overall measured strain noise, 
 red: strain noise contribution from currently operational linear controller for \chardp{}, \
 blue: strain noise contribution from RL policy as run on the LIGO Livingston Observatory on Dec 5, 2024 (mean, 10 and 90\
 Dashed green indicates the control design goal derived from the quantum back-action limit by applying a design margin of 10x; the control noise should drop below this curve. A detailed accounting of technical noise sources is available in \cite{O4InstrumentPaper}.
 }
 \label{fig:rl_vs_lin_strain_coords}
 \label{fig:cosmo_v_mass} 
\end{figure}

The gravitational wave (GW) detectors LIGO and Virgo have revolutionized astrophysics by detecting mergers of exotic objects such as black holes (BHs) and neutron stars (NSs) \cite{AbEA2016a,AbEA2017d,AbEA2021a}. 
Currently, most of the detectable signal lies in the 30--2000\,Hz band, leaving the low-frequency band (10--30\,Hz) largely unexplored. Enhancing sensitivity in this band could lead to a substantialincrease in cosmological reach, and thus in the scientific capabilities of LIGO (Fig. \ref{fig:cosmo_v_mass}A). The 10--30\,Hz band is also important for the early (pre-merger) detection of binary neutron stars (BNS), potentially doubling the warning time, to enable real-time observation of observe neutron star collisions, and thus the creation of heavy elements, and the birth of black holes~\cite{Hang:Early:2021,Banerjee2023,Tohuvavohu_2024}.
However, such sensitivity improvements are currently partially limited by injected control noise on the interferometer mirrors.  Furthermore, as the control noise is a bottleneck to overall sensitivity improvements, there is currently little to be gained from improvements to other noise sources.
We address this challenge with a novel tailored Reinforcement Learning (RL) method, and improve the alignment control of the LIGO mirrors. We lower the injected control noise on the most demanding feedback control loop, the common-hard-pitch (\chardp{}) loop of the Livingston Observatory, below the quantum back-action limit. By eliminating the harmful noise from this critical representative controller, we pave the path to improve LIGO's sensitivity.

The space-time strain associated with even the loudest GW signals produces a signal equivalent to only $\approx$10$^{-19}$ meters of mirror motion.  
As a comparison, the environmental disturbance, due to Earth tides and seismic vibration, is roughly 13 orders of magnitude larger.
To measure the weak GW signals, laser-interferometric GW detectors have hundreds of opto-mechanical degrees of freedom that require stabilization.
Active control is used to achieve precise stabilization in the face of complex mirror dynamics and inherently unstable degrees of freedom.
More specifically, the opto-mechanical response of the interferometer (i.e., the plant) is subject to dynamic variations: even low absorption of the high-power laser beam ($\sim$300--500\,kW) causes thermal distortions in the mirrors, leading to offsets in sensor signals and changes in opto-mechanical resonant frequencies. In addition, the high power laser also creates substantial forces and torques on the suspended mirrors, leading to opto-mechanical instabilities of several mechanical eigenmodes \cite{SiSi2006, Bar2010, BRAGINSKY2001331}. 
These resonances are stabilized using feedback control, but any noise injected by the feedback controllers into the GW readout harms the peak astrophysical sensitivity and drowns out the GW signals themselves.

In simplified terms, the main control design challenge is that larger control action in lower frequencies provides better disturbance rejection, but injects higher noise into the observation band. Conversely, lowering the control action reduces injected noise, but results in insufficient disturbance rejection and possible loss of stability. Linear control systems theory shows fundamental limits to this trade-off \cite{astrom_feedback_2021,Stein:unstable} under certain assumptions about the controller design. The ultimate aim of controller design is to shape the ‘closed loop’ behavior, i.e., the performance of the designed controller acting in a feedback loop with the plant.

There are many classical methods to achieve the desired closed-loop behavior.
Early methods, i.e., the classic (open) loop shaping methods, exploit the direct relationship between the open- and closed-loop transfer functions to design the controller. Since the 1980s, the focus of design has shifted from open loop design to directly shaping closed loop transfer functions, i.e. the sensitivity functions \cite{bode1945,zames1981},  usually through optimization (e.g. convex optimization \cite{barratt1993}, $H_{\infty}$ \cite{doyle1989}). These methods are more general and can take into account a larger variety of design goals and constraints. Yet they still require strong assumptions such as convexity and linearity. For GW detectors like LIGO, progress using traditional approaches has come to a plateau.
Machine learning for interferometer control was presented in \cite{Mukuetal2023}, with the primary aim to improve the contrast of the interferometer, rather than optimizing closed loop performance.

Here, we present a novel control design method, Deep Loop Shaping (DLS), to design controllers that satisfy specific demands on the system’s frequency-domain behavior. DLS has no constraints regarding the use of nonlinear models and control structures. It exploits the machinery of deep reinforcement learning to directly optimize frequency-domain properties and shape the closed loop behavior.
We demonstrate DLS's utility on the critical LIGO \chardp{} control loop, achieving state-of-the-art feedback control performance. The injected control noise was reduced by up to two orders of magnitude while maintaining mirror stability. Applying deep loop shaping more widely on LIGO can improve sensitivity. Furthermore, the method has wide applicability to control engineering: for example, highly unstable systems, vibration suppression, and noise cancellation all have strong frequency-dependent control demands.

\begin{figure}
    A \includegraphics[width=\textwidth]{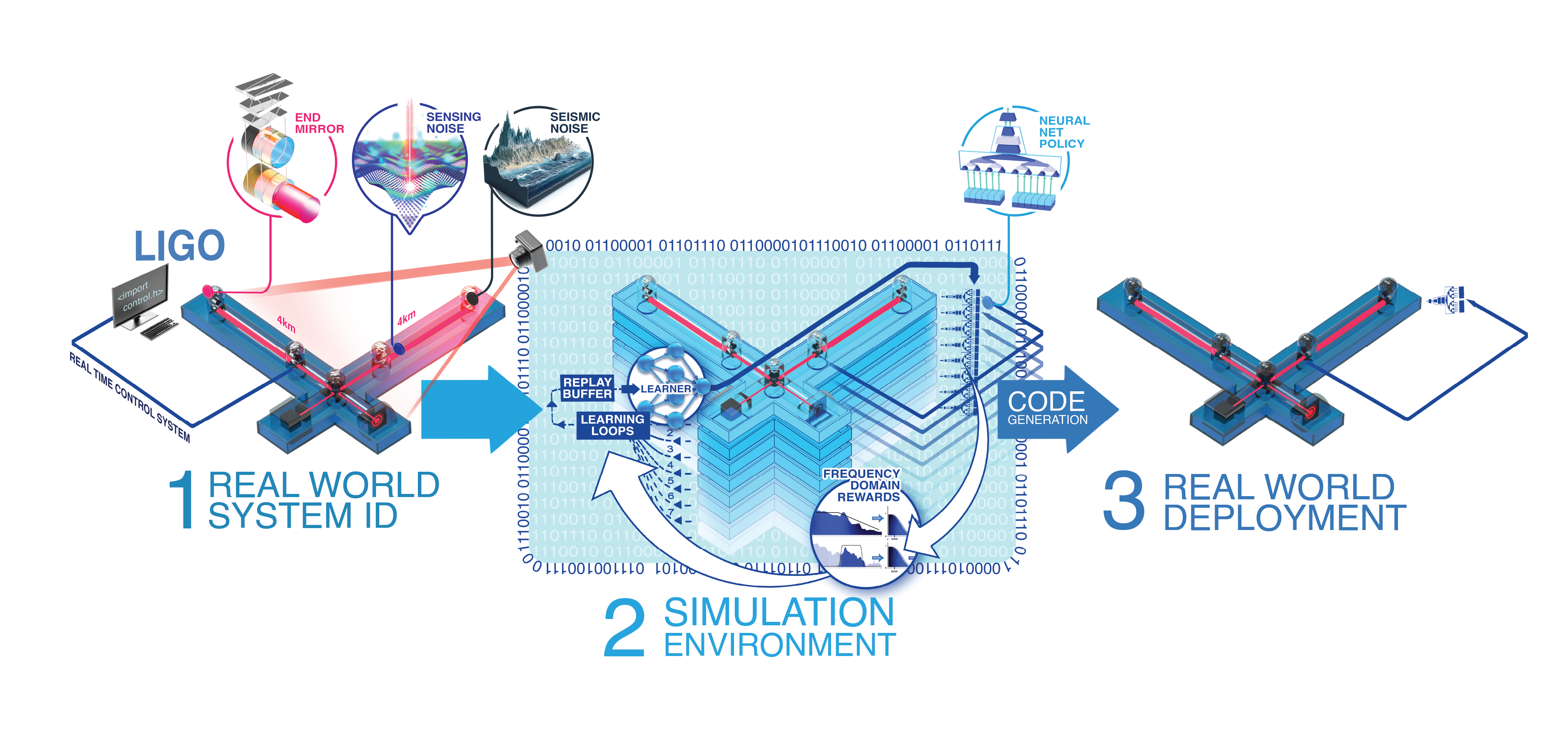}\\
    B \includegraphics[width=0.8\textwidth]{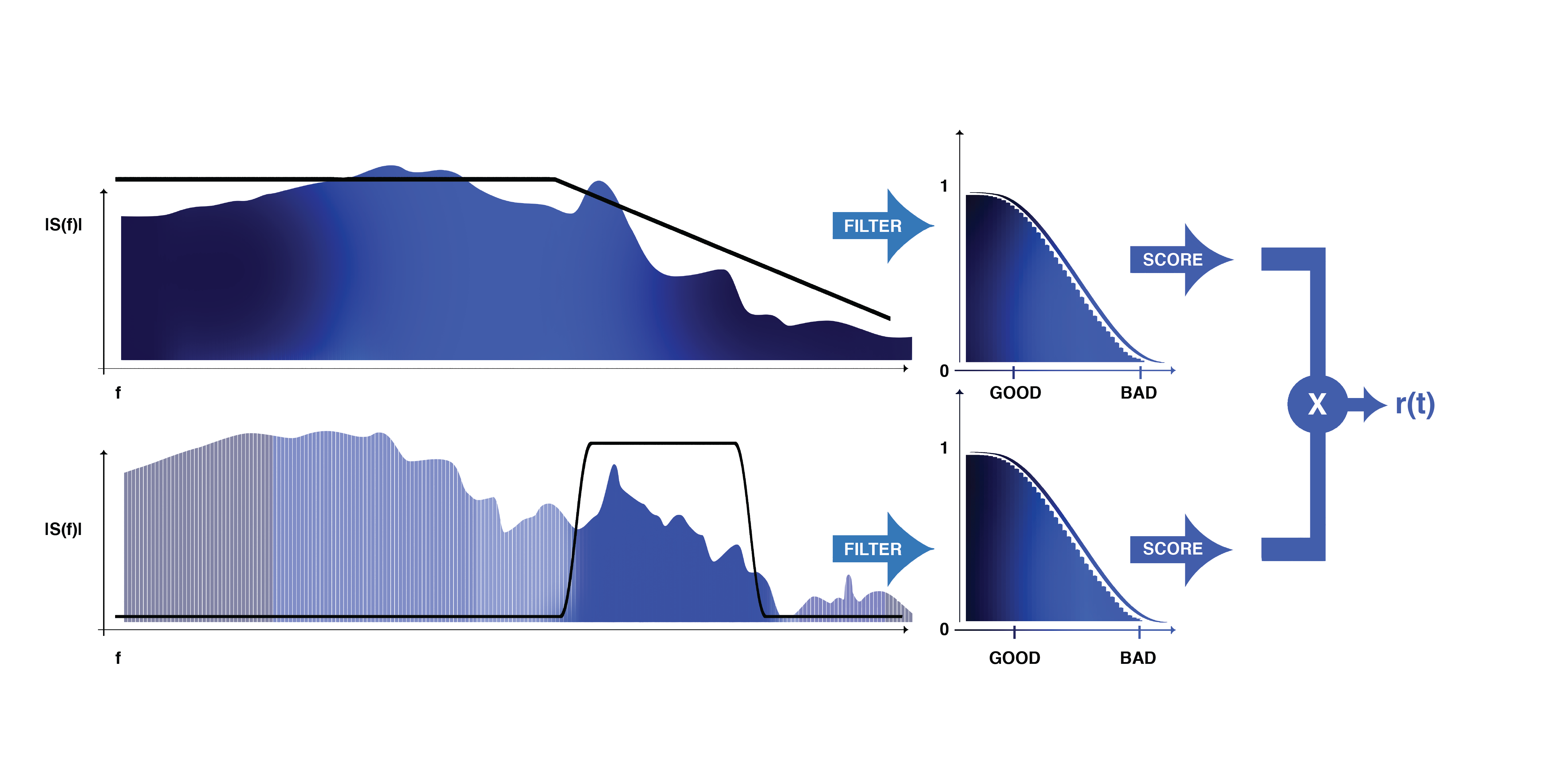}
    \caption{\textbf{Deep Loop Shaping -- Reinforcement learning with Frequency Domain Rewards}
     (A) (1) A model is identified from plant measurements. (2) The identified model is used as a learning environment. Frequency-domain rewards are used to compute rewards. (3) The optimized control policy is deployed on the plant. 
     (B) Illustration of the frequency rewards and the multiplicative scoring. 
    }
    \label{fig:freq_rewards}
    \label{fig:concept_sketch}
\end{figure}

\subsection*{The LIGO Controls Challenge}
\label{s:LIGOcontrols}

Angular Sensing and Control (ASC) is the challenge of maintaining the orientation of the interferometer mirrors. Stabilization is accomplished through a hybrid active-passive isolation system. Passive stabilization happens through a series of pendulums, from which the optics are suspended. These pendulums suppress seismic disturbances above 10\,Hz by several orders of magnitude. However, active stabilization is required to reject seismic disturbances below $\sim$3\,Hz. This stabilization is accomplished by a set of actuators that produce a torque on the suspended optics at the penultimate stage of the suspension system.
Additionally, there are disturbances caused by the radiation-pressure forces of the high-power laser beam \cite{Dool2013} that couple the angular motions of the cavity mirrors. 
The sensors used to measure the angular motion have a good signal-to-noise ratio below a few Hz to enable active stabilization, but in the 10--30\,Hz band, the sensor noise is orders of magnitude larger than the signal related to angular motion, and so motion in this band is not visible from the angular sensor, and is only seen in the interferometer strain spectrum. 
The active controller injects sensing noise in this frequency band through the actuator to the test mass. This effect is the primary cause of test mass angular motion above 10\,Hz~\cite{Bar2010,YuEA2017}.
Avoiding the injection of noise as much as possible and at the same time guaranteeing rejection of seismic disturbances is the main design goal for ASC controllers. We address this problem, which comprises the most challenging control loops in a GW observatory, with Deep Loop Shaping.

\paragraph{The common-hard-pitch (\chardp{}) loop}
\label{s:chardp}
In this work, we primarily focus on the ASC control loop `common-hard-pitch' (\chardp). `Hard pitch' refers to the stiffer of the two opto-mechanical pitch eigenmodes of the arm cavities. `Common' signifies a relation of modes between the cavities of the two arms~\cite{Bar2010}. The \chardp\,degree of freedom is the most difficult of the entire ASC system to stabilize and optimize. 
Reducing the control noise in this loop well below the quantum limit would remove this source of noise as a blocking issue for improved astrophysical sensitivity.

\subsubsection*{Closed loop shaping as Reinforcement Learning problem}
\label{s:deeploopshaping}

Here, we formulate \chardp{} closed-loop control as an optimal control problem, and find approximate solutions through reinforcement learning (RL). ASC requirements are naturally expressed as functions of the system response in the frequency domain; i.e., as desired spectra of state-space signals. 
We introduce a novel reward scheme 
based on frequency domain behavior 
to enforce the desired closed-loop shaping of the control policy in such a way that RL can discover an effective control policy. 
It is similar to traditionally used methods of shaping sensitivity functions, but RL removes restrictions on the reward definition and system dynamics. RL can also discover nonlinear policies represented with deep neural networks  
that can serve as drop-in replacements for the existing hand-crafted controllers, and, as we will show, enables improved performance without compromising robustness.

RL designs controllers by adapting a parameterized state-action mapping. Our specific choice of learning algorithm is Maximum-a-Posteriori-Policy Optimization (MPO) \cite{abdolmaleki2018mpo}. We use a small multi-layer-perceptron (MLP) with a dilated convolution input layer for the policy network, which executes sufficiently fast for control. The critic network is a Long-Short-Term-Memory network (LSTM) with input and output MLPs, as the critic is not needed for deployment.

\subsection*{Frequency Domain Rewards}
\label{s:FD_rewards}
RL naturally lends itself to reward descriptions formulated in the time domain, e.g., scoring events that happen at certain times. 
Instead, we directly formulate the ASC requirements as
rewards in the frequency domain. To do so, we design linear filters for the \chardp{} response signal
whose transfer functions each select a certain frequency band of the signal. We use a low-pass filter to reward pitch alignment, a band-pass filter to reduce control action in the 8\,--\,30\,Hz band, and an additional band-pass filter for frequencies $>$40\,Hz to avoid high-frequency artifacts. 
A high output from a filter at a given timestep corresponds to a large historic response in the measured frequency band. These per-timestep response measures can then be used to construct a reward for RL.
Specifically, we compute the RL reward by passing the filter outputs through
a sigmoid function
to compute a (per-filter) score in [0,1], with a value of $1$ if
the specification is fulfilled and fading to $0$ as the response worsens. These individual filter scores are multiplied to yield the per-timestep reward, then used by the RL method to choose a policy that minimizes the discounted sum of this reward over time.
This formulation of multiplying rewards can loosely be understood as a soft logical-AND; i.e. we want all properties to be fulfilled for high reward.

\subsection*{Training and Deployment}
\label{s:deployment}

We train nonlinear control policies with 
RL against a linear stochastic state-space simulation of the plant dynamics (i.e., opto-mechanical response of the interferometer), identified from measurement data of the plant. 
We use domain randomization to add robustness to the learned policies. Specifically, at the beginning of each episode, we randomize the angular instability pole frequency and sample variations in the seismic noise, including the overall noise strength.

At the conclusion of training, we perform several steps to ready the policy for hardware testing. First, a deterministic policy is created by using only the mean of the policy Gaussian. 
Second, we validate this deterministic policy across a selected set of disturbances and non-nominal plant parameters. We examine the reward achieved, as well as measure key performance criteria such as RMS of the control effort in the observation band (10--30 $\unit{Hz}$), and visually inspect the error and control spectra. With performance confirmed, we ``export'' the policy for the hard real-time control required for execution on LIGO without further training or adaptation on the plant.

We deploy the control policies directly in the existing control infrastructure of interferometer~\cite{advligorts_2021}. As such, the RL-trained policies are drop-in replacements of the existing SISO controllers. In particular, the LIGO control system uses somewhat arbitrary `counts' as the units for ASC inputs and outputs, and we adopt these conventions for the controller and the controller-simulator interface. We also report our results in these units for technical reasons. 

\subsection*{Deployment on gravitational wave observatory hardware}
\label{s:results}

We ran the deployed policies on the LIGO Livingston Observatory (LLO). In the experiments, the control of \chardp{} was under the sole authority of a neural-network-based control policy. We measured the ASC noise during policy execution, as well as comparison spectra from the standard controller before and after the nonlinear policy. In Fig.~\ref{fig:rl_vs_lin_strain_coords}B, we compare the performance of the neural network policy against the standard controller for a $> 10\,\unit{min}$ stretch on Dec 5, 2024. The figure shows the projection of the measured angular control noise into the GW readout. Additional details of this experiment are shown in the Supplementary Materials.

We find excellent performance for the neural network policy. In the crucial 3\,--\,30\,Hz band, we see a reduction of noise of up to 2 orders of magnitude.
At the same time, the neural network policy shows similar control authority as the linear controller in the control band ($< 3\,\unit{Hz}$). The control noise added by the neural network policy is well below the fundamental thermodynamic and quantum back-action limit in the whole band of interest. These results show that the neural network policy has effectively removed the issue of noise injected by active control as a limit to the astrophysical sensitivity.

In the supplementary materials
we present additional results from April and August 2024, with total time on the instrument of well over 1 hour. The sustained control of the unstable \chardp{} mode demonstrates robustness of the neural network policy to normal seismic activity.
We see 
a good match between the training simulation and the real plant under the tested conditions for frequencies $> 0.1\,\unit{Hz}$, which increases confidence in our results. 
We additionally compare the control policy against the incumbent linear controller in terms of statistical measures such as non-Gaussianity and non-stationarity. We find that while the policy does exhibit some non-stationarity, the overall reduction in noise still leads to a clear benefit for signal detection.

For comparison, we derived controllers with convex optimization and show a series of simulation-based results in the supplementary materials.
While these optimized linear controllers have similar predicted performance, they are not fit for deployment in the high-stakes environment of the real observatory. In particular, they are open-loop unstable, and their disturbance rejection behavior is highly aggressive, in contrast to the neural network policies.
In addition to experiments on LLO, we used the same methodology on the mode-cleaner of the Caltech prototype and similarly find that DLS is capable of reducing noise in a band of interest while maintaining overall control.

\clearpage 

\fancyhead{} 
\fancyhead[RO,LE]{Published in Science 10.1126/science.adw1291}

\bibliography{main} 

\begin{thebibliography}{10}
\providecommand{\url}[1]{\texttt{#1}}
\expandafter\ifx\csname urlstyle\endcsname\relax
  \providecommand{\doi}[1]{doi:\discretionary{}{}{}#1}\else
  \providecommand{\doi}{doi:\discretionary{}{}{}\begingroup
  \urlstyle{rm}\Url}\fi

\bibitem{AbEA2016a}
B.~P. Abbott, \emph{et~al.}, {Observation of Gravitational Waves from a Binary
  Black Hole Merger}. \emph{Phys. Rev. Lett.} \textbf{116}, 061102 (2016),
  \doi{10.1103/PhysRevLett.116.061102},
  \url{http://link.aps.org/doi/10.1103/PhysRevLett.116.061102}.

\bibitem{AbEA2017d}
B.~P. Abbott, \emph{et~al.}, {GW170817: Observation of Gravitational Waves from
  a Binary Neutron Star Inspiral}. \emph{Phys. Rev. Lett.} \textbf{119}, 161101
  (2017), \doi{10.1103/PhysRevLett.119.161101},
  \url{https://link.aps.org/doi/10.1103/PhysRevLett.119.161101}.

\bibitem{AbEA2021a}
R.~Abbott, \emph{et~al.}, {GWTC-3: Compact Binary Coalescences Observed by LIGO
  and Virgo during the Second Part of the Third Observing Run}. \emph{Phys.
  Rev. X} \textbf{13}, 041039 (2023), \doi{10.1103/PhysRevX.13.041039},
  \url{https://link.aps.org/doi/10.1103/PhysRevX.13.041039}.

\bibitem{Hang:Early:2021}
H.~Yu, R.~X. Adhikari, R.~Magee, S.~Sachdev, Y.~Chen, Early warning of
  coalescing neutron-star and neutron-star-black-hole binaries from the
  nonstationary noise background using neural networks. \emph{Phys. Rev. D}
  \textbf{104}, 062004 (2021), \doi{10.1103/PhysRevD.104.062004},
  \url{https://link.aps.org/doi/10.1103/PhysRevD.104.062004}.

\bibitem{Banerjee2023}
{Banerjee, Biswajit}, \emph{et~al.}, Pre-merger alert to detect prompt emission
  in very-high-energy gamma-rays from binary neutron star mergers: Einstein
  Telescope and Cherenkov Telescope Array synergy. \emph{A\&A} \textbf{678},
  A126 (2023), \doi{10.1051/0004-6361/202345850},
  \url{https://doi.org/10.1051/0004-6361/202345850}.

\bibitem{Tohuvavohu_2024}
A.~Tohuvavohu, \emph{et~al.}, Swiftly Chasing Gravitational Waves across the
  Sky in Real Time. \emph{The Astrophysical Journal Letters} \textbf{975}~(1),
  L19 (2024), \doi{10.3847/2041-8213/ad87ce},
  \url{https://dx.doi.org/10.3847/2041-8213/ad87ce}.

\bibitem{khanFrequencydomainGravitationalWaves2016}
S.~Khan, \emph{et~al.}, Frequency-Domain Gravitational Waves from Nonprecessing
  Black-Hole Binaries. {{II}}. {{A}} Phenomenological Model for the Advanced
  Detector Era. \emph{Physical Review D} \textbf{93}~(4), 044007 (2016),
  \doi{10.1103/PhysRevD.93.044007}.

\bibitem{O4InstrumentPaper}
E.~Capote, \emph{et~al.}, Advanced {{LIGO}} Detector Performance in the Fourth
  Observing Run (2024), \doi{10.48550/ARXIV.2411.14607}.

\bibitem{SiSi2006}
J.~A. Sidles, D.~Sigg, {Optical torques in suspended Fabry-P\'erot
  interferometers}. \emph{Physics Letters A} \textbf{354}~(3), 167 -- 172
  (2006), \doi{http://dx.doi.org/10.1016/j.physleta.2006.01.051},
  \url{http://www.sciencedirect.com/science/article/pii/S0375960106001381}.

\bibitem{Bar2010}
L.~Barsotti, M.~Evans, P.~Fritschel, Alignment sensing and control in advanced
  LIGO. \emph{Classical and Quantum Gravity} \textbf{27}~(8), 084026 (2010),
  \doi{10.1088/0264-9381/27/8/084026},
  \url{https://iopscience.iop.org/article/10.1088/0264-9381/27/8/084026}.

\bibitem{BRAGINSKY2001331}
V.~Braginsky, S.~Strigin, S.~Vyatchanin, Parametric oscillatory instability in
  Fabry–Perot interferometer. \emph{Physics Letters A} \textbf{287}~(5),
  331--338 (2001), \doi{https://doi.org/10.1016/S0375-9601(01)00510-2},
  \url{https://www.sciencedirect.com/science/article/pii/S0375960101005102}.

\bibitem{astrom_feedback_2021}
K.~Åström, R.~Murray, \emph{Feedback {Systems}: {An} {Introduction} for
  {Scientists} and {Engineers}, {Second} {Edition}} (Princeton University
  Press) (2021), \url{https://books.google.com/books?id=l50DEAAAQBAJ}.

\bibitem{Stein:unstable}
G.~Stein, Respect the unstable. \emph{IEEE Control Systems Magazine}
  \textbf{23}~(4), 12--25 (2003), \doi{10.1109/MCS.2003.1213600}.

\bibitem{bode1945}
H.~W. Bode, \emph{Network Analysis and Feedback Amplifier Design} (D. Van
  Nostrand Company, New York, NY, USA) (1945).

\bibitem{zames1981}
G.~Zames, Feedback and optimal sensitivity: Model reference transformations,
  multiplicative seminorms, and approximate inverses. \emph{IEEE Transactions
  on Automatic Control} \textbf{26}~(2), 301--320 (1981).

\bibitem{barratt1993}
C.~Barratt, S.~Boyd, Closed-Loop Convex Formulation of Classical and Singular
  Value Loop Shaping, in \emph{Control and Dynamical Systems: Digital and
  Numeric Techniques and Their Applications in Control Systems, Part 1}, C.~T.
  Leondes, Ed. (Academic Press, San Diego, CA, USA), vol.~55, pp. 1--24 (1993).

\bibitem{doyle1989}
J.~C. Doyle, K.~Glover, P.~P. Khargonekar, B.~A. Francis, State-space solutions
  to standard {$H_2$} and {$H_\infty$} control problems. \emph{IEEE
  Transactions on Automatic Control} \textbf{34}~(8), 831--847 (1989).

\bibitem{Mukuetal2023}
N.~Mukund, \emph{et~al.}, Neural sensing and control in a kilometer-scale
  gravitational-wave observatory. \emph{Physical Review Applied} \textbf{20},
  064041 (2023), \doi{10.1103/PhysRevApplied.20.064041},
  \url{https://link.aps.org/doi/10.1103/PhysRevApplied.20.064041}.

\bibitem{Dool2013}
K.~Dooley, \emph{et~al.}, Angular control of optical cavities in a
  radiation-pressure-dominated regime: the Enhanced LIGO case. \emph{Journal of
  the Optical Society of America A} \textbf{30}, 2618--26 (2013),
  \doi{10.1364/JOSAA.30.002618},
  \url{https://opg.optica.org/josaa/abstract.cfm?uri=josaa-30-12-2618}.

\bibitem{YuEA2017}
H.~Yu, \emph{et~al.}, Prospects for Detecting Gravitational Waves at 5 Hz with
  Ground-Based Detectors. \emph{Physical Review Letters} \textbf{120}~(14),
  141102 (2017), \doi{10.1103/PhysRevLett.120.141102},
  \url{https://journals.aps.org/prl/abstract/10.1103/PhysRevLett.120.141102}.

\bibitem{abdolmaleki2018mpo}
A.~Abdolmaleki, \emph{et~al.}, Maximum a Posteriori Policy Optimisation, in
  \emph{International Conference on Learning Representations} (2018).

\bibitem{advligorts_2021}
R.~Bork, \emph{et~al.}, advligorts: {The} {Advanced} {LIGO} real-time digital
  control and data acquisition system. \emph{SoftwareX} \textbf{13}, 100619
  (2021), \doi{10.1016/j.softx.2020.100619},
  \url{https://www.sciencedirect.com/science/article/pii/S2352711020303320}.

\bibitem{hoffman2020acme}
M.~Hoffman, \emph{et~al.}, Acme: A research framework for distributed
  reinforcement learning. \emph{arXiv preprint arXiv:2006.00979}  (2020).

\bibitem{yang2021launchpad}
F.~Yang, \emph{et~al.}, Launchpad: A Programming Model for Distributed Machine
  Learning Research. \emph{arXiv preprint arXiv:2106.04516}  (2021).

\bibitem{dm_env2019}
A.~Muldal, \emph{et~al.}, dm\_env: A python interface for reinforcement
  learning environments (2019), \url{http://github.com/deepmind/dm_env}.

\bibitem{haiku2024github}
T.~Hennigan, T.~Cai, T.~Norman, L.~Martens, I.~Babuschkin, {H}aiku: {S}onnet
  for {JAX} (2024), \url{http://github.com/deepmind/dm-haiku}.

\bibitem{cassirer2021reverb}
A.~Cassirer, \emph{et~al.}, Reverb: A Framework For Experience Replay (2021),
  \url{https://arxiv.org/abs/2102.04736}.

\bibitem{lightsaber_git}
J.~Harms, Lightsaber, \url{https://github.com/janosch314/Lightsaber} (2025).

\bibitem{ligo_scientific_collaboration_2025_15793015}
{LIGO Scientific Collaboration}, Identified Plant Model \& Selected ASC
  Experimental Data of LIGO Livingston (2025), \doi{10.5281/zenodo.15793015},
  \url{https://doi.org/10.5281/zenodo.15793015}.

\bibitem{AndHar2021}
T.~Andric, J.~Harms, Lightsaber: A Simulator of the Angular Sensing and Control
  System in LIGO. \emph{Galaxies} \textbf{9}~(3) (2021),
  \doi{10.3390/galaxies9030061}, \url{https://www.mdpi.com/2075-4434/9/3/61}.

\bibitem{And2023}
T.~Andric, \emph{Low-frequency sensitivity limitations in current and future
  gravitational-wave detectors}, {PhD} dissertation, Gran Sasso Science
  Institute (2023).

\bibitem{Frit1998}
P.~Fritschel, \emph{et~al.}, Alignment of an Interferometric Gravitational Wave
  Detector. \emph{Applied optics} \textbf{37}~(28), 6734--47 (1998),
  \doi{10.1364/AO.37.006734},
  \url{https://opg.optica.org/ao/abstract.cfm?uri=ao-37-28-6734}.

\bibitem{MaEA2014}
F.~Matichard, \emph{et~al.}, {Advanced LIGO two-stage twelve-axis vibration
  isolation and positioning platform. Part 1: Design and production overview}.
  \emph{Precision Engineering} \textbf{40}~(0), -- (2014),
  \doi{http://dx.doi.org/10.1016/j.precisioneng.2014.09.010},
  \url{http://www.sciencedirect.com/science/article/pii/S0141635914001561}.

\bibitem{MaEA2015}
F.~Matichard, \emph{et~al.}, {Advanced LIGO two-stage twelve-axis vibration
  isolation and positioning platform. Part 2: Experimental investigation and
  tests results}. \emph{Precision Engineering} \textbf{40}, 287 -- 297 (2015),
  \doi{http://dx.doi.org/10.1016/j.precisioneng.2014.11.010},
  \url{//www.sciencedirect.com/science/article/pii/S0141635914002098}.

\bibitem{AbEA2016b}
B.~P. Abbott, \emph{et~al.}, {GW150914: The Advanced LIGO Detectors in the Era
  of First Discoveries}. \emph{Physical Review Letters} \textbf{116}, 131103
  (2016), \doi{10.1103/PhysRevLett.116.131103},
  \url{http://link.aps.org/doi/10.1103/PhysRevLett.116.131103}.

\bibitem{aLIGO:PEM:2021}
P.~Nguyen, \emph{et~al.}, Environmental noise in advanced LIGO detectors.
  \emph{Classical and Quantum Gravity} \textbf{38}~(14), 145001 (2021),
  \doi{10.1088/1361-6382/ac011a},
  \url{https://dx.doi.org/10.1088/1361-6382/ac011a}.

\bibitem{Sey2017}
B.~C. Seymour, M.~Kasprzack, A.~Pelé, A.~Mullavey, \emph{Characterization of
  Nonlinear Angular Noise Coupling into Differential Arm Length of the LIGO
  Livingston Detector. LIGO-T1700343-v1}, Tech. rep., LIGO (2017),
  \url{https://dcc.ligo.org/LIGO-T1700343/public}.

\bibitem{siegman1986lasers}
A.~E. Siegman, \emph{Lasers} (University science books) (1986).

\bibitem{BE2011}
L.~Barsotti, M.~Evans, \emph{Modeling of Alignment Sensing and Control for
  Advanced LIGO. LIGO-T0900511-v4}, Tech. rep., LIGO-MIT (2011),
  \url{https://dcc.ligo.org/LIGO-T0900511/public}.

\bibitem{aLIGO:TCS}
A.~F. Brooks, \emph{et~al.}, Overview of Advanced LIGO adaptive optics.
  \emph{Appl. Opt.} \textbf{55}~(29), 8256--8265 (2016),
  \doi{10.1364/AO.55.008256},
  \url{https://opg.optica.org/ao/abstract.cfm?URI=ao-55-29-8256}.

\bibitem{Bui2020}
A.~Buikema, \emph{et~al.}, Sensitivity and performance of the Advanced LIGO
  detectors in the third observing run. \emph{Physical Review D} \textbf{102},
  062003 (2020), \doi{10.1103/PhysRevD.102.062003},
  \url{https://link.aps.org/doi/10.1103/PhysRevD.102.062003}.

\bibitem{Hang2019}
H.~Yu, \emph{Astrophysical signatures of neutron stars in compact binaries and
  experimental improvements on gravitational-wave detectors}, {PhD}
  dissertation, Massachusetts Institute of Technology (2019).

\bibitem{Alloca2020}
A.~Allocca, \emph{et~al.}, Interferometer Sensing and Control for the Advanced
  Virgo Experiment in the O3 Scientific Run. \emph{Galaxies} \textbf{8}, 85
  (2020), \doi{10.3390/galaxies8040085},
  \url{https://www.mdpi.com/2075-4434/8/4/85}.

\bibitem{Morrison:94}
E.~Morrison, B.~J. Meers, D.~I. Robertson, H.~Ward, Automatic alignment of
  optical interferometers. \emph{Appl. Opt.} \textbf{33}~(22), 5041--5049
  (1994), \doi{10.1364/AO.33.005041},
  \url{https://opg.optica.org/ao/abstract.cfm?URI=ao-33-22-5041}.

\bibitem{AdvLigo2015}
P.~Fritschel, the LIGO Scientific~Collaboration, Advanced LIGO. \emph{Classical
  and Quantum Gravity} \textbf{32}~(7), 074001 (2015),
  \doi{10.1088/0264-9381/32/7/074001},
  \url{http://dx.doi.org/10.1088/0264-9381/32/7/074001}.

\bibitem{Dool2011}
K.~L. Dooley, \emph{Design and performance of high laser power interferometers
  for gravitational-wave detection}, {PhD} dissertation, University of Florida
  (2011).

\bibitem{Mart2015}
D.~Martynov, \emph{Lock Acquisition and Sensitivity Analysis of Advanced LIGO
  Interferometers}, {PhD} dissertation, California Institute of Technology
  (2015).

\bibitem{Cahillane_2022}
C.~Cahillane, G.~Mansell, Review of the Advanced LIGO Gravitational Wave
  Observatories Leading to Observing Run Four. \emph{Galaxies} \textbf{10}~(1),
  36 (2022), \doi{10.3390/galaxies10010036},
  \url{http://dx.doi.org/10.3390/galaxies10010036}.

\bibitem{Wardeal2008}
R.~Ward, \emph{et~al.}, dc readout experiment at the Caltech 40m prototype
  interferometer. \emph{Classical and Quantum Gravity} \textbf{25} (2008),
  \doi{10.1088/0264-9381/25/11/114030},
  \url{https://iopscience.iop.org/article/10.1088/0264-9381/25/11/114030}.

\bibitem{tanioka2020angularresponsetriangularoptical}
S.~Tanioka, \emph{et~al.}, Angular response of a triangular optical cavity
  analyzed by a linear approximation method (2020),
  \url{https://arxiv.org/abs/2002.02703}.

\bibitem{franklin2002feedback}
G.~F. Franklin, J.~D. Powell, A.~Emami-Naeini, J.~D. Powell, \emph{Feedback
  control of dynamic systems}, vol.~4 (Prentice hall Upper Saddle River)
  (2002).

\bibitem{barton2010models}
M.~Barton, \emph{et~al.}, Models of the Advanced LIGO Suspensions in
  Mathematica™. \emph{Internal Technical Document T020205-02D (LIGO, 2006)}
  (2010).

\bibitem{shapiro2014noise}
B.~Shapiro, \emph{et~al.}, Noise and control decoupling of Advanced LIGO
  suspensions. \emph{Classical and Quantum Gravity} \textbf{32}~(1), 015004
  (2014).

\bibitem{Matichard_2015}
F.~Matichard, \emph{et~al.}, Seismic isolation of Advanced LIGO: Review of
  strategy, instrumentation and performance. \emph{Classical and Quantum
  Gravity} \textbf{32}~(18), 185003 (2015),
  \doi{10.1088/0264-9381/32/18/185003},
  \url{https://dx.doi.org/10.1088/0264-9381/32/18/185003}.

\bibitem{smith1992floating}
L.~M. Smith, B.~Bormar, R.~Joseph, G.-J. Yang, Floating-point roundoff noise
  analysis of second-order state-space digital filter structures. \emph{IEEE
  Transactions on Circuits and Systems II: Analog and Digital Signal
  Processing} \textbf{39}~(2), 90--98 (1992).

\bibitem{Mnih2015}
V.~Mnih, \emph{et~al.}, Human-level control through deep reinforcement
  learning. \emph{Nature} \textbf{518}~(7540), 529--533 (2015),
  \doi{10.1038/nature14236}.

\bibitem{degrave2022}
J.~Degrave, \emph{et~al.}, Magnetic control of tokamak plasmas through deep
  reinforcement learning. \emph{Nature} \textbf{602}~(7897), 414--419 (2022).

\bibitem{jax}
J.~Bradbury, \emph{et~al.}, {JAX}: composable transformations of
  {P}ython+{N}um{P}y programs (2018), \url{http://github.com/jax-ml/jax}.

\bibitem{oord2016wavenetgenerativemodelraw}
A.~van~den Oord, \emph{et~al.}, WaveNet: A Generative Model for Raw Audio
  (2016), \url{https://arxiv.org/abs/1609.03499}.

\bibitem{Bruce:chi_square}
B.~Allen, ${\ensuremath{\chi}}^{2}$ time-frequency discriminator for
  gravitational wave detection. \emph{Phys. Rev. D} \textbf{71}, 062001 (2005),
  \doi{10.1103/PhysRevD.71.062001},
  \url{https://link.aps.org/doi/10.1103/PhysRevD.71.062001}.

\bibitem{2020SciPy-NMeth}
P.~Virtanen, \emph{et~al.}, {{SciPy} 1.0: Fundamental Algorithms for Scientific
  Computing in Python}. \emph{Nature Methods} \textbf{17}, 261--272 (2020),
  \doi{10.1038/s41592-019-0686-2}.

\bibitem{gwpy}
D.~M. {Macleod}, J.~S. {Areeda}, S.~B. {Coughlin}, T.~J. {Massinger}, A.~L.
  {Urban}, {GWpy: A Python package for gravitational-wave astrophysics}.
  \emph{SoftwareX} \textbf{13}, 100657 (2021),
  \doi{10.1016/j.softx.2021.100657},
  \url{https://www.sciencedirect.com/science/article/pii/S2352711021000029}.

\bibitem{Drigg2006}
J.~Driggers, \emph{Optomechanical Alignment Instability in LIGO Mode Cleaners.
  LIGO document T060240-00-R}, Tech. rep., LIGO Laboratory (2006),
  \url{https://dcc.ligo.org/LIGO-T060240/public}.

\bibitem{Tsang2024Hinf}
T.~{Tsang}, \emph{et~al.}, {H-infinity Optimization for Active Seismic
  Isolation Systems in Gravitational-Wave Detectors}. \emph{arXiv e-prints}
  arXiv:2407.15972 (2024), \doi{10.48550/arXiv.2407.15972}.

\bibitem{barrattExampleExactTradeoffs1989}
C.~Barratt, S.~Boyd, Example of Exact Trade-Offs in Linear Controller Design.
  \emph{IEEE Control Systems Magazine} \textbf{9}~(1), 46--52 (1989),
  \doi{10.1109/37.16750}.

\bibitem{boydLinearControllerDesign1991}
S.~Boyd, C.~Barratt, \emph{Linear {{Controller Design}}: {{Limits}} of
  {{Performance}}} (Prentice Hall, Englewood Cliffs, N.J.) (1991).

\bibitem{barrattInteractiveLoopshapingDesign1992}
C.~Barratt, S.~Boyd, Interactive Loop-Shaping Design of {{MIMO}} Controllers,
  in \emph{{{IEEE Symposium}} on {{Computer-Aided Control System Design}}}
  (IEEE, Napa, CA, USA) (1992), pp. 76--81, \doi{10.1109/CACSD.1992.274446}.

\bibitem{diamondCVXPYPythonEmbeddedModeling2016}
S.~Diamond, S.~Boyd, {{CVXPY}}: {{A Python-Embedded Modeling Language}} for
  {{Convex Optimization}}. \emph{Journal of Machine Learning Research}
  \textbf{17}~(83), 1--5 (2016), \doi{10.48550/ARXIV.1603.00943}.

\bibitem{agrawalRewritingSystemConvex2018}
A.~Agrawal, R.~Verschueren, S.~Diamond, S.~Boyd, A Rewriting System for Convex
  Optimization Problems. \emph{Journal of Control and Decision} \textbf{5}~(1),
  42--60 (2018), \doi{10.1080/23307706.2017.1397554}.

\bibitem{seilerIntroductionDiskMargins2020}
P.~Seiler, A.~Packard, P.~Gahinet, An {{Introduction}} to {{Disk Margins}}
  [{{Lecture Notes}}]. \emph{IEEE Control Systems} \textbf{40}~(5), 78--95
  (2020), \doi{10.1109/MCS.2020.3005277}.

\end{thebibliography}
\bibliographystyle{sciencemag}

\subsection*{Acknowledgments}

We thank Jeff Dean for strategic help and inspiration at the start of the project.

\paragraph*{Funding:}

The authors gratefully acknowledge the support of the United States National Science Foundation (NSF) for the construction and operation of the LIGO Laboratory and Advanced LIGO as well as the Science and Technology Facilities Council (STFC) of the United Kingdom, and the Max-Planck-Society (MPS) for support of the construction of Advanced LIGO. Additional support for Advanced LIGO was provided by the Australian Research Council. LIGO was constructed by the California Institute of Technology and Massachusetts Institute of Technology with funding from the National Science Foundation and operates under Cooperative Agreement No. PHY-18671764464. Advanced LIGO was built under Grant No. PHY-18680823459.

\paragraph*{Author contributions:}

RXA, JBu, SC, JH, MR, and BT conceived the project. 
RXA, JBu, CD, AH, JH, and BT led the project. 
TA, RXA, IB, JBu, JBe, CD, GD, JD, AG, JH, ML, BT, GT, and CW developed the physics simulations. 
TA, IB, JBu, YHJC, CD, JD,  ML, BT, and CW integrated the physics simulations with the learning framework.
AA, JBu, RH, SH, ML, WM, and BT developed the learning framework and performed learning experiments. 
CD, TN, JR, and CW developed the real-time neural network interface. 
RXA, JBe, JBu, CD, AG, BT, and CW integrated the real-time neural network with the control system and ran experiments on LLO and Caltech’s 40m prototype. 
CB, JBu, YHJC, CD, YD, OG, ML, and CW developed data curation tools.
RXA, IB, JBu, YHJC, BT, and CW developed and ran the data analysis.
LF, PK, HO, and MR consulted for the project.
TA, RXA, JBu, JH, BT, and CW wrote the manuscript.
The LIGO Instrument Team maintains and runs the LIGO Observatory.

\paragraph*{Competing interests:}
None to declare.

\paragraph*{Data and materials availability:}

The learning algorithm used in the actor-critic RL method is MPO~\cite{abdolmaleki2018mpo}, a reference implementation of which is available under an open-source license~\cite{hoffman2020acme}. Additionally, the software libraries launchpad~\cite{yang2021launchpad}, dm\_env~\cite{dm_env2019}, Jax/Haiku~\cite{haiku2024github}, and reverb~\cite{cassirer2021reverb} were used, which are also open source. Simulations are implemented in Lightsaber~\cite{lightsaber_git}
and advLigoRTS~\cite{advligorts_2021}.
The identified LLO model and experimental data are available at \cite{ligo_scientific_collaboration_2025_15793015}.

\subsection*{Supplementary materials}
Supplementary Text:
\begin{itemize}
\setlength\itemsep{-1em}
\item Members of the LIGO Instrument Team
\item Modeling and Simulation 
\item Machine learning
\item {\tt squidward } RL policy details 
\item {\tt spongebob }RL policy details 
\item Additional results
\item  Baselines
\end{itemize}
Figures S1 to S16\\
Tables S1 to S3\\
References \textit{(27-\arabic{enumiv})}\\

\newpage

\renewcommand{\thefigure}{S\arabic{figure}}
\renewcommand{\thetable}{S\arabic{table}}
\renewcommand{\theequation}{S\arabic{equation}}
\renewcommand{\thepage}{S\arabic{page}}
\setcounter{figure}{0}
\setcounter{table}{0}
\setcounter{equation}{0}
\setcounter{page}{1} 
\renewcommand{\thesection}{S\arabic{section}}
\setcounter{section}{0}

\begin{center}
\section*{Supplementary Materials for\\ \scititle}

Jonas~Buchli$^{\ast\dagger}$,
Brendan~Tracey$^{\dagger}$,
Tomislav~Andric$^{\dagger}$,
Christopher~Wipf$^{\dagger}$,\\
Yu~Him~Justin~Chiu$^{1\dagger}$,
Matthias~Lochbrunner$^{\dagger}$,
Craig~Donner$^{\dagger}$,
Rana~X~Adhikari$^{\ast\dagger}$,\\
Jan~Harms$^{\ast\dagger}$,
Iain~Barr,
Roland~Hafner,
Andrea~Huber,\\
Abbas~Abdolmaleki,
Charlie~Beattie,
Joseph~Betzwieser,
Serkan~Cabi,\\
Jonas~Degrave,
Yuzhu~Dong,
Leslie~Fritz,
Anchal~Gupta,
Oliver~Groth,\\
Sandy~Huang,
Tamara~Norman,
Hannah~Openshaw,
Jameson~Rollins,\\
Greg~Thornton,
George~van~den~Driessche,
Markus~Wulfmeier,\\
Pushmeet~Kohli,
Martin~Riedmiller,
The LIGO Instrument Team \\
\small$^\ast$Corresponding author. Email: buchli@google.com, rana@caltech.edu, jan.harms@gssi.it, pushmeet@google.com\\
\small$^\dagger$These authors contributed equally to this work.
\end{center}

\subsubsection*{This PDF file includes:}
Supplementary Text:
\begin{itemize}
\setlength\itemsep{-1em}
\item Members of the LIGO Instrument Team
\item Modeling and Simulation 
\item Machine learning
\item {\tt squidward } RL policy details 
\item {\tt spongebob } RL policy details 
\item Additional results
\item  Baselines
\end{itemize}
Figures S1 to S16\\
Tables S1 to S3\\

\newpage

\section{Members of the LIGO Instrument Team}

R.~Abbott$^{1}$,  
I.~Abouelfettouh$^{2}$,  
R.~X.~Adhikari\,$^{1}$,  
A.~Ananyeva$^{1}$,  
S.~Appert$^{1}$,  
S.~K.~Apple\,$^{3}$,  
K.~Arai\,$^{1}$,  
N.~Aritomi$^{2}$,  
S.~M.~Aston$^{4}$,  
M.~Ball$^{5}$
% ...
\small$^{1}$LIGO Laboratory, California Institute of Technology, Pasadena, USA 
\\
$^{2}$LIGO Hanford Observatory, Richland, USA 
\\
$^{3}$University of Washington, Seattle, USA 
\\
$^{4}$LIGO Livingston Observatory, Livingston, USA 
\\
$^{5}$University of Oregon, Eugene, USA 
\\
% ...

\begin{figure}
    \centering
    \includegraphics[width=\textwidth]{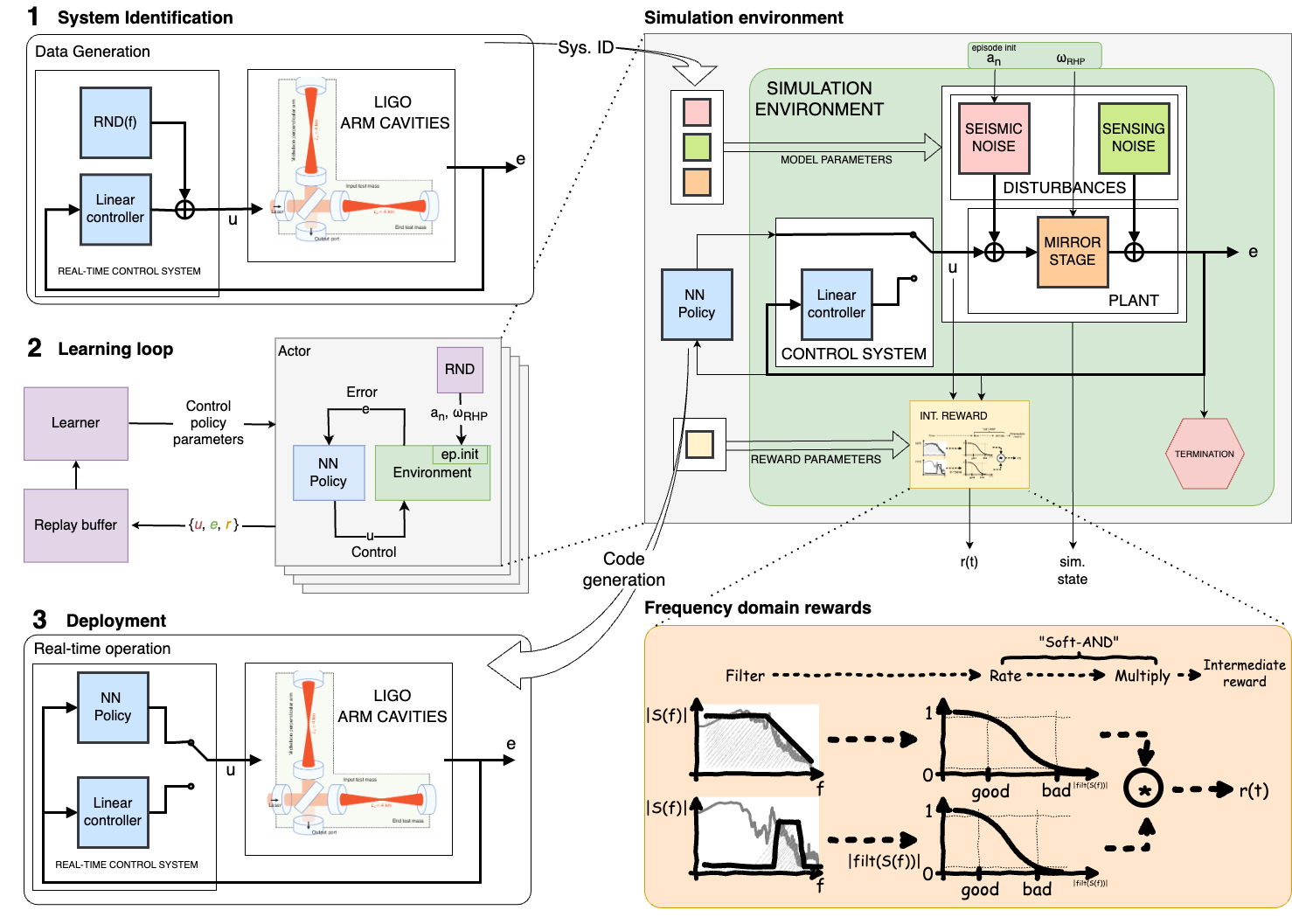}
    \caption{\textbf{Detailed illustration of Deep Loop Shaping.} (1) A model is identified from plant measurements (Section\ref{s:modelling})
    (2) The identified model is used as a learning environment in a distributed actor-critic setup (Section \ref{s:ML}). 
    We show details of the learning environment, and an illustration of frequency-domain rewards to compute intermediate reward r(t) in three main steps: apply a filter, score with sigmoid, multiply for soft-AND (Section \ref{s:a_rewards}).
    (3) The optimized control policy is exported for hard-real-time deployment on the plant using code generation.
    }
\end{figure}

\section{Modeling and Simulation}
\label{s:modelling}
\label{s:units}

To train controllers using a Reinforcement Learning approach, we require a simulation based on a model that is rich and precise enough to capture the relevant dynamics for the agent to learn from, and ideally as computationally efficient as possible.
The results in this work show that linear time-invariant (LTI), single-input-single-output (SISO) models fit these requirements.  In the following, we briefly review the relevant dynamics, starting with the nonlinear, multi-dimensional physics for the ASC problem. We give an indication of why these simpler models are adequate and motivate some of the important limits and control objective specifications.
Some additional results on the Caltech 40\,m prototype input mode cleaner were obtained with a nonlinear MIMO (multi-input-multi-output) time domain simulation (cf. Section \ref{s:40m}), and this section covers these models.

\paragraph{A note on units}

The input and output units on the LIGO real-time control system are somewhat arbitrary Digital-Analog-Converter `counts', and we adopt these conventions for the control policy and the simulator interface, and report our results in these units. This choice is appealing as it truly represents the existing controller interface, and we can abstract the plant and all its interface as a `black box' over which we identify the dynamics and establish the control policy. 

In that light, the system ID of the plant model underlying the simulator is also done in units of counts. Therefore, for modeling and training, we do not need a specific, precise knowledge of the calibration factors that would be required to convert to SI units. In particular, the estimation of the conversion factor of control signal counts to torques is cumbersome and not readily available in a robust way. Therefore, stating SI units would not be meaningful for this quantity. 

For all these reasons,  `counts' are therefore our ground truth and we report our discussion, parameters, and results directly in counts, with one exception. The derivation of the control specification is rooted in first-principle physics modeling, and therefore we use radians for this and use an estimated plant calibration factor to translate into counts (see Table \ref{t:physicstable}). 

\subsection{Modeling and System Identification}

A lot of control design methods, including ours, are dependent on accurate plant models.
Developing accurate models of the LIGO plant is complicated by the challenges of system identification (Sys-Id). Accurate Sys-Id is hindered by time constraints and the limited dynamic range available for experimentation, leading to uncertainties in the models. Furthermore, the plant parameters can exhibit slow time variations due to environmental changes and thermal effects. These factors necessitate a conservative approach to control, relying on manual tuning that is both labor-intensive and challenging to adapt when conditions change, such as during shifts in seismic activity or mirror absorption. Our method improves the workflow in that, once an identified plant model exists, we can fully automate the control design, even taking into account nonlinear plant dynamics and allowing nonlinear control architectures.  
 The coupling between this angular noise and the GW readout is inherently nonlinear, due to the optomechanical couplings between the high-power cavity laser beam and the last two stages of the suspension (cf Fig.~\ref{fig:asc_sketch1} and Section \ref{s:relphysmodellASC}). In our work, we develop and use both nonlinear and linear time-domain models, which include the radiation-pressure dynamics \cite{AndHar2021,And2023}. They were initially validated on the input-mode cleaner of the 40\,m detector prototype at Caltech (cf. Section \ref{s:40m}). To develop controllers for the arm cavities' \chardp{} loop, of the LIGO Livingston observatory, we use an identified linear model (Section \ref{s:SISOmodel}).

\subsection{Relevant physics modeling for ASC control \& derivation of control specifications}
\label{s:relphysmodellASC}
The suspended optics of the LIGO detectors need to be aligned with respect to each other for the laser interferometer to operate robustly at high sensitivity~\cite{Frit1998}. 
The alignment procedure can be divided into two phases: 
an initial alignment to produce first light in the optical resonators, and a continuous high-precision alignment to reach and maintain the high-sensitivity state. 
This study is concerned with the second phase and optimization of detector sensitivity.

\begin{figure}
    \centering\includegraphics[width=0.8\textwidth]{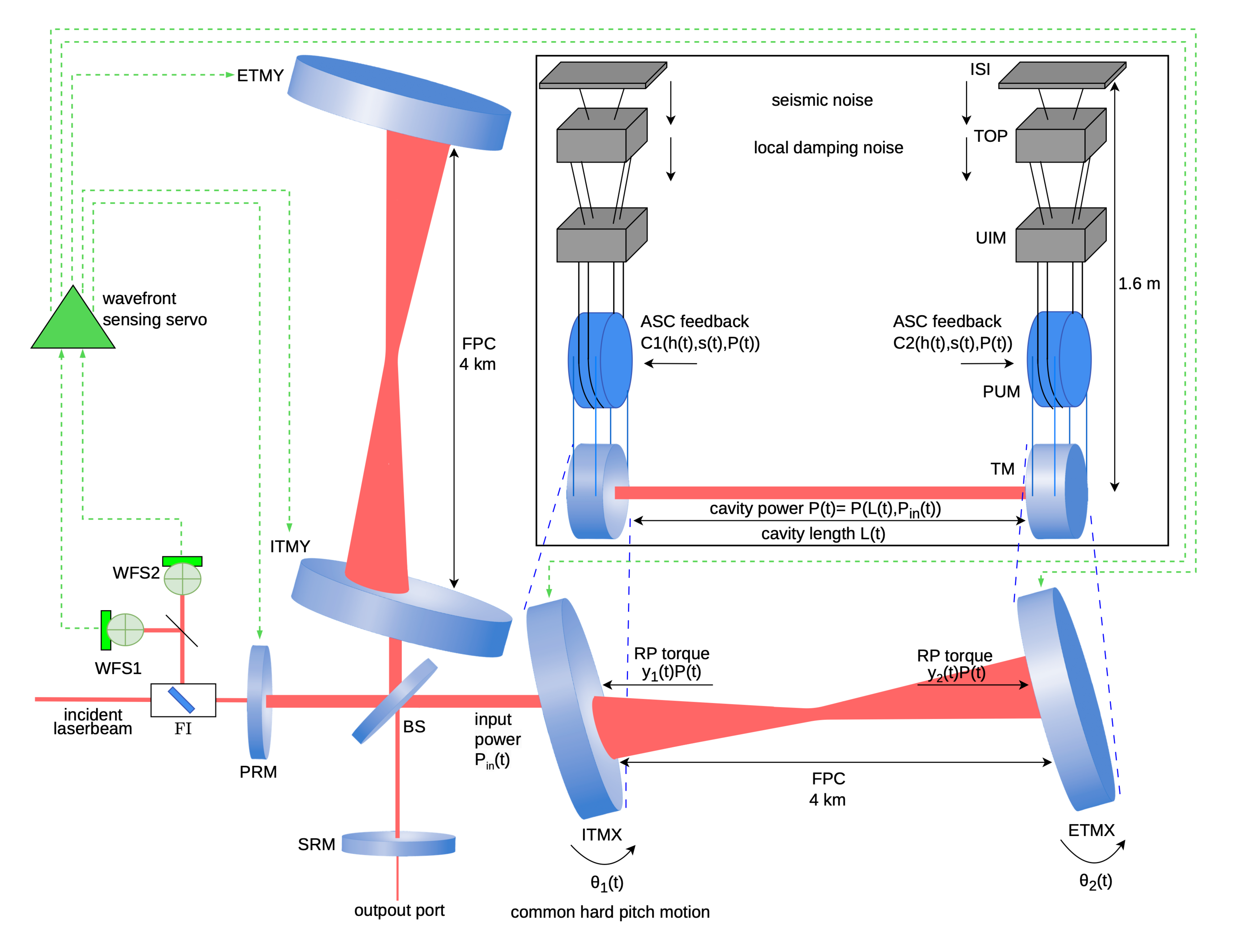}
    \caption{\textbf{LIGO Arm cavity and mirror stages.} The diagram shows the dual-recycled Fabry–Perot Michelson interferometer (IFO) layout with the laser beams, the simulated optomechanical system with mirror suspension system, and the integrated ASC system. ITMX and ITMY represent the input test masses (ITMs) for the x- and y-arms, respectively. Fabry-Perot cavities (FPCs) are established by the ITMs and end test masses (ETMs) in each arm. Additionally, the power-recycling (PRM) and signal-recycling mirrors (SRM), along with the beam splitter (BS) and ITMs, form the power-recycling cavity (PRC) and signal-recycling cavity (SRC), respectively. The Faraday Isolator (FI) extracts the interferometer reflection and sends it to a set of wavefront sensors (WFSs). Common hard (\chardp{}) motion mode is shown. It is one of the geometric modes of the mirrors that form the basis of the control modes. Gravitational wave observatories measure changes in the gravitational field by detecting changes in the interference of a laser beam split into two orthogonal arms of an interferometer. In the top right box, the diagram illustrates the simulated optomechanical system, consisting of the high-power cavity laser beam, the main input noises, the schematic of the quadruple pendulum stage (QUAD) in LIGO, and the associated control system. QUAD is suspended from the Internal Seismic Isolation (ISI) platform. The ASC signal is fed back to the PUM stage. This image is not to scale.}
    \label{fig:asc_sketch1}
\end{figure}

The isolation and control system of the LIGO detectors is highly complex and must protect a sub-attometer displacement measurement against environmental noises that are many orders of magnitude larger~\cite{MaEA2014, MaEA2015, AbEA2016b, aLIGO:PEM:2021}. 
The observed degrees of freedom are correlated and often depend on the global state of the interferometer. 
There are many forms of actuation, including mechanical forces and torques acting on the mirror suspension system, or changing optical parameters like the refractive index and geometry of transparent components. 
A simplified representation of the LIGO angular sensing and control scheme (for \chardp{}) is shown in Fig.~\ref{fig:asc_sketch1}.
The main elements of such a system can be categorised as follows:
\begin{itemize}
    \item Optomechanical dynamics of the plant
    \item Noise inputs: sensor noise, environmental vibrations, diffuse scattered light, actuator quantization (DAC) noise
    \item Controller: sensing matrix, control filter, actuation matrix
\end{itemize}
The test masses (TM), i.e., the main interferometer optics, together with the penultimate masses (PUM), form the final monolithic unit of the LIGO suspensions. The fused-silica TM and PUM are connected by sub-millimeter-thick fused-silica fibers. The lowest-frequency mechanical resonances (pendulum, pitch, yaw motions of the TM) are in the 0.3--1\,Hz band~\cite{AbEA2016b}. 
In addition to the vibration noise entering through the suspensions, radiation-pressure (RP) forces produced by the high-power laser beam act on the TMs \cite{Dool2013}. 
When the beam is miscentered on a TM, the RP force produces a torque on the mirrors. 
The primary RP coupling modeled in the simulations is the torque exerted by light on the suspended mirrors~\cite{Sey2017}
\begin{equation}
\tau_{\rm RP}(t)=\frac{2 P_a(t)}{c} y(t),
\label{eq:torque}
\end{equation}
where $y(t)$ is the beam spot motion, and $P_a(t)$ is the arm-cavity power. 
These optomechanical dynamics couple the angular motion of the two TMs that form the Fabry-Perot cavity, which leads to the introduction of global angular degrees of freedom called the ``hard'' mode (associated with a rotation of the beam axis) and ``soft'' mode (approximately associated with a shift of the beam axis)~\cite{SiSi2006}. 
The torsional stiffness for the soft and hard modes are
\begin{equation}
  \tau_{S, H}= \kappa_{\rm RP} \frac{g_{1}+g_{2} \pm \sqrt{\left(g_{1}-g_{2}\right)^{2}+4}}{2},
\end{equation}
where the plus sign corresponds to the soft mode and the minus sign to the hard mode and
\begin{equation}
  \kappa_{\rm RP}=\frac{2 P_a L_a}{c\left(g_{1} g_{2}-1\right)},\quad g_{1, 2}=1 - \frac{L_a}{R_{\rm ITM, ETM}},
  \label{eq:gfactor}
\end{equation}
\begin{table}[h]
\centering
\caption{List of input parameters and calculated values used in the analysis.} 
\label{tab:parameters}
\begin{tabular}{@{} l c c @{}}
\hline
\textbf{Parameter Name} & \textbf{Symbol} & \textbf{Value with Unit} \\
\hline
Arm-cavity power           & \(P_a\)          & 308 \unit{kW}         \\
Arm-cavity length          & \(L_a\)          & 3994.5\unit{m}     \\
ITM radius of curvature & \(R_{\rm ITM}\)      & 1934\unit{m}\\
ETM radius of curvature & \(R_{\rm ETM}\)          & 2245\unit{m}       \\
Suspension's restoring torque for pitch  & \(\tau_p\)          & 9.72 Nmrad$^{-1}$ \\
TM's equivalent moment of inertia for pitch & \(I_p\)          & 0.76 kgm${^2}$ \\
Mechanical free pitch resonance & \(f_p\)          & 0.57\unit{Hz} \\
Eigenfrequency for pitch soft mode & \(f_S\)          & 0.43\unit{Hz} \\
Eigenfrequency for pitch hard mode & \(f_H\)          & 1.86\unit{Hz} \\
Beam angle-to-offset coefficient for hard mode & \(\left.\frac{\mathrm{d} y}{\mathrm{~d} \theta}\right|_{\mathrm{H}}\)          & 2000\unit{mrad^1} \\
\chardp{} sensor calibration factor &
\(\gamma_{cp}\) &
2.5 nrad/counts \\
Hard mode's sensing noise    & \(\nu_H\)       & $1 \cdot 10^{-13}$\unit{radHz^{-1/2}} \\
\hline
\end{tabular}
\label{t:physicstable}
\end{table}
and the values are given in Table \ref{tab:parameters}. 
With these values and the corresponding g-factors (as you can see in Eq. \ref{eq:gfactor}, it is a dimensionless factor related to the optical cavity's geometric stability~\cite{siegman1986lasers}, where $g_1$ is the g-factor of ITM, and $g_2$ is the g-factor of ETM), the soft mode is not problematic~\cite{BE2011}. 
The angular optomechanics gives rise to the so-called Sidles-Sigg instability, whose control is an important requirement of LIGO's ASC system. 
The optical stability condition for a Fabry-Perot cavity requires  $0<g_{1} g_{2}<1$ \cite{SiSi2006}. 
The eigenfrequencies of each of the optomechanical modes can then be written as:
\begin{equation}
  f_{S, H}=\frac{1}{2 \pi} \sqrt{\frac{\tau_{p}+\tau_{\rm S, H}}{I_p}},
\end{equation}
all the needed parameters and calculated values are given in Table \ref{tab:parameters}. 
As shown in Eq. \ref{eq:gfactor}, the g-factor (and thereby the optomechanical stiffness) depends strongly on the radius of curvature for these marginally stable cavities. The radius of curvature of the mirrors changes due to thermal expansion of the mirror driven by laser beam heating and the LIGO thermal compensation system~\cite{aLIGO:TCS}.
The changes of radii of curvature are up to 15\,m, which results in $\sim$10\

\begin{figure}[ht!]
  \centering
    \includegraphics[width=0.35\textwidth]{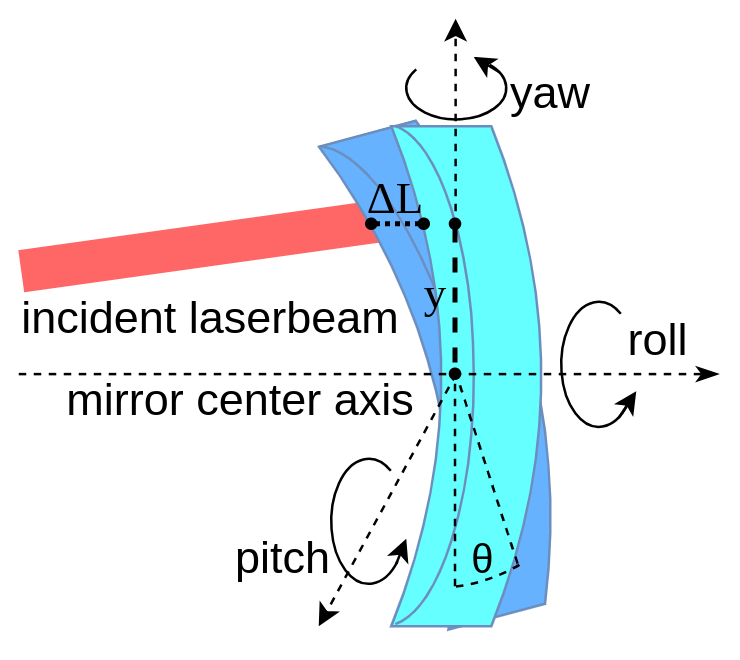}
    \unskip\ \vrule\   \includegraphics[width=0.63\textwidth]{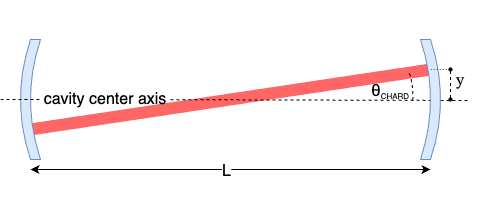}
  \caption{(left) Angle-to-length coupling due to beam-spot miscentering and visualization of the roll-pitch-yaw angles for the TM (right) Illustration of the cavity angle and related geometry used to derive control specifications.}
  \label{fig:cavity_axis_angle}
  \label{fig:dl}
  \label{fig:pitchyaw1}
\end{figure}

Angular motion of the TMs leads to a change in distance between the two TMs as seen by the laser beam, which means that it becomes a limitation of the LIGO sensitivity to GWs. Actually, the most important nonlinearity to be addressed in angular controls is that the strain noise coming from angular motion is a product of beam-spot motion and angular motion. The beam-spot motion is relatively slow (mainly below 0.5\,Hz), while angular motion is relevant at frequencies higher than 10\,Hz \cite{Bar2010, Sey2017, Bui2020}. 
One of the problems here is that the frequency components of spot position and angular motion beat, creating hard-to-subtract noise. 
The formula describing this bilinear process in the time domain using local angles is \cite{BE2011}
\begin{equation}
  \Delta L_a(t)= y(t) \times \theta(t).
  \label{eq:deltaL}
\end{equation}
This coupling is easy to understand geometrically, as shown in Figure \ref{fig:dl}. 
If the beam-spot position does not coincide with the mirror’s axis of rotation, a length signal is created \cite{Sey2017}. 
To evaluate the angular noise coupling into strain noise, we calculate and add the length variation contributed by each mirror. 
Given the long arms of the interferometer, even slight angular motion of one test mass (TM) can produce substantial beam-spot movement on the opposite TM. 
These values are also given in Table \ref{tab:parameters}. 
This coefficient is defined as follows \cite{Hang2019, AndHar2021},
\begin{equation}
  \left.\frac{\mathrm{d} y}{\mathrm{~d} \theta}\right|_{\mathrm{S, H}}=\frac{L_a}{2} \frac{\left(\mathrm{g}_{\mathrm{2}}+\mathrm{g}_{\mathrm{1}}\right) \pm \sqrt{\left(\mathrm{g}_{\mathrm{2}}-\mathrm{g}_{\mathrm{1}}\right)^{2}+4}}{\left(\mathrm{~g}_{\mathrm{2}} \mathrm{g}_{\mathrm{1}}-1\right)},
\end{equation}
having in mind the RMS of hard mode angles of 250\,nrad, the corresponding beam-spot motions are approximately 4.5\,mm. The resulting beam-spot motion places a direct requirement on the accuracy of the error signal, as ideally, this motion is driven solely by the error signal.
However, one of the primary metrics for assessing the efficacy of the ASC system is its contribution to differential arm length (DARM) fluctuations. Since DARM is the most sensitive degree of freedom to gravitational waves, the ASC must be optimized to minimize its noise contribution \cite{Alloca2020}. \\
The target is to balance the suppression of beam-spot motion and angular noise such that the product of these two factors remains below the DARM noise threshold.
In practical terms, the 250\,nrad requirement on the error signal alone is driven not by the achievable gain in the hard-mode control loop. Systematic offsets in some of the other angular and beam centering control systems leads to low-frequency beam spot motions of $\sim$1\,mm. So, reducing the error signal RMS below this level has diminishing returns.
Instead, to meet DARM’s sensitivity goals, we aim to balance error suppression across angular modes. Insufficient control of either mode could lead to instability, while excessive control could introduce excess noise. Since DARM and ASC interactions are essentially nonlinear, achieving this balance is crucial as coherence between DARM and ASC signals is generally low. Our RL control approach ensures that residual DARM motion is suppressed to meet the interferometer’s sensitivity needs. \\
The calculation of beam-spot motion is done through this equation
\begin{equation}
  \mathbf{y}=\left[\begin{array}{l}
    y_{ITM} \\
    y_{ETM}
    \end{array}\right]=\frac{L_a}{1-g_{1} g_{2}}\left[\begin{array}{cc}
    g_{2} & 1 \\
    1 & g_{1}
    \end{array}\right]\left[\begin{array}{l}
    \theta_{ITM} \\
    \theta_{ETM}
    \end{array}\right].
\label{eq:bsm}
\end{equation}

\subsubsection{Derived control specifications}
\label{s:control_specs}

In summary, the design goal of the control system is to minimize the effective length change in the optical path in the mirror cavity. As seen in Eq. \ref{eq:deltaL}, the effective length caused by the mirror angles has two major components that multiply together in a bilinear coupling, namely (1) the beam spot motion $y(t)$ and (2) the mirror angle $\theta(t)$. The first term is mostly influenced by the low-frequency errors, and the second term by motions in the observation frequency band of interest. 

\paragraph{Derivation of total error RMS specification}

We derive the error RMS specification starting from the bilinear coupling in Eq. \ref{eq:deltaL}.
Looking at Fig.~\ref{fig:cavity_axis_angle} it's easy to see that
\begin{equation}
    \Delta y = \frac{L_a}{2} \Delta \theta
\end{equation}
combining with Eq. \ref{eq:deltaL} therefore
\begin{equation}
  \Delta  L_a(t)= \frac{L_a}{2} \left( \Delta \theta (t) \right)^2
  \label{eq:lattheta2}
\end{equation}
The angle error can be split into three components
\begin{equation}
    \Delta \theta = \Delta \theta_{sys} + \Delta \theta_{LF} +\Delta \theta_{HF}
\end{equation}
where
 $\Delta \theta_{sys}$ stems from offsets in some centering sensors in the system and is effectively about 250 \unit{nrad},
 $\Delta \theta_{LF}$ is unsuppressed seismic noise (mostly 0-1\,\unit{Hz}) and 
 $\Delta \theta_{HF}$ are the errors in the GW observation band.
 The first two errors are fairly large, whereas the third one is tiny.
 
 We can now express
 \begin{equation}
     \left(\Delta \theta \right)^2= 
     \Delta\theta^2_{sys} +       \Delta\theta^2_{LF} +
     2 \Delta \theta_{sys} \Delta \theta_{LF} +
     2 \Delta \theta_{sys} \Delta \theta_{HF} +
      2 \Delta \theta_{LF} \Delta \theta_{HF} +
      \Delta\theta^2_{HF} 
 \end{equation}
The first three terms are approx. DC components and are compensated by the DARM control loop. 
The last square term is small and negligible. 
The two terms that matter for the GW band strain are $2 \Delta \theta_{sys} \Delta \theta_{HF}$ 
and $2 \Delta \theta_{LF} \Delta \theta_{HF}$.

For the HF performance of the controller, we want therefore $\theta_{LF} < \theta_{sys} = 250\,\unit{nrad}$ which establishes our error RMS control specification for the linear controller. 
The length change specifications are $\Delta L < 10^{-20}\,\unit{m}$ and using Eq. \ref{eq:lattheta2} we have therefore a spec for the GW band contributions of $\Delta \theta_{HF}<10^{-16}\,\unit{rad}$ in the 10--30\,Hz band.
In that band, we also require the $\Delta \theta_{HF}$ to contribute strain noise below the quantum limit and a 10x design margin, which is the more stringent specification, and therefore the one we use and illustrate in the results. 

\paragraph{Derivation GW observation band control specification}
In order to derive a goal (cf. Fig.~\ref{fig:rl_vs_lin_strain_coords}) for the control noise PSD, we use the estimated quantum limit and \chardp{} to strain coupling function from the LIGO O4 noise budgets \cite{O4InstrumentPaper}. 
We illustrate this limit in the presentation of our results, e.g. Fig.~ \ref{fig:control_error_spectra}.

\paragraph{Summary of control specifications}
In summary, quantitatively, the design goals for the RL control policies are:

\begin{itemize}
    \item Based on requirements for beam spot motion on the mirror, the root-mean-square (RMS) of the AC component of the mirror cavity axis angle must be smaller than $250\,\unit{nrad}$ to limit the bilinear coupling.
    \item The injected control noise in the observation band (10--30\,Hz) should remain below the quantum back-action limit for a 40\,\unit{kg} mass and 1\,\unit{MW} of laser power with a given design margin of 10x.
    In practice, the second requirement translates into lowering the control noise in the observation band by at least an order of magnitude in comparison with the currently operational controller (cf. Fig.~\ref{fig:rl_vs_lin_strain_coords}).
\end{itemize}

\subsubsection{Sensing system}
The fundamental objective of the ASC system is to suppress the angular motion of the mirrors. In addition to the seismic disturbance, it must also mitigate instabilities caused by radiation pressure, all while avoiding the reintroduction of noise into the observational band \cite{Bar2010, Hang2019}. For angular sensing, quadrant photodiodes (QPD) are used to measure the beam positions. Some of the quadrant photodiodes are also demodulated at RF frequencies corresponding to other fields in the interferometer; this RF QPD system is called a wavefront sensor (WFS) \cite{Morrison:94}.
Multiple WFS and QPD sensors are combined using a fixed matrix to diagonalize the system in the opto-mechanical eigenmode basis. 
The analog WFS signals are passed through a spectral whitening filter before being digitized. Below $\sim$30\,Hz, the sensor noise is a combination of photon shot noise, quantization noise in the ADC, and seismic/acoustic vibrations of the sensor. The approximate high-frequency sensing noise levels are given in Table \ref{tab:parameters}. The soft mode loops have a smaller bandwidth than the hard mode loops, resulting in less noise contribution to the GW channel \cite{Dool2013, Alloca2020, Hang2019, AdvLigo2015}.

The collected signal is filtered and fed back to the Penultimate Mass (PUM) via four electromagnetic actuators, which produce torque to align the mirrors. 
This process, including the final two stages of suspension in LIGO, is illustrated in Fig.~\ref{fig:asc_sketch1}.
Actuating on this upper pendulum stage also has the added benefit of mechanically filtering out the quantization noise in the actuator's DAC.

\subsection{Linear Angular Control}
\label{s:linearASC}
To maintain its high-sensitivity resonant condition, the IFO must correct for angular motions, typically achieved through linear negative feedback designed using a combination of intuition and quantitative methods. Designing a stable filter that effectively suppresses noise across a wide range of input power levels is a significant challenge. The controller, functioning as a linear filter, adopts a low-pass filter shape with the Unity Gain Frequency (UGF) tuned to reduce mirror motion at lower frequencies (below 1\,Hz). A steep cut-off filter is necessary to prevent the injection of sensing noise into the observational band, albeit introducing phase lag (and potential instability) with every pole used to achieve a steeper drop-off. The downside is that the control filter does not roll off quickly enough to satisfy LIGO's noise requirements in the 10--30\,Hz range. This trade-off between servo stability and sensing noise presents a limitation on the achievable loop bandwidth \cite{Dool2011, Mart2015}. 
While controlling the soft mode is relatively straightforward, managing the unstable hard mode presents a greater challenge, as its natural frequency increases with power, potentially rendering the dynamics of the closed control loop unstable \cite{YuEA2017}. Below $\sim0.1$\,Hz, most of the control is offloaded to the upper stages of the mirror suspension system.

To summarize, LIGO's ASC system currently uses linear feedback error controllers, which are typically designed ``by-hand'' to satisfy a number of requirements:
\begin{enumerate}
    \item Control band RMS below 250\,nrad
    \item Minimize GW observation band noise injection
    \item stability during thermal changes of the optomechanical plant
    \item suppression of rare, large outliers: this may be large impulses nearby, small earthquakes across the world, or a slow increase in microseism beyond the 3$\sigma$ level.
\end{enumerate}

\subsubsection{Loop requirements and current performance of the linear \chardp{} controller}
\label{s:linctrlperf}

The control challenge in the \chardp{} loop stems from a resonant eigenmode of the opto-mechanical dynamics ('optical spring') caused by the light-pressure induced torque and the demanding loop shaping specifications where a greater than 200\,dB/decade roll off in the control spectrum is required (cf. Fig.~\ref{fig:control_error_spectra}).
The difficulty of achieving these specifications is reflected by the current linear control design, which, despite more than a decade of improvements,  approaches but does not fully meet these specifications. The control performance is close to the limit of robust stability with $\approx$15--20\,deg phase-margin and 6\,dB gain margin, and further improvements with traditional methods are increasingly hard to come by.
So far, sensitivity improvements related to the control systems noise in the low-frequency band had to be achieved by hard commissioning work with gradual progress over the course of years \cite{AdvLigo2015,Cahillane_2022}.
In addition, to ensure reliable operation, the LIGO control system is designed to prioritize stability over optimal performance. 
Unlike typical laboratory experiments, where short-term optimal control is feasible, the need for continuous astronomical observations means that reliability is paramount. 
As a result, the control loops are intentionally sub-optimal, favoring a stable, ``safe'' configuration that avoids the risk of destabilizing the system during observation runs.

\subsubsection{Modeling of the Input Mode Cleaner at the Caltech 40\,m prototype}
\label{s:IMC-model}

To verify our deployment methodology, we used a subsystem of the Caltech 40\,m prototype, the so-called Input Mode Cleaner (IMC) \cite{Wardeal2008, tanioka2020angularresponsetriangularoptical}. 
The IMC is a 3 mirror ring cavity, with a 25\,m perimeter. This system was used for initial prototyping -- it is well-understood, has a high uptime, and was generally available for controls prototyping.
For the IMC we used a nonlinear MIMO model of the plant based on existing models and identified plant characteristics implemented in Lightsaber \cite{AndHar2021}. 
The model is described in more detail in \cite{And2023}.

The Lightsaber-IMC serves as the time-domain simulator for the ASC system of the IMC at the Caltech 40\,m prototype facility. 
It incorporates linear couplings based on state-space models, encompassing aspects such as the suspension system, local damping, and feedback angular controls. 
The primary mechanical degree of freedom simulated in Lightsaber-IMC is the pitch motion of the IMC mirrors. Along with static models, nonlinear optomechanical couplings are incorporated. 
The bilinear noise coupling for the IMC is essentially the same as for the main LIGO IFO.
The readout of the mirrors' pitch motion involves sensing/electronics noise, with the main feedback control filters implemented in the sensor basis.
Filtered signals from control feedback and local damping loops are directly fed back to the suspended mirrors \cite{And2023}. 
To prevent an unnecessarily large dimension of the second-order section models, these input noises are simulated using spectral methods. The IMC has an input power of 1\,W, which gives an average cavity power of 500\,W.

\subsection{Linear SISO Model}
\label{s:SISOmodel}
For training the \chardp{} policies for the LLO arm cavities, we used a linear, time-invariant (LTI) model, converting the dynamics from continuous time into discrete time using the Modified Matched Pole-Zero bilinear transformation~\cite{franklin2002feedback} where required.
The time-domain model of the opto-mechanical plant includes the dynamics of the quadruple pendulum (modified by the radiation pressure dynamics), the optical gain of the sensors, and the baseline linear controller.
The simulation further contains two independent, non-white, Gaussian noise processes, representing the seismic disturbance and the sensor noise, respectively. 
The parameters for the system and the noise are identified from data (see next section).
The mechanical pendulum model is based on a state space model \cite{barton2010models}, which has been fit and verified against stand-alone measurements of the suspensions, including the noise of the sensors \cite{shapiro2014noise}. 
The \chardp{} degree of freedom includes the four mechanical suspensions corresponding to each of the 4 test mass mirrors. 
For this SISO model, we represent the 4 mirror opto-mechanical plant with a single transfer function (cf. Section \ref{s:relphysmodellASC}). 
The frequency responses of the sensors and actuators are nearly independent of frequency from 0--100\,Hz, so we model them as simple gain blocks.
For simplicity, we have modeled the actuator as a fourth-order system, rather than including the full pendulum dynamics. 
Below $\sim$0.05\,Hz, the feedback control signal is offloaded to the upper stages of the suspension. This discrepancy limits the accuracy of our simulation, and therefore the NN policy, below $\sim$0.1\,Hz.
The only other non-modeled component of note is the overall time delay (or 'latency') due to the control system, which is estimated at $\sim$1\,ms and can therefore be neglected for our purpose.  

The simulated noise sources are implemented by generating random noise from the inverse FFT of the noise PSD identified in Section \ref{s:sysID}. This results in a time series where the noise in each frequency bin has a Gaussian distribution.

\begin{figure}
    \centering
    \includegraphics[width=0.8\columnwidth]{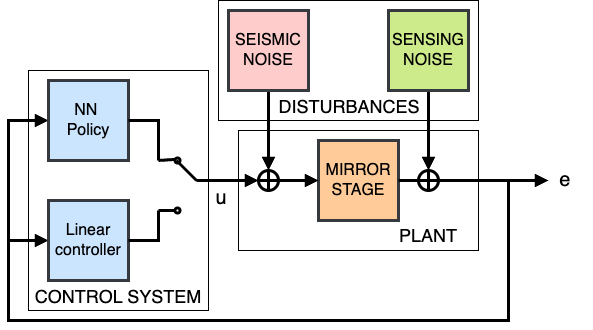}
    \caption[SISO Model]{The \chardp{} degree of freedom is modeled as a SISO system. The opto-mechanical plant is subjected to Gaussian noise processes as force and sensor noise. The noisy sensor output is then fed to the baseline linear controller and the NN policy. Either output can be selected to control the interferometer.}
    \label{fig:siso_model}
\end{figure}

\subsection{System Identification}
\label{s:sysID}
The transfer function of the physical plant is measured with the linear feedback loop closed; the common hard modes of the LIGO interferometers are not open-loop stable.
Frequency-weighted random noise is driven into the actuator, and the frequency response is estimated.

\begin{figure}
    \centering
 \includegraphics[width=0.75\columnwidth]{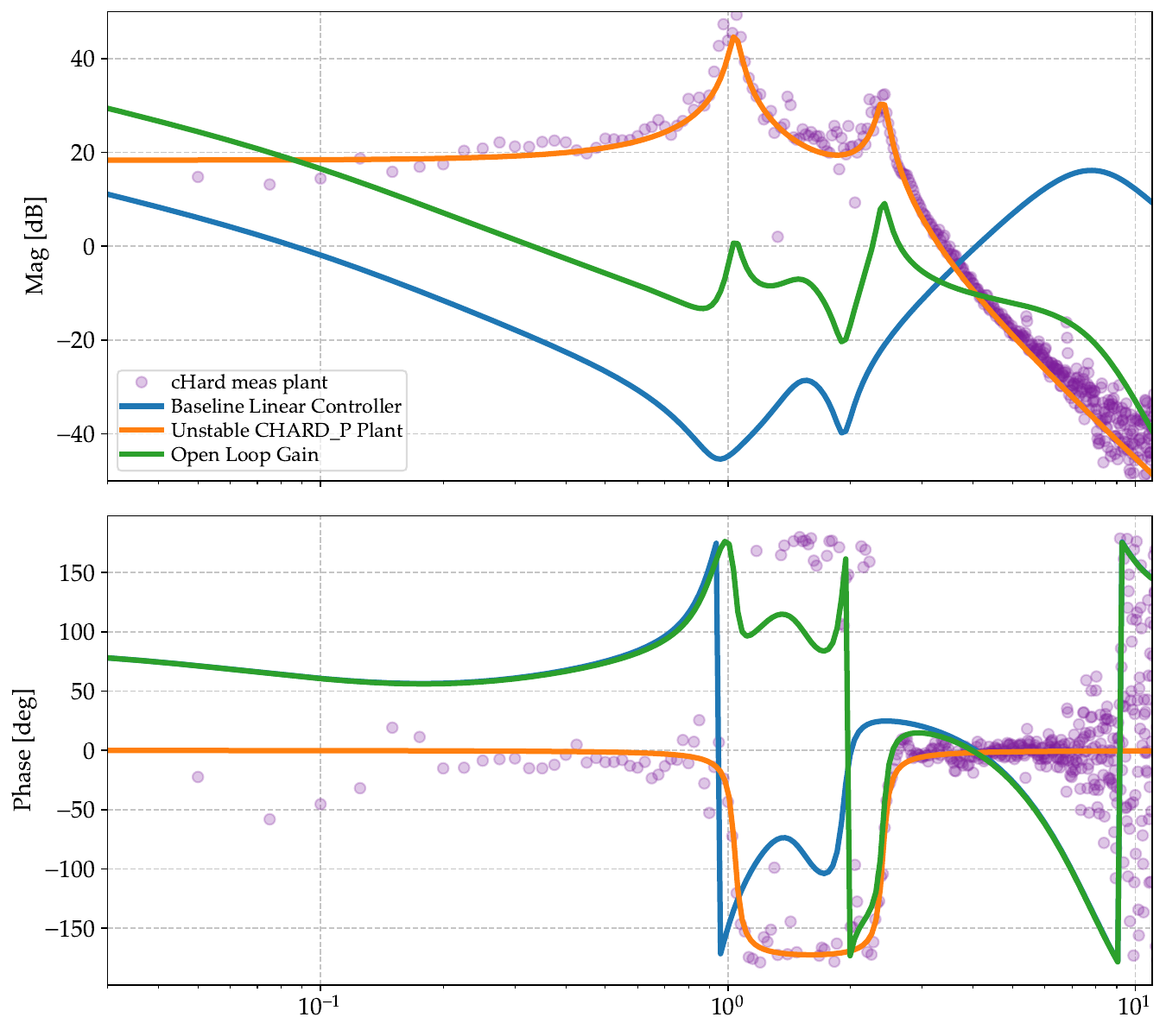}
 \caption[CHARD$_P$ Linear Control]{Transfer functions of the (modeled) opto-mechanical plant, linear controller, and total open loop gain. Dots represent the SysId measurement of the plant. The ``hard'' mode resonance at $\sim$2.3\,Hz is unstable as can be seen from the phase lead around the resonance. The linear controller has multiple unity gain crossings in the 1--3\,Hz band due to resonances in the plant which are not inverted in the controller. The discrepancy between measurement and model below 0.1\,Hz is due to the upper stages of the pendulum, which are neglected in this simple actuator model.}
    \label{fig:chardp_olg_model}
\end{figure}

In order to design a more optimal controller, it is beneficial to have a precise model of the plant. In order to make this measurement with high precision, one must overpower the ambient noise sources or integrate for a long time.
The LIGO system is fragile in the sense that one must be careful while measuring to avoid driving the interferometer into a nonlinear regime. Overdriving the system weakly can bias the measurement. Over-driving more severely can cause the interferometer to move out of resonance with the laser field, necessitating a $\sim$30--90\,min servo lock reacquisition sequence. The result is a $\sim$20\

The strength of the angular opto-mechanical springs also varies with time. The torsional stiffness depends on the circulating laser power and the radius of curvature of the mirrors. The curvature of these mirrors is driven by thermal gradients in the mirrors. Roughly 1\,ppm of the incident laser power is absorbed in each mirror. The thermal equilibrium time constant for these changes is a few hours.  In order to have accurate plant models, it would be necessary to allow the system to come to equilibrium and then take some long measurements. These measurements are not compatible with astrophysical observations, and so the time available for them is limited.
As a result, we must design controllers able to handle substantial plant variations.

\begin{figure}
    \centering
 \includegraphics[width=0.75\columnwidth]{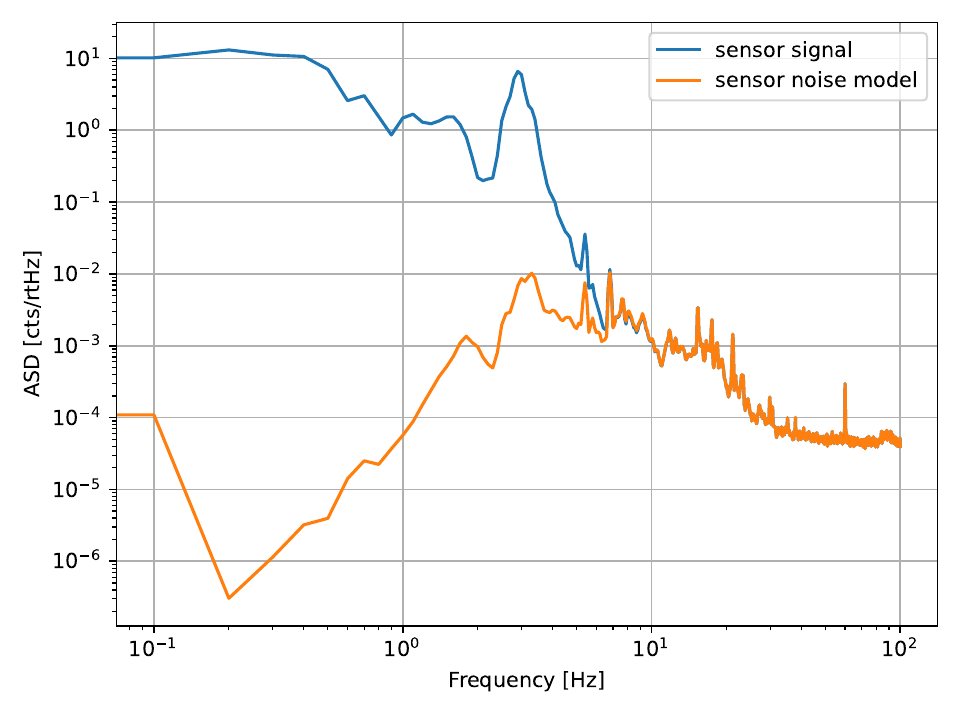}
 \caption[CHARD$_P$ Sensor Noise]{Simulated noise spectra at the \chardp{} sensor. At frequencies larger than $\sim$8\,Hz, the dominating noise is assumed to be from the sensor.}
    \label{fig:chardp_sensnoise}
\end{figure}

The noise affecting each of the angular loops can be separated into two categories:
(i) acceleration noise, and
(ii) sensing noise.
The acceleration noise sources are mainly external seismic disturbances
(filtered by the LIGO seismic isolation platforms \cite{Matichard_2015})
and cross-coupling from other control loops.
The sensing noise is mostly due to diffuse light, photodetector electronics, vibration of the sensors, and photon shot noise.
For purposes of simulation, we use a simple cutoff frequency, $f_{cut}$, assuming all noise below $f_{cut}$ is an external acceleration noise, and that all of the noise above $f_{cut}$ is from the sensing (as shown in Fig.~~\ref{fig:chardp_sensnoise}).
Our noise modeling of the system suggests that this is a fair approximation.
The sensor noise is not white noise, but rolls up below 10\,Hz due to some DAQ signal conditioning electronics, and the peak around 20\,Hz is due to vibration of the sensor itself. 
We then use the servo error and control noise PSD to estimate the acceleration and sensing noise PSD for the simulation using the following loop algebra:
\begin{align}
    x_{seis} = x_{ctrl} \times \frac{1 - G}{G} && x_{sens} = x_{err} \times (1 - G)
\label{eq:loop_undo}
\end{align}
where $x_{ctrl}$ and $x_{err}$ are the control and error signals, respectively, of the feedback loop, and $G$ is the open-loop gain of the feedback loop \cite{astrom_feedback_2021}. The noise generated in this way is random and Gaussian and does not capture the non-Gaussianity present in the seismic and acoustic noise.

\section{Machine learning}
\label{s:ML}

\subsection{Loop shaping as Reinforcement Learning problem}

We assume that we can approximate the system dynamics using a Markov-decision-process (MDP) for which we solve the continuous state action optimal control problem of the form
\begin{eqnarray}
    \pi^{*} = \argmax_{a \sim \pi} \, J = \mathbb{E}_{\pi} \left[ \sum_{t=0}^{\infty} \gamma^t r(s_t, a_t, s_{t+1}) \right]  \\
    \textrm{subject to} \quad p(s_{t+1} = s' | s_t = s, a_t = a) = f(s', s, a)
\end{eqnarray}
where $r(s_t, a_t, s_{t+1}) = r(t)$ is a suitable chosen intermediate reward,  $\gamma < 1$ a discount factor, and $f(s', s, a)$ and is the (generally unknown) stochastic system dynamics. We make the assumption that the system is sufficiently close to being Markov by using an observed state consisting of N previous observations $s' = s_t, \ldots, s_{t-N}$, where $s'$ are measured ('observed') quantities of the real system (or their simulated counterparts). Values of N are listed in the results.

\subsection{Angular control problem as RL environment }
\label{sec:asc_rl_env}

To build the simulation and the training environment used in the RL framework, we use the discrete time model derived in Section \ref{s:SISOmodel}. In order to use the time domain model to train RL policies, we extend it by adding clipping at action limits (+/- 3000 counts), facilities to randomize key parameters, and termination on large errors.
The control policy is evaluated at 256\,Hz and provided to the integrated model with a simple sample-and-hold. The model is integrated by stepping the discrete time model at a frequency of 2048\,Hz for numerical stability. An optional additional pure time delay at the input is used in some cases, e.g. in the simulation of the 40m IMC plant.
The dynamic range needed to represent quantities within the LIGO control system and observatory dynamics exceeds that of the Direct Form-II recipe usually used for digital filters, and so a low-noise biquad implementation (cf. Appendix B of \cite{Mart2015} and Eq. 51 of \cite{smith1992floating}) is used to reduce the amount of signal-dependent round-off noise.
The simulation is stepped in closed loop with the control policy for a set number of steps (typically 262144 steps, corresponding to 1024\,s) to complete a so called episode. The episode is stopped early if a termination criterion is met (e.g., errors larger than a critical level).

\subsection{Distributed Actor-Critic learning}

We use a state-of-the-art distributed actor critic method to train the policy based on the maximum a posteriori policy optimization (MPO) algorithm with key metaparameters listed in Table \ref{tab:mpoparams}.

In this setup, a number of `actor processes' are generating experience data by running exploration policies against the simulator introduced in Section \ref{sec:asc_rl_env}. The trajectories generated by these simulations are stored in a so-called replay buffer, from where a `learner' draws trajectory chunks at random to optimize the policy parameters using the MPO algorithm. The exploration policies used by the actors are regularly updated to be close to the currently best policy found by the learner (cf. Fig.~\ref{fig:concept_sketch}).

The deep actor critic method uses two neural network based function approximators to represent the `state action value function Q function' (`critic') and the policy (`actor').
Of these, only the policy has to run at the time of deployment and as such is subject to real-time limitations as well as the limitation to only rely on data available as measurements on the real plant. The critic, on the other hand, has no such limitations, it can thus be much larger (i.e., an asymmetric actor-critic setup) and can, in principle, consume privileged information. 

We typically use 50 actor processes running on CPU, in conjunction with 1 learner process running on a GPU or TPU (e.g., Nvidia A100) per experiment metaparameter configuration (`seed').

\subsection{Critic and policy network architectures}
\label{s:networks}

We use a multi-layer-perceptron (MLP) with a dilated convolution input layer for the policy network. MLPs do not have state (e.g., long-term memory), which helps for hardware transfer as it limits the ability to memorize long-term simulation effects. In addition, we use a small MLP so it is fast to compute and can meet the 2048 \unit{Hz} timing requirement.
A disadvantage to MLPs is that they can struggle to learn dynamics. Therefore, we choose to represent the critic network with a Long-Short-Term-Memory network (LSTM) alongside input and output MLPs. Only the policy network is needed for deployment, so the extra power in the critic helps to fit an accurate value function without harming transfer.
We use ``stacking'' for both the policy and critic networks, where the past N observations are sent as one to the network input. Standard RL theory assumes that the system can be modeled as a Markov-Decision-Process (MDP). While this assumption often fails for practical systems, improving the system observability typically helps learning.
We believe that the requirement for the length of history in our case, which is notably 1-2 orders longer than typical RL control applications (e.g., in robotics), is dictated by the signal processing needs rather than, directly, the memory properties of the physical noise processes in the interferometer. 
Furthermore, given that the policy network has no state, and the observation is not rich enough to reconstruct the Markov representation of the system, the measurements need to be augmented with a finite history of past measurements. We found the required stacking length empirically.

The critic network typically consists of an LSTM with a single hidden layer with 512 nodes, the LSTM is pre-pended with an encoder of size [128, 256, 512] and appended with a decoder of size [512, 512, 256, 2], which produces the estimated Q value. For the actor network, we used a variety of architectures, listed below.

The policy and critic network consume the (angular) errors and the output of the policy of the last step. The policy network produces control actions to be sent to the simulated environment or the LLO real-time control software via the standard LIGO real-time control software stack \cite{advligorts_2021}. The observations are scaled and stacked in-graph. The networks have a scaling layer that scales the inputs to be in the order of 1. The outputs are limited with a hyperbolic tangent to be in [-1, 1] and we then apply a scalar scaling to map it to [-2700, 2700] counts.

\subsection{Rewards}
\label{s:a_rewards}

Even though we use nonlinear control policies, from considerations motivated by control theory, i.e., considerations of sensitivity functions and Bode's sensitivity integral, we do not expect that we can suppress the control noise over the whole frequency spectrum arbitrarily. We therefore define a band of frequencies of primary scientific interest in the measurement problem, where we prioritise low noise performance. In our case this is the band where GW events of interest are expected to occur. In that light, we define and choose an `observation' band corresponding to 8--30\,Hz. We assume it is in this band only that suppression of the control noise really matters and we attempt to leave the control signal in the other bands as unrestricted as possible (however, see the high frequency auxiliary penalty below).

In RL the reward is the sole way to specify the desired property of the learned control policy. We need to formulate rewards that ensure that the solutions fulfill the control performance specification in Section \ref{s:control_specs}.
To compute the intermediate reward $r(i)$ at step $i$, we use reward components that score how well certain desired properties are fulfilled. These components have values in [0, 1], where 0 indicates a timestep where the property is not at all fulfilled (`bad') and 1 a solution where the property is fulfilled at that timestep (`good'). We compose the reward components from a function computing a frequency domain property of interest, in our case linear filters, and a nonlinear transform to ensure output between [0,1]. 
\begin{equation}
    c_i = s(|f_i(t)|,g,b)\,
\end{equation}
where $f$ is the output of the filter (e.g., an IIR bandpass filter), $s$ is the nonlinear transform.

For the nonlinear transform, here we use the scaled sigmoid to `squash' the output of the filter to lie within [0, 1]. The scaled sigmoid takes two parameters $g$ (`good') and $b$ (`bad') at which we would like the output to be at 0.95 and 0.05, respectively.
\begin{align}
    c = s(v)(g,b) &= \frac{1}{1+ e^{-k(v)}(g,b)} \\
    k(v) &= l - (v -b) \frac{l-h}{g-b}
\end{align}
where l,h  are $ + / - ln(19)$ respectively to satisfy the anchoring of $b$ and $g$ at 0.05 / 0.95 respectively.

The single reward components are multiplied to yield the total intermediate reward $r(i)$. Intuitively, one can think of this combination as a soft-logic AND; i.e., we want all single properties to be fulfilled for a good overall score. At the input of the sigmoid, we use a suitable function extracting the signal property of interest. As we are interested in scoring frequency domain properties, in our case, these are a variety of filters to extract signal energies in chosen bands.
We use the following scoring function reward components:
\begin{description}
\item [``RMS penalty'':] This reward aims at pushing the RMS of the signal below a certain level. To do so we are penalizing the absolute value of the error at each timestep (i.e., the scoring function is an all-pass):
\begin{equation}
c_e(i) = s(|o(i)|)(g,b)
\end{equation}

\item [``Band penalty''] With this component, we aim at penalizing the signal energy in a given band by using a band-pass filter
\begin{equation}
c_{MB}(i) = s(f_{BP}(o(i)))(g,b)
\end{equation}
There are a wide variety of filter implementations and settings that can be used here, and we list details for the experiments in the result sections. The general consideration is that the pass-band SNR is high enough such that the band pass is selective enough for the given signal, taking into account the roll-off of the signal. To be more specific, in our problem, the observation band energy of the control signal is in the order of 4-5x lower than in the control authority band (<3Hz). Therefore, for the BP to be selective, we require a stop-band gain of -120\unit{dB} or more.

A variant of this penalty uses a low-pass filter if the signal below a given cut-off frequency should be penalized.

\end{description}
One or several such components can be used with varying settings and properties to produce $N$ single components $c_n$.
Finally, as outlined above, we calculate the total intermediate reward as the product of the component rewards
\begin{equation*}
    r(i) = \prod_n c_n(i)
\end{equation*}
The choice of elementary components, the `bad'/`good' level setting, as well as some filter-specific details vary over the presented results, and we therefore list these values in Section \ref{s:results}.

\paragraph{Filter choice and design} In principle there are a wide variety of filters that can be used to implement frequency domain rewards. Important considerations are numerical stability, filter order relative to achieved signal-to-noise ratio, band selectivity, among others.
We use the sums-of-squares implementation of linear infinite impulse response filters throughout our work. The type of filters used are Elliptic and Chebyshev, designed using SciPy's {\tt iirdesign} function.
Furthermore, for the observation band filter, we need enough difference between stop and passband, as otherwise, due to the steep roll-off of the control signal, the penalty will also score parts of the spectrum that do not belong to the observation band.
Other possibilities not used in the presented results is to use windowed or short-term FFTs for online scoring or FFTs of the whole episode for post-episode scoring.
Better understanding the respective advantages of the various choices is future work and clearly beyond the scope of this work.

\paragraph{Penalizing excess noise in the observation band}

As outlined, the primary objective is to suppress strain noise contributions of the ASC loop for which the mirror angle error is a direct correlate, and it would thus be natural to formulate the penalties used in the reward on these signals. 
However, the level of strain noise contribution is not directly observable due to the noise floor in the angle measurements. On the other hand, by the design of the mirror stage as a high-order mechanical low-pass filter (cf. Appendix \ref{s:relphysmodellASC}) it is ensured that the main control noise in the observation band directly corresponds to the control effort in that band. Hence, we can use the control signal to formulate penalties for excess noise in the observation band.

\subsection{Training and deployment}

Beyond the specifics mentioned here, the algorithmic and training setup is a fairly standard distributed actor critic setup as used in a lot of previous work, e.g., such as \cite{Mnih2015, degrave2022}. 

We use the linear state space model driven by two Gaussian noise processes, representing the seismic disturbance and the sensor noise, respectively described in Section \ref{s:SISOmodel} to generate training data for the learning algorithm.
The simulation allows to run a control policy in the loop where we can use either a linear controller (e.g. the currently employed linear controller) or a neural network policy.
The policy or controller take the simulated error signal (e.g., common-hard pitch angle error) as an input and compute a scalar control action that is sent back to the simulation.

When training the control policies, we randomize the frequency of the pole related to the fluctuation of the laser cavity power, by choosing the pole frequency in a uniform distribution in +/-20\

We train the policies in an episodic fashion, where a random set of parameters for the frequency and seismic noise level, and optionally a new policy parameter set is chosen and then kept constant for 1024\,s of simulated real time.
We typically use 50 actor processes running on CPU, in conjunction with 1 learner process running on a GPU or TPU (e.g., Nvidia A100) per experiment metaparameter configuration (`seed').

While the training is running, we monitor the performance of the control policies via the total discounted reward of so-called greedy evaluators that use the expected value of the control policy instead of a random sample like in the exploration policies of the actor processes. Using the expected value mimics the way the policy is used on the real plant. When training has sufficiently progressed, we select a well-performing policy based on the reward of the greedy evaluators.
The candidate policy is further subjected to additional robustness testing in simulation, assessing robust performance under plant variations, noise variations, disturbances, among others.

While the main training routines are based on a flexible software stack, taking advantage of autodifferentiable code via the JAX library \cite{jax} among others,
when running the policy against the real system, we run the inference with hard real-time constraints to guarantee real-time properties of the overall control loop. To achieve this, we convert the neural network graph of the control policy into ANSI C code with minimal dependency on external libraries, deterministic runtime complexity, and only static memory allocation using a custom code generator. This code is subsequently compiled and loaded as a Linux kernel module in the LIGO control system~\cite{advligorts_2021}.
As in simulation, the control policy is stepped at 256\,Hz and control signals are provided to the downstream control system with a simple sample-and-hold.
 We use the existing SISO controller infrastructure, in particular, the same input and output conventions. The input to the policies is the error in the respective control loop as acquired via the DAQ electronics (e.g., the common hard mode of the mirror pitch error for \chardp{}) and the output is the actuator command sent to the EM actuators via DA converters.

\section{{\tt squidward } RL policy details }
\label{s:squidward_details}

In the following section we list the policy {\tt squidward}, training, and environment details for the \chardp{} RL control policy we have been running on LLO on December 5, 2024. It is currently the best-performing control policy that has been deployed on LLO.

\begin{description}
\item[Error RMS penalty:] We use a Butterworth Lowpass filter with corner frequency at 3\,Hz and designed via a stop band specification of -100\,dB at 8\,Hz using scipy's iirdesign method. The good/bad settings are $\text{good} = 2 \times 10^1$, $\text{bad} = 3 \times 10^2$.

The good and bad levels are normalized on the scalar noise level multiplier set at random at each episode start.

\item[Observation band penalties:] The signal used is the control action $u(i)$. We use two Elliptic IIR filter with settings for pass band, the values for pass- and stop-band corner frequencies, as well as the scoring, are given in the table below. The two penalties are designed at the primary objective of frequency suppression in the observation band of interest.
We use two filters to account for the $1/f^4$ roll off of the natural dynamics of the system, i.e., the filter for the lower part of the observation band spectrum has higher good/bad values in comparison to the filter for the upper part of the band.

\item [High Frequency penalty:] While using solely the overall error and band penalty are enough to train policies with outstanding RMS and GWD band performance, the policies sometimes show `high frequency artifacts', i.e., increased control energy in the frequency bands beyond the measurement band. To counteract these artifacts and encourage policies with low HF control energy, we add a second band pass-based penalty with somewhat less stringent `bad'/`good' settings. Here, the idea is to give a fairly relaxed reward to regularize the actions at high frequencies. Thus, as per our discussion on the GW detection band above, this is an auxiliary reward. Settings are listed in the table below. Note that we oversample the signal by a factor of 2 such that we can effectively penalize signal energy at the Nyquist frequency of 128\,Hz.

\item[Intermediate reward:] All individual scores are combined into the final, scalar intermediate reward $r(t)$ by multiplication.

\begin{equation*}
    r(i) = c_e(i) \cdot c_{mb}(i) \cdot c_{hf}(i)
\end{equation*}

\end{description}

Note that we oversample the IIR filters by a factor of 2 ($f_s=512$) to allow for scoring effectively up to the Nyquist frequency of the controller ($f_s/2 = 128$\,Hz).

\begin{table}[h]
\centering
\caption{Band penalties, IIR filter, and scoring settings.}
\label{tab:band_penalties}
\begin{tabular}{|c|c|c|c|c|c|c|}
\hline
{Filter} & \multicolumn{2}{c|}{Passband ($\omega_p$) [Hz]} & \multicolumn{2}{c|}{Stopband ($\omega_s$) [Hz]} & \multicolumn{2}{c|}{Scoring} \\
\cline{2-7}
 & Lower & Upper & Lower & Upper & Good & Bad \\
\hline
Obs. band low  & 8.0 & 20.0 & 5.0 & 35.0  & \(10^{-1}\) & 2.5 \\
\hline
Obs. band high & 20.0 & 30.0 & 5.0 & 45.0 & \(10^{-2}\) & 1  \\
\hline
High frequency & 30.0 & 130.0 & 15.0 & 150.0 & \(5\cdot10^{-1}\) & \(10^{2}\)  \\
\hline
\end{tabular}
\end{table}

\subsection{Network architectures}

\begin{description}
    \item[Policy network:] The network representing the control policy has dilated convolution layers \cite{oord2016wavenetgenerativemodelraw} layers (1, 2, 4, 8, 16, 32, 64, 128) and thus an input length of $N=256$.

    The output of all convolution layers is fed to an MLP with layers (256, 256, 128) with a Gaussian Head with a $\mathrm{tanh}$ on the mean.
   \item[Critic network:] The network representing the Q-value function, i.e., the `critic' is a LSTM based network with encoders and decoders before and after the LSTM core, respectively. The encoder has layer sizes (128, 256, 512) and the decoder (512, 512, 256). The LSTM core has a single layer of 512 units.

\end{description}

\subsection{Common training and meta-parameters}

Training metaparameters for the RL policies are listed in Table \ref{tab:env_training_params}.

\begin{table}[h]
\centering
\caption{Environment and Training Setup Parameters.}
\label{tab:env_training_params}
\begin{tabular}{@{} l l @{}}
\hline
\textbf{Environment Parameters} & \\
\hline
Scalar noise multiplier & Random uniform in [1.0, 5.0] \\
\(\omega_{RHP}\) variation & Log uniform in [0.8, 1.2] \\
Randomization strategy & episodic \\
Hard action limit & 3000 counts \\
Soft action limit & 2700 counts \\
Reward discount factor & 0.999 \\
\hline
\textbf{Training Setup} & \\
\hline
Number of actors & 50 \\
Episode length & 1024 s / 262144 steps \\
Maximum replay size & \(20 \cdot 10^6\) \\
Network target update period (critic/actor) & 100 \\
Exploration noise, initial value standard deviation & 0.0125 \\
(in normalized network output) & \\
Batch size & 256 \\
Activation function & ELU \\
MPO $\epsilon$ & 0.1 \\
MPO $\epsilon$ KL mean & 0.001 \\
MPO $\epsilon$ KL covariance & \(10^{-6}\) \\
Optimizer & Adam \\
Learning rate (actor) & \(10^{-4}\) \\
Learning rate (critic) & \(5 \cdot 10^{-5}\) \\
Learning rate (alpha) & \(10^{-4}\) \\
\hline
\end{tabular}
\label{tab:mpoparams}.
\end{table}

\section{{\tt spongebob }RL policy details }
\label{s:spongebob_details}

In the following section, we list the policy, training, and environment details for the \chardp{} RL control policy we have been running on LLO on April 26, 2024, and August 8, 2024.  The setup and parameters are the same as listed in Section \ref{s:squidward_details} unless noted here. 
It is the control policy that currently has seen the longest deployment on LLO (>1h).

\subsection{Rewards}
\label{s:spongebob_rewards}

\begin{description}
\item[Error RMS penalty:]
In place of the filter, we use the absolute value of the error signal $|e(t)|$, this conceptually corresponds to using an all-pass in place of the filter. The good and bad settings for the sigmoid scoring function are $\text{good} = 2 \times 10^0$ and $\text{bad} = 2 \times 1.5 \times 10^1$, respectively.

The good and bad levels are normalized on the scalar noise level multiplier set at random at each episode start.

\item[Observation and high frequency band penalties] as in Section \ref{s:spongebob_rewards}.

\end{description}

\subsection{Network architectures}
\label{s:spongebob_nets}

\begin{description}
    \item[Policy network:] The network representing the control policy is a diamond shaped MLP with layers (128, 256, 512, 512, 256, 128) with a Gaussian Head with a $\mathrm{tanh}$ on mean.
    \item[Critic network:] Same as for the {\tt squidward} policy  Section \ref{s:squidward_details}. 
\end{description}

At the input of both actor and critic network, we stack the $N=128$ past observations.

\begin{figure}
    \centering
    \includegraphics[width=0.45\textwidth]{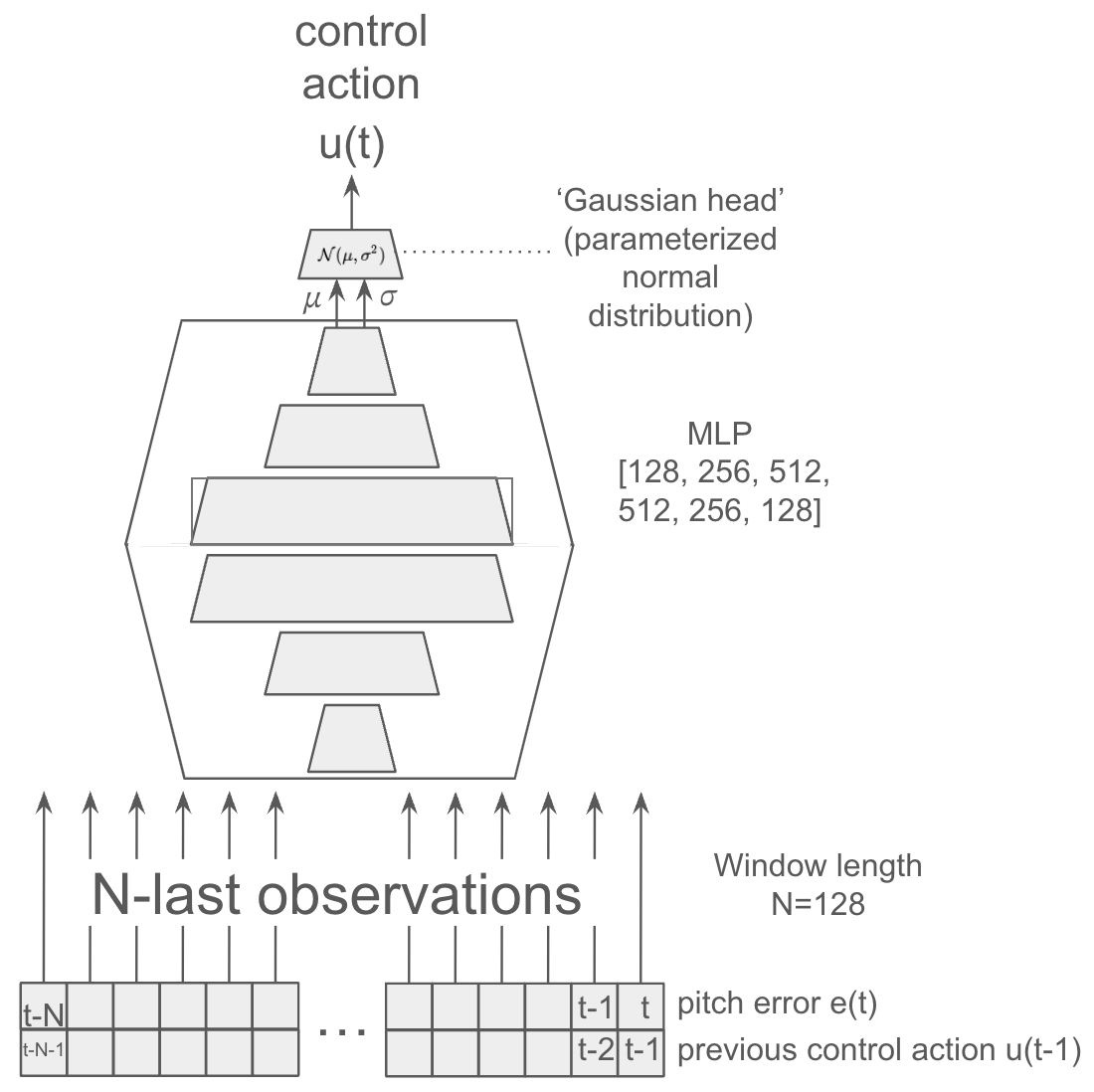}
    \includegraphics[width=0.5\textwidth]{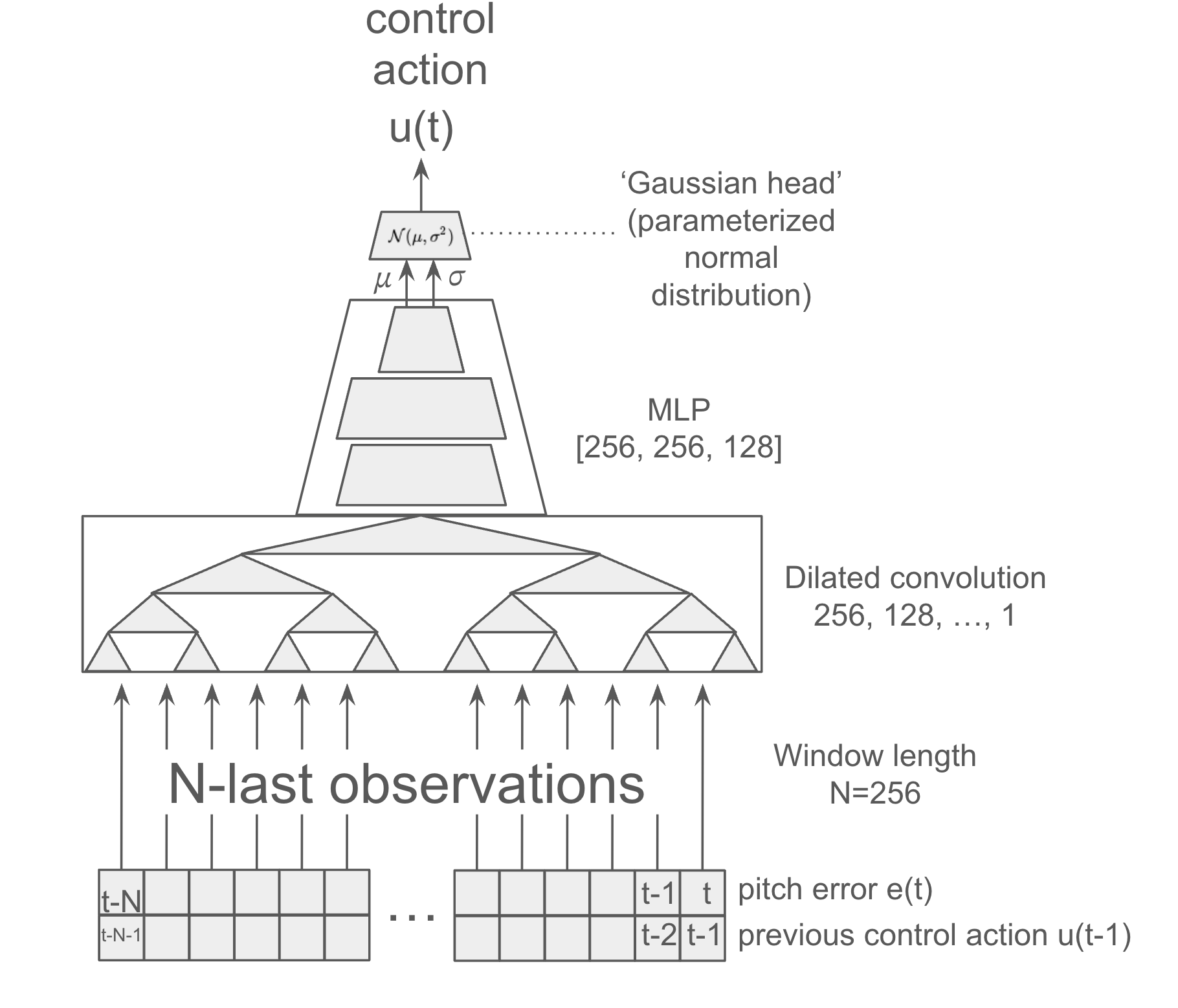}
    \caption{Policy network architectures: (left) spongebob (right) squidward. }
    \label{fig:my_label}
\end{figure}

\section{Deployment on hardware - experimental results}

\begin{figure}
    \centering
         \includegraphics[width=0.49\textwidth]{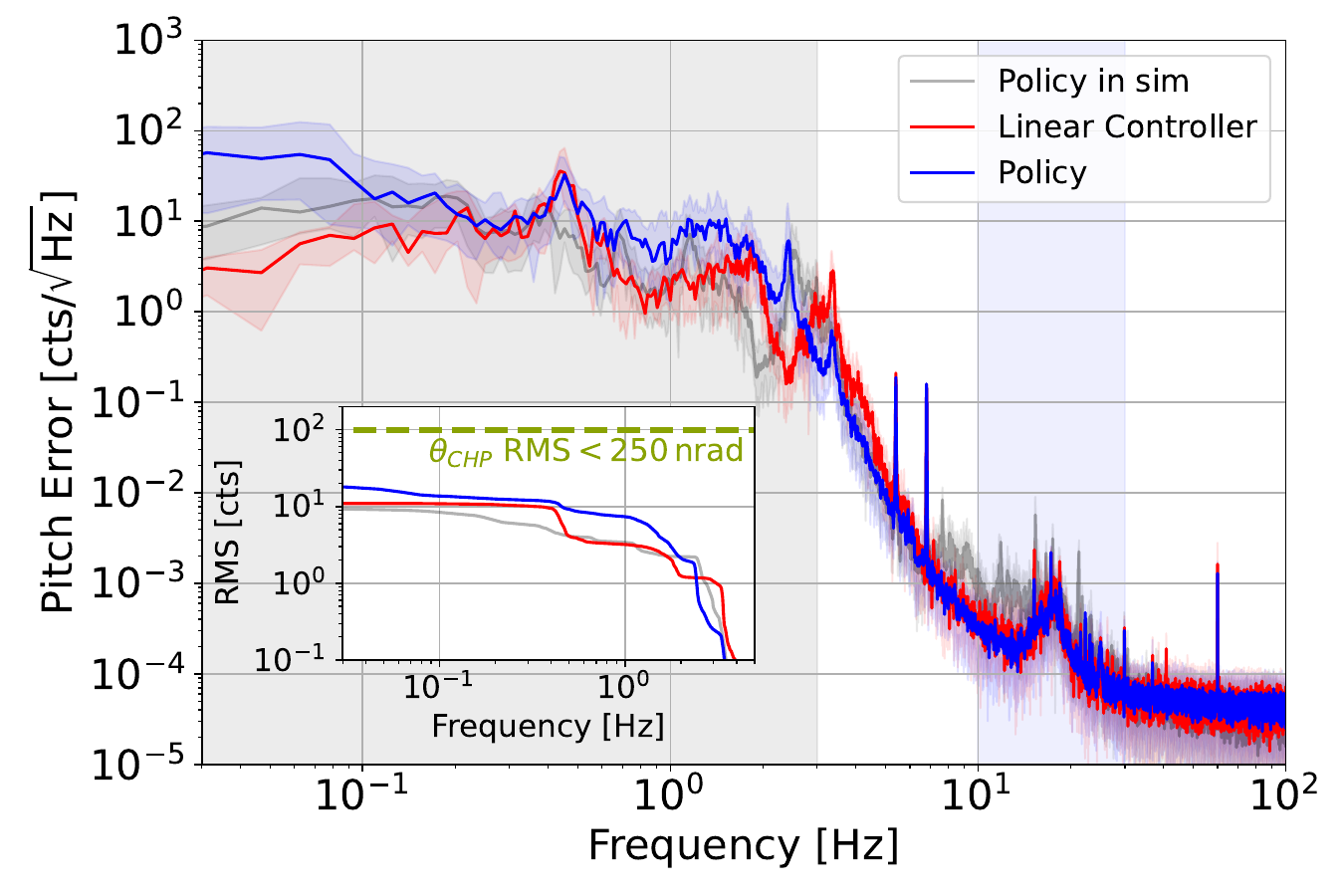}
         \includegraphics[width=0.49\textwidth]{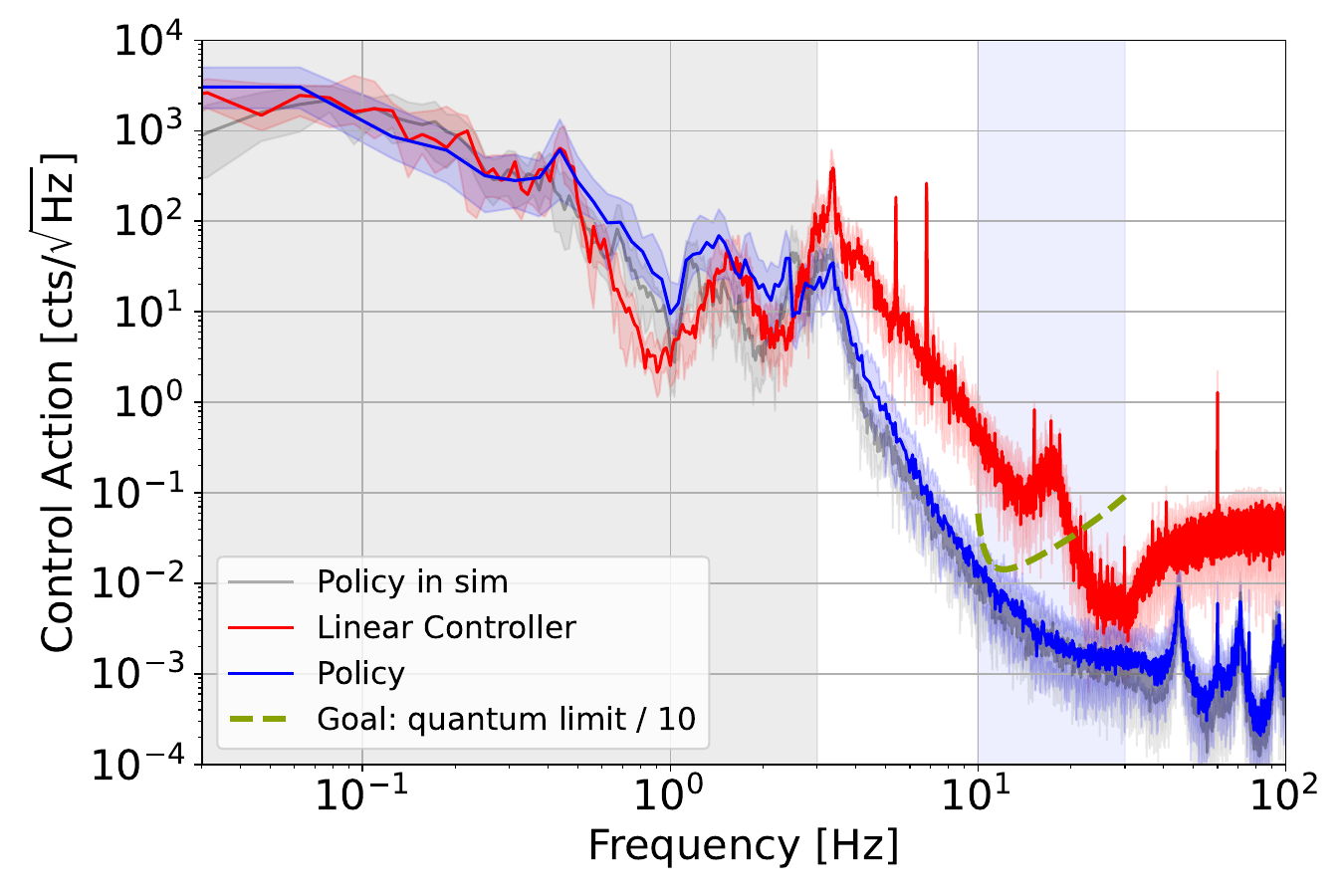} \includegraphics[width=0.49\textwidth]{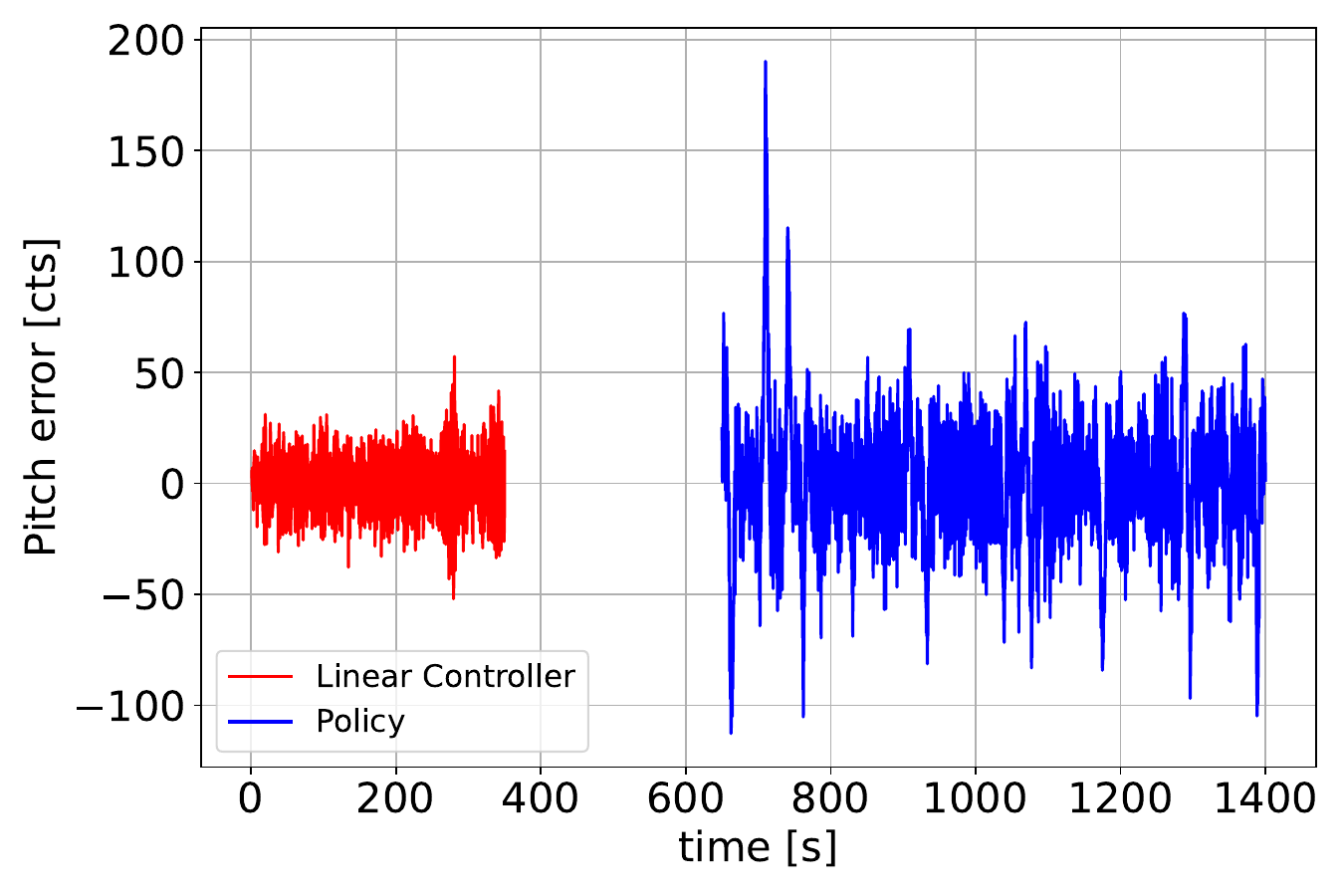}
         \includegraphics[width=0.49\textwidth]{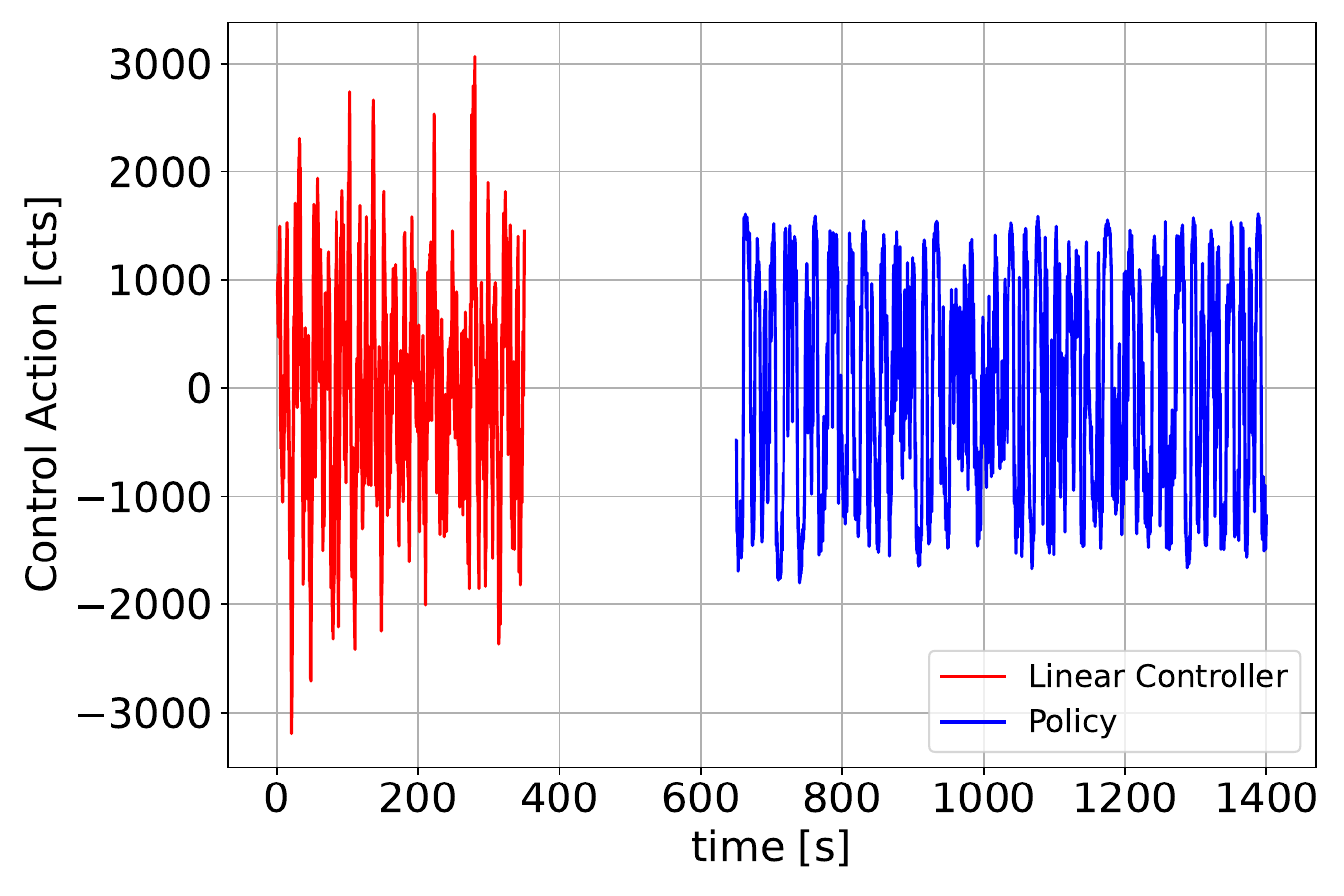}
    \caption{\textbf{Amplitude spectral density of control and error signals.} Comparison of the in-loop residual motion and the required feedback control action for the baseline linear controller and the Deep Loop Shaping control policy. 
    (left) Pitch Error
    (right) Control Action/Torque.
    The top row shows the amplitude spectral density (ASD) of the signals, computed using Welch's method and a 64\,s FFT window. 
    The blue shading indicates the targeted observation band 10--30\,Hz and the grey shading the control band $< 3\,\unit{Hz}$.
   The bottom row shows the time series of the signals. The times indicated are relative to GPS time 1417457187. We have selected 350 seconds of data just before the experiment for the comparative evaluation of the linear controller, and show 750 seconds of data from the time when the RL policy is controlling \chardp. We measure a pitch error RMS of 33\,counts, and a band-limited control action RMS in 10--30\,Hz of $2.5\cdot10^{-2}$\,counts for the policy, compared to 11.3 / $7.9\cdot 10^{-1}$\,counts for the incumbent linear controller.
   The noise of the angular sensor dominates the error spectrum above $\sim$8\,Hz, and so the improvements in the control action are not visible in this sensor.
   The inset compares the cumulative (integrated from the right) in-loop RMS of \chardp{} vs a control design goal. 
   See additional results in Fig.~\ref{fig:rl_vs_lin_addl}, and Appendix~\ref{s:sysID} for more details on the noise and plant identification.}
    \label{fig:control_error_spectra}
\end{figure}

 In Fig. \ref{fig:control_error_spectra} we show additional details of the result shown in the main text, where we compare the performance of the neural network policy {\tt squidward}  against the standard controller for a $> 10\,\unit{min}$ stretch on Dec 5, 2024. 

\subsection{Additional results: Diamond shaped MLP policy ('spongebob') on LLO}
\label{s:spongebob_results}

\begin{figure}
    \centering
         \includegraphics[width=0.49\textwidth]{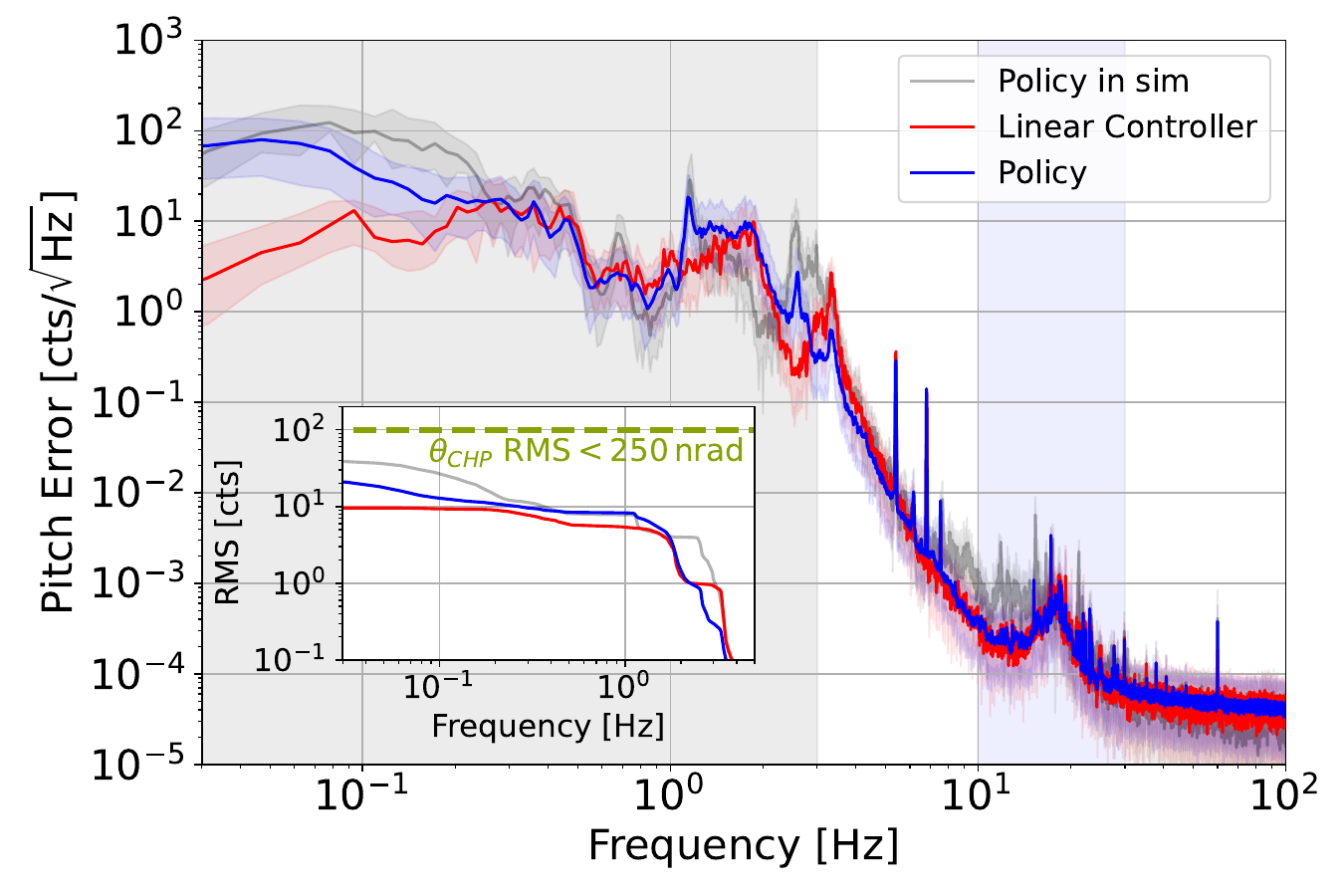}
         \includegraphics[width=0.49\textwidth]{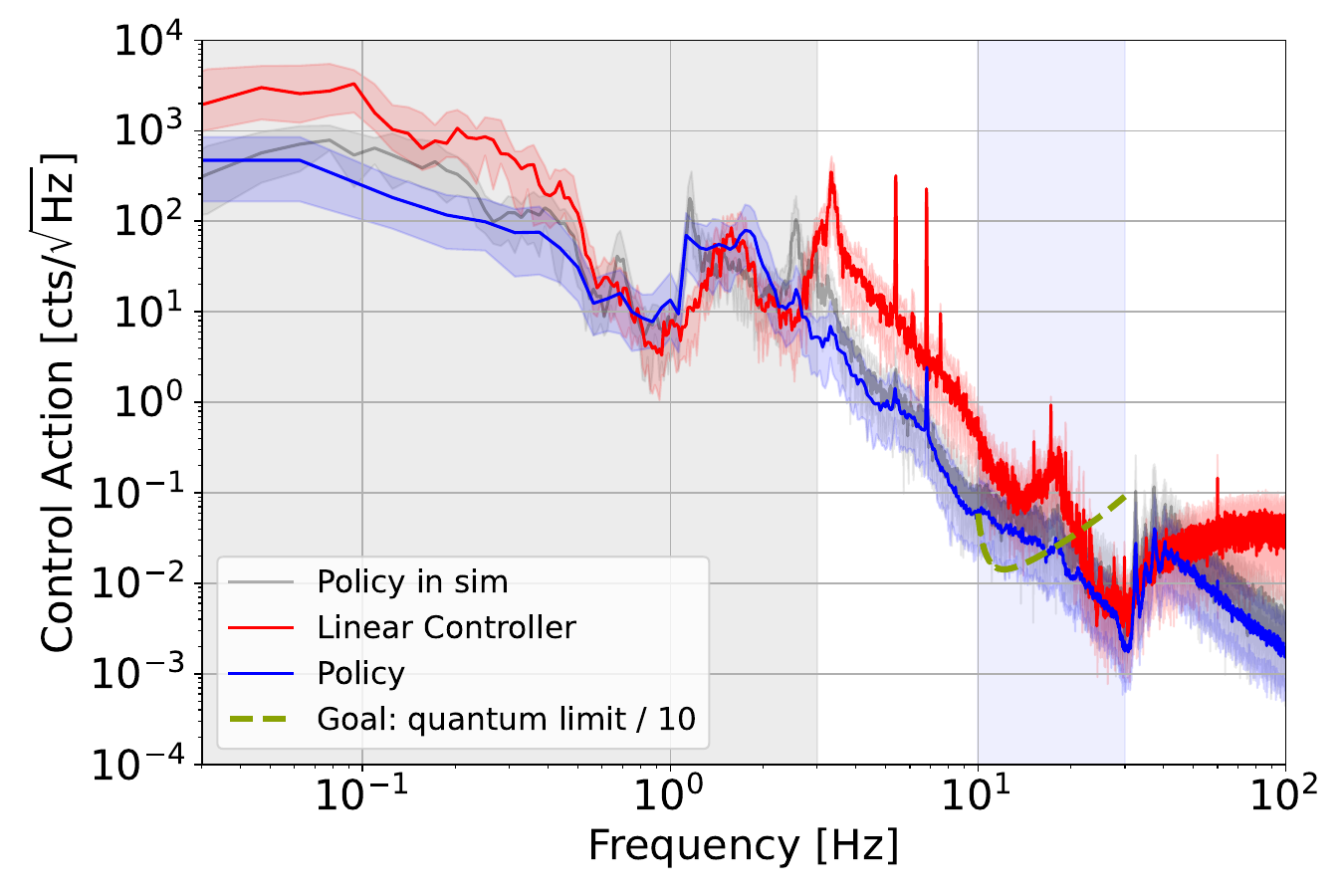} \includegraphics[width=0.49\textwidth]{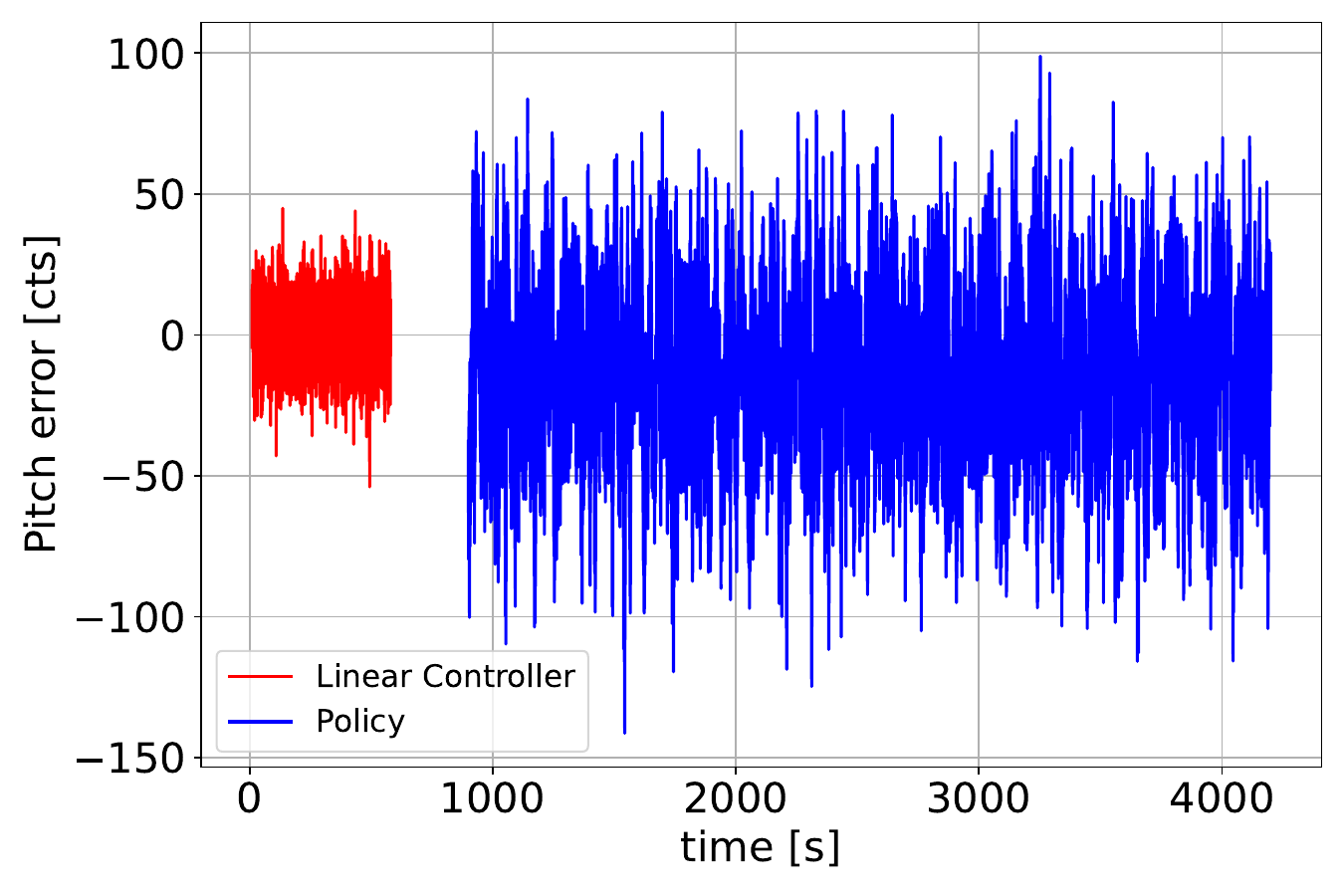}
         \includegraphics[width=0.49\textwidth]{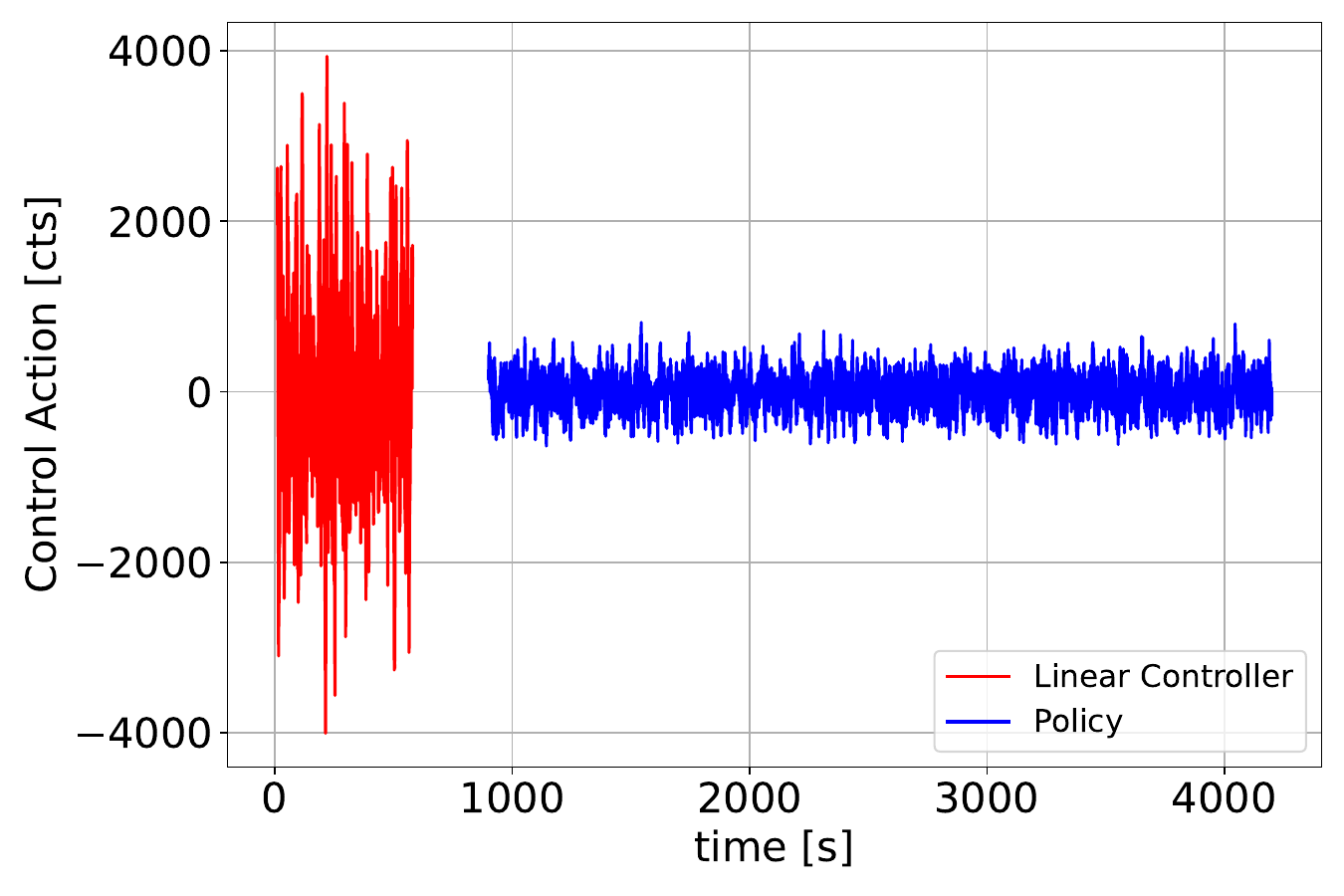}
    \caption{Comparison of the in-loop residual motion and the required feedback control action for the baseline linear controller and the deep loop shaping based control policy. (left) Pitch error (right) Control action.
    (A,B) Amplitude spectral density (ASD) of the signals computed using a spectrogram with a time window of 64\,s. The blue shading indicates the targeted observation band 10--30\,Hz and the grey shading the control band $< 3\,\unit{Hz}$.
   The bottom row shows the time domain plots of the signals. The times indicated are relative to GPS time 1407160711. We have selected 560 seconds of data just before the experiment for the comparative evaluation of the linear controller, and show 3300 seconds of data from the time when the RL policy is controlling \chardp{}.
   We measure Error RMS 33.5 counts and bandlimited control signal RMS in 10--30\,Hz of $1.6 \cdot10^{-1}$ counts for the policy and 11.3 / $6.8 \cdot 10^{-1}$ counts for the incumbent linear controller.
    See additional results in Fig.~\ref{fig:rl_vs_lin_addl}.}
    \label{fig:control_error_spectra_spongebob}
\end{figure}

See figure \ref{fig:control_error_spectra_spongebob}.

\subsubsection{Quality of transfer from simulation to real system (sim2real)}
\label{s:sim2real}

\begin{figure}
    \centering
    \includegraphics[width=0.7\textwidth]{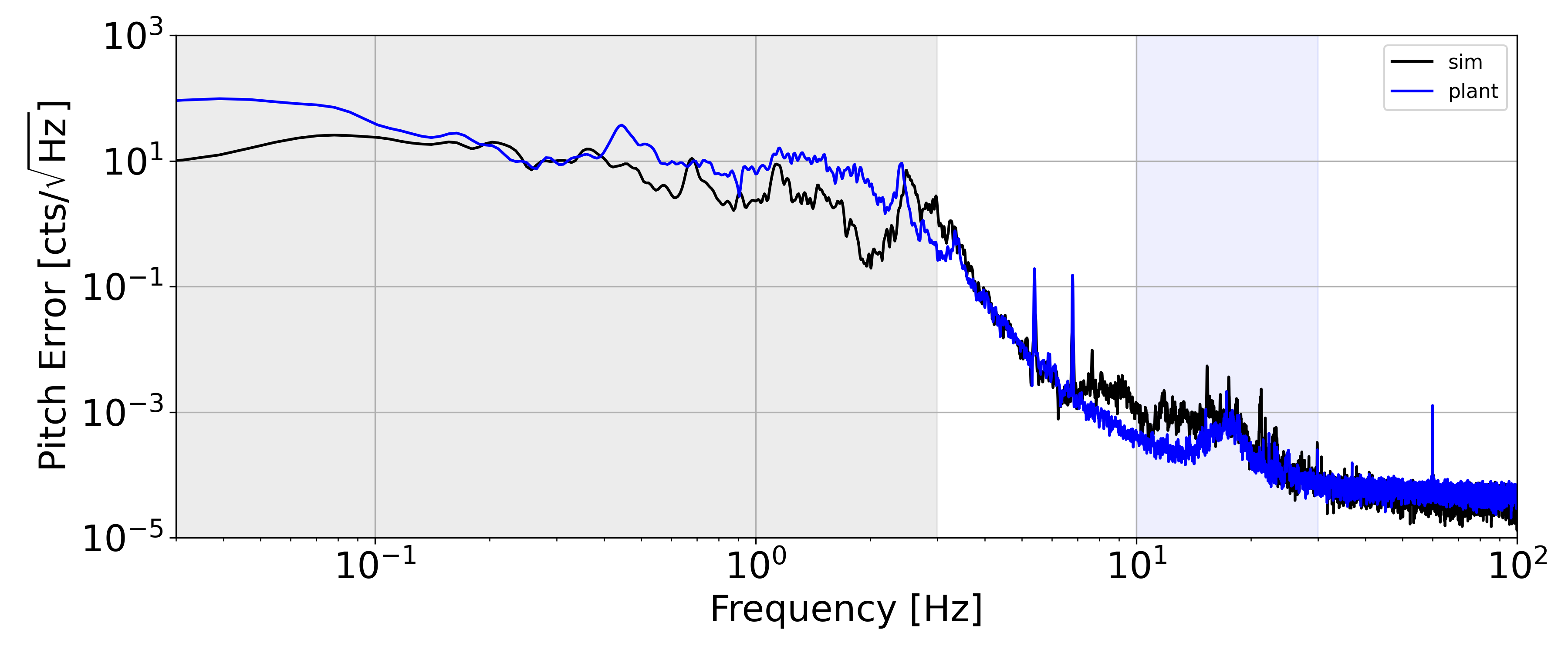}
    \includegraphics[width=0.7\textwidth]{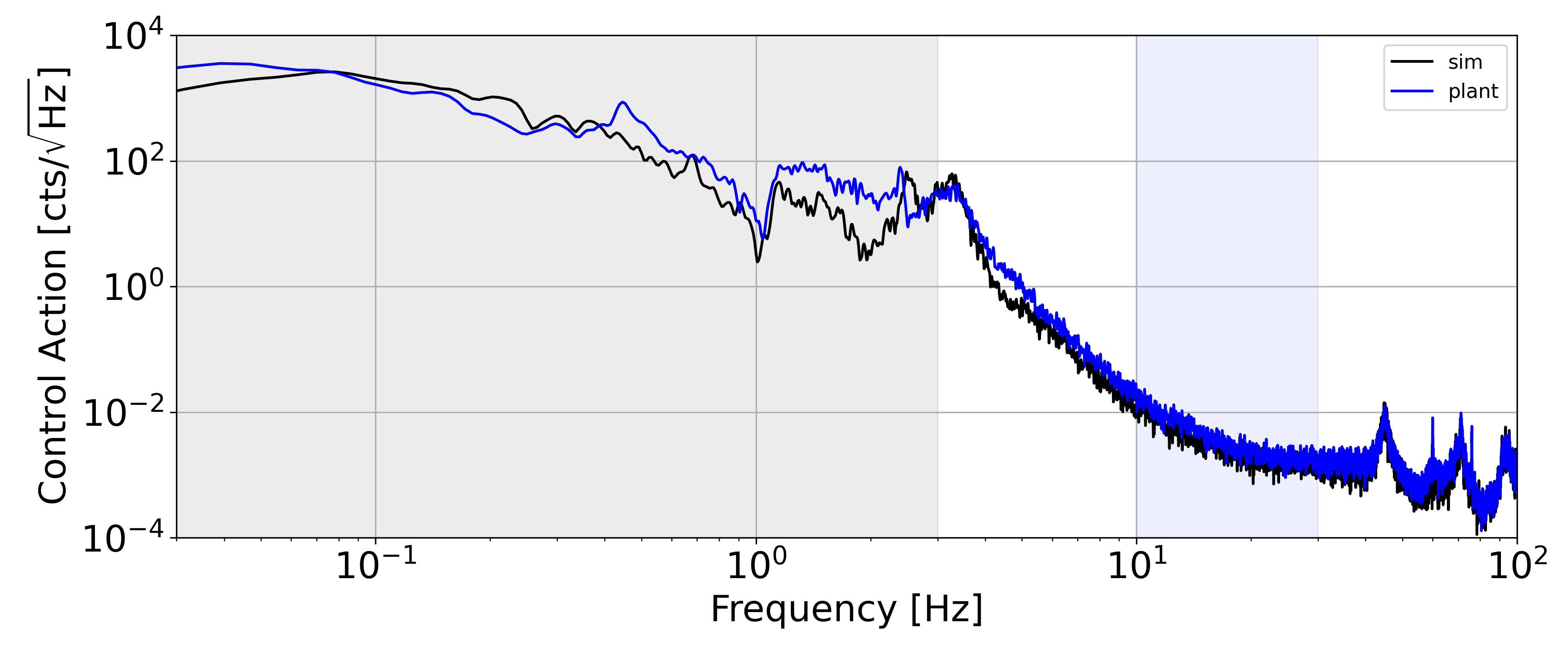}
    \rule[0.5ex]{0.6\textwidth}{0.55pt} \\
    \vspace{0.25cm}
    \includegraphics[width=0.7\textwidth]{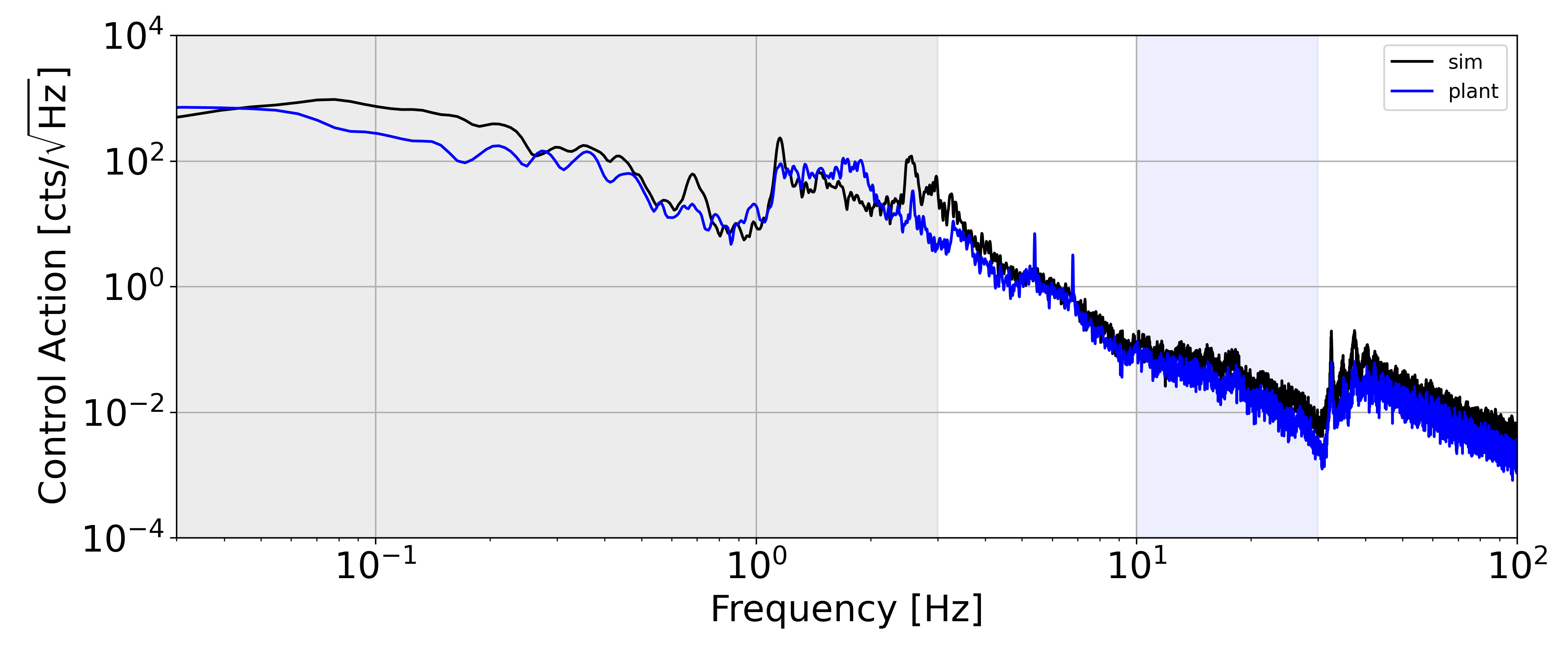}\\
    \includegraphics[width=0.7\textwidth]{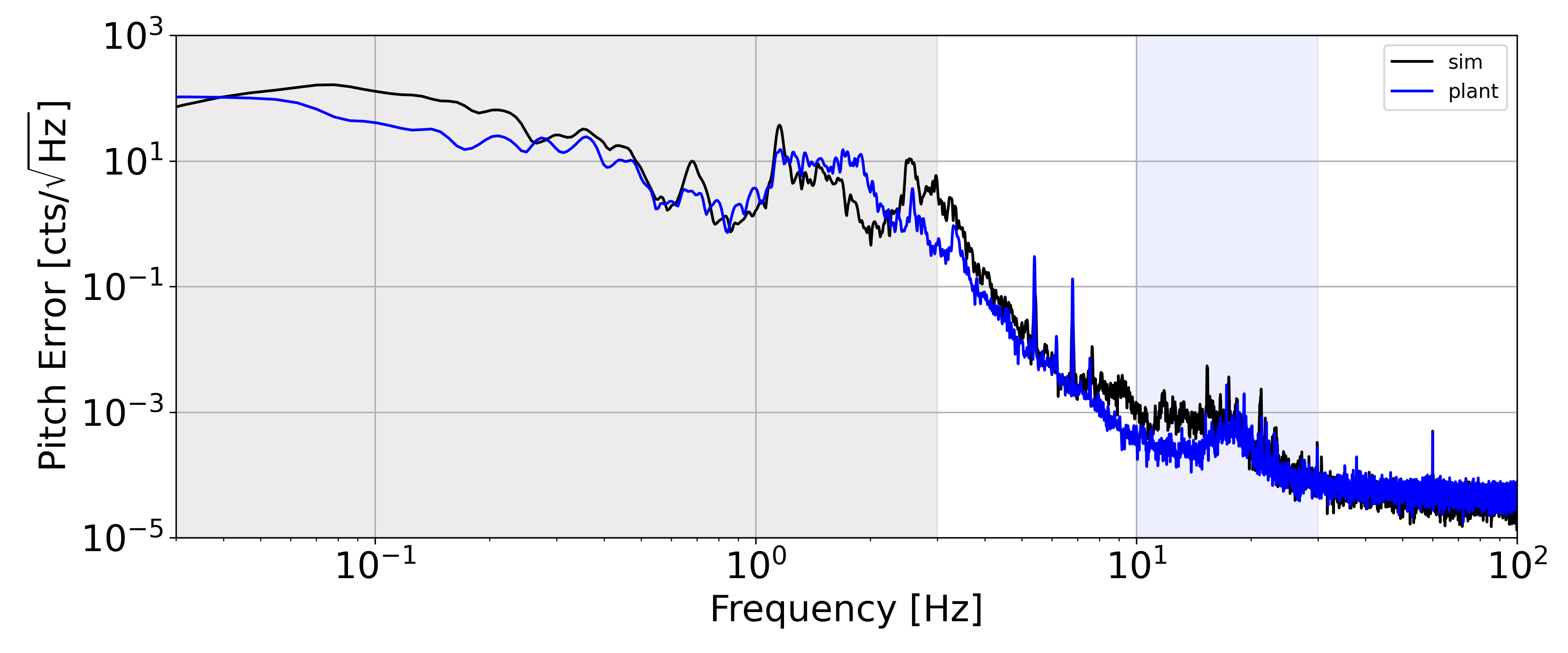}
    \caption{ Results of the RL policies running against simulation vs real plant data: (top) {\tt squidward} (bottom) {\tt spongebob}. See text for discussion.}
    \label{fig:sim2real}
\end{figure}

In Fig.~\ref{fig:sim2real}, we show a comparisons between both the {\tt squidward} and {\tt spongebob} policy running against the linear state-space simulation and data from the real system. We note that the match at critical frequencies of interest for the observation band is good, whereas below 0.1\,Hz there is an increased mismatch when compared to the results from spongebob (below).

\subsection{Non-Gaussian Analysis}
\label{s:rayleigh}

The Matched Filter algorithm, which is used to detect mergers of black holes and neutron stars, is optimal in the case that the background noise is stationary and Gaussian~\cite{Bruce:chi_square}.
It has usually been the case that the noise in the 10--100\,Hz band is less Gaussian than the noise at higher frequencies (which is dominated by thermal noise and shot noise). The non-stationary behavior results in an increased false alarm rate and, therefore, a higher detection threshold.
The RL policies described in this manuscript do an excellent job of reducing the PSD of the control signal in the desired frequency band, but the noise must also be stationary, so that the background is concomitantly reduced.

\begin{figure}
 \includegraphics[width=0.5\columnwidth]{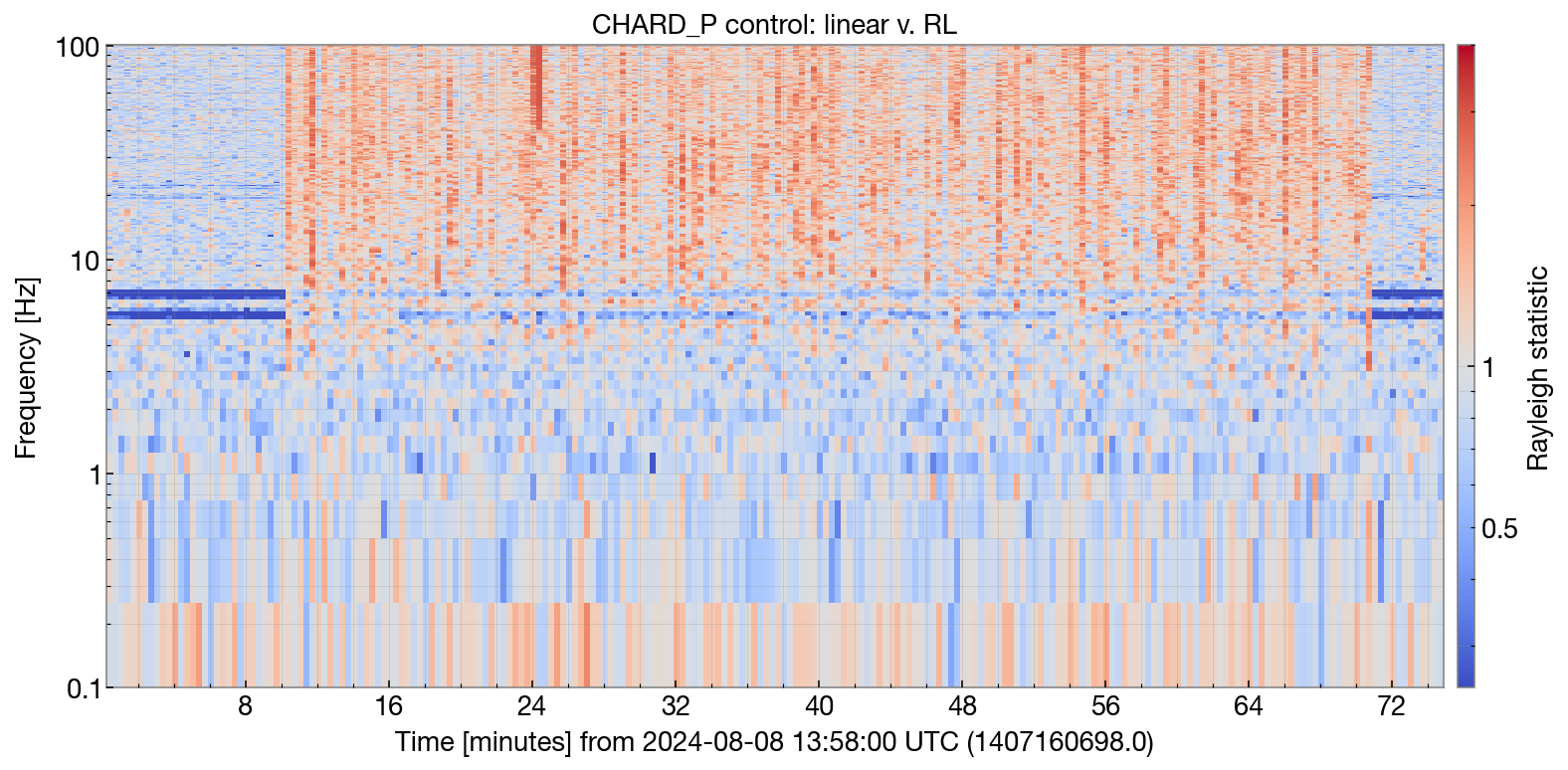}
 \includegraphics[width=0.5\columnwidth]{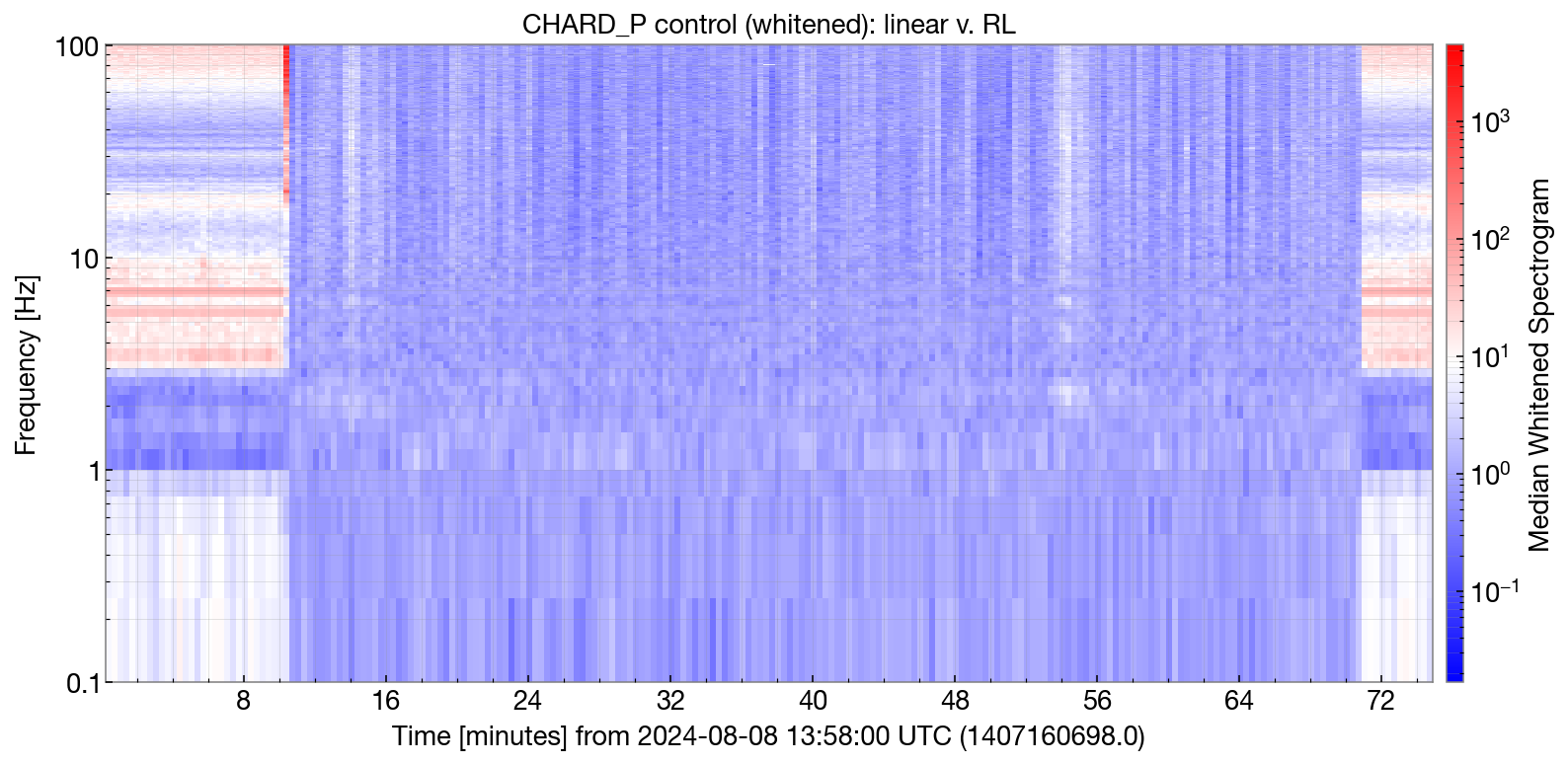}
 \includegraphics[width=0.5\columnwidth]{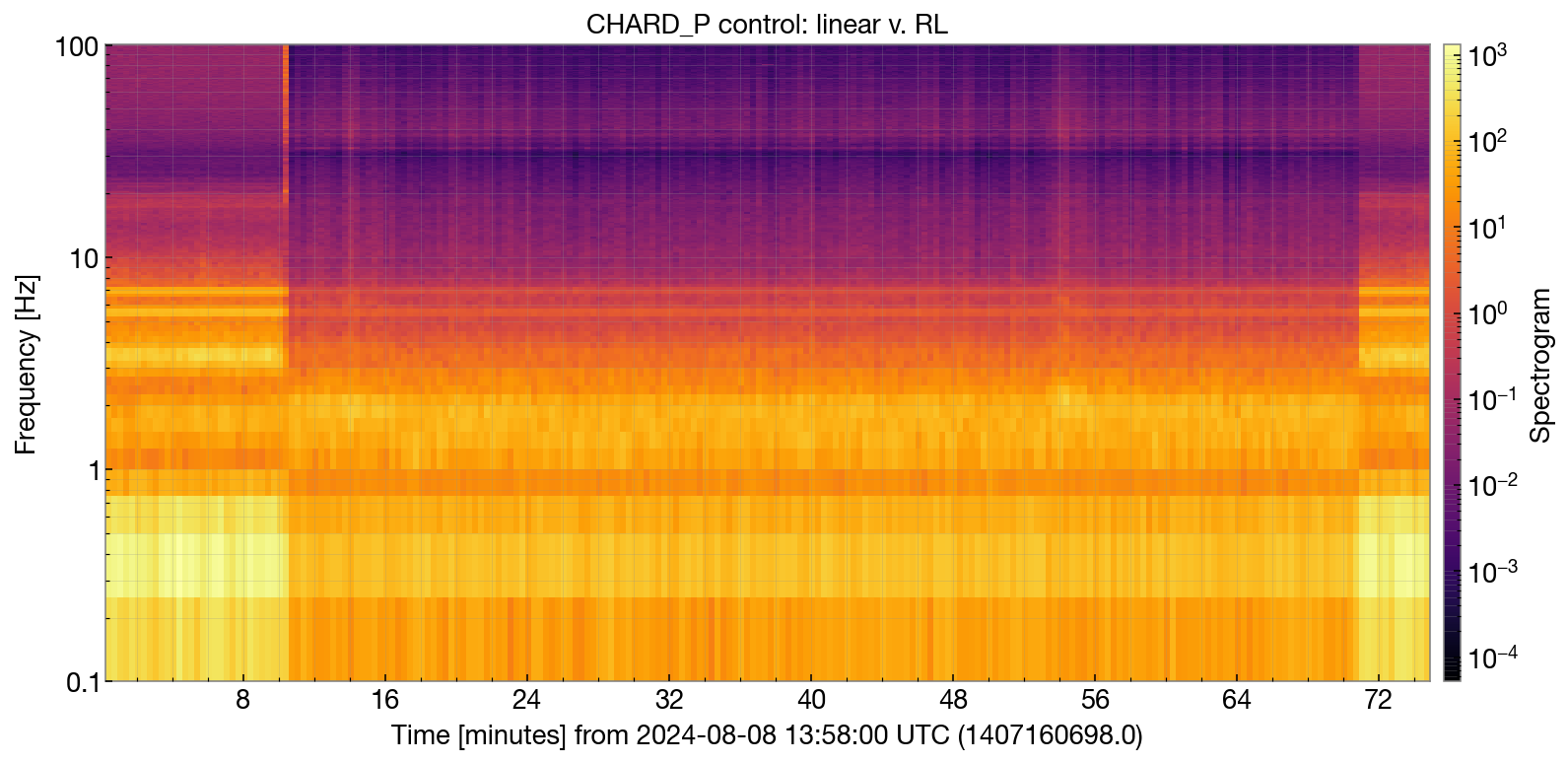}
 \includegraphics[width=0.5\columnwidth]{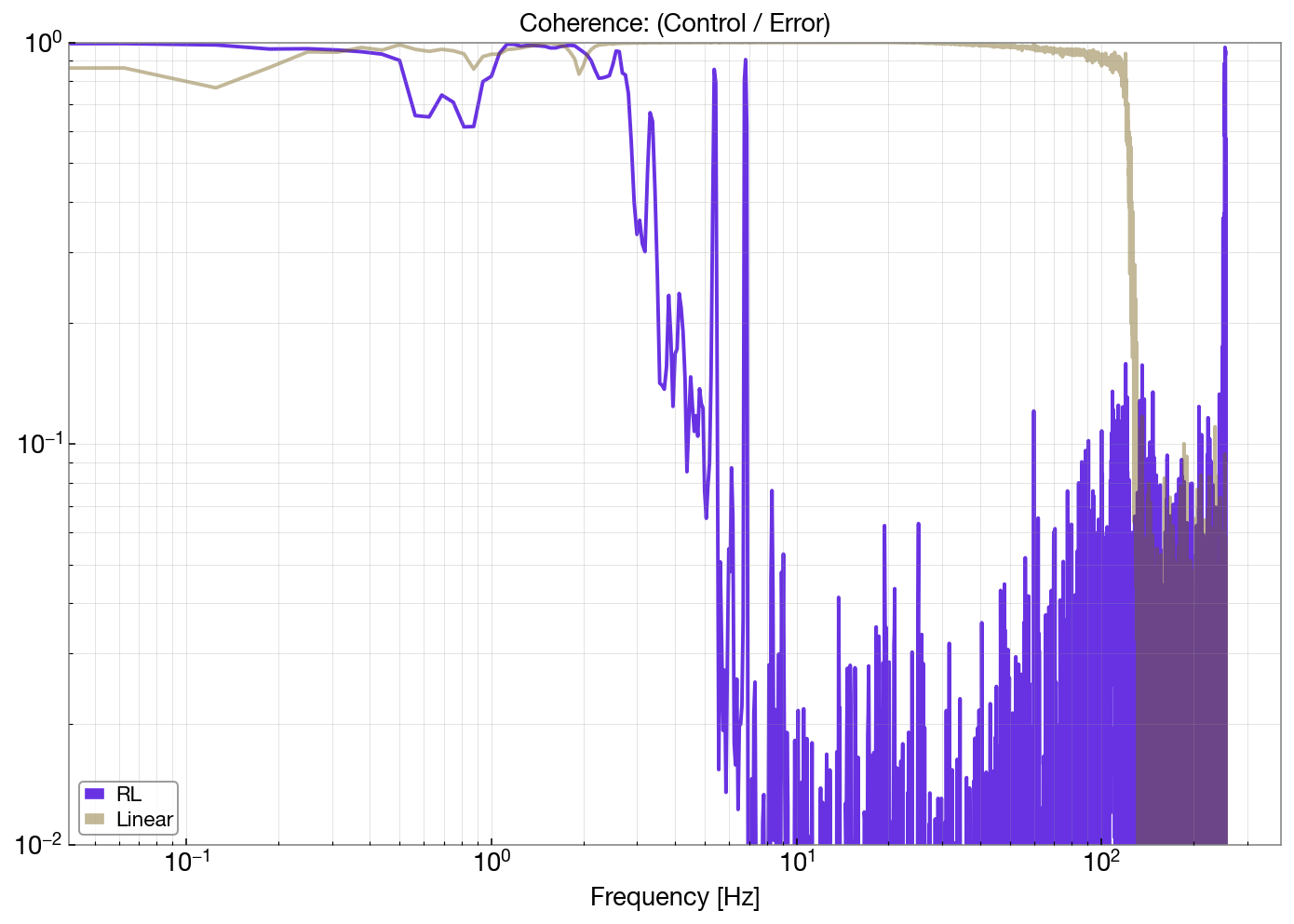}
\caption[Rayleigh Analysis]{(bottom left) raw spectrogram of the \chardp{} feedback signal, showing the reduced noise above 10\,Hz during the 60 minutes that the RL policy is in control. (upper right) spectrogram plot normalized by the median PSD for the whole stretch, highlighting small changes in the noise. (upper left) The Rayleigh spectrogram as described in the text. (lower right) Coherence of the output of the controller with the input (error) signal. The linear controller is highly coherent as expected, whereas the RL policy has an evidently nonlinear response above $\sim$2.5\,Hz.}
\label{fig:rl_vs_lin_addl}
\end{figure}

In Fig.~\ref{fig:rl_vs_lin_addl}, in addition to the PSD comparisons above, we show analysis of the RL controller and compare it to the linear baseline controller.

The upper left panel of Fig.~\ref{fig:rl_vs_lin_addl} is a ``Rayleigh-gram''. Similar to a classic spectrogram or waterfall plot, it shows noise as a function of time and frequency. The typical spectrogram \cite{2020SciPy-NMeth} uses Welch's periodogram method to estimate the PSD for each time slice. For the Rayleigh-gram \cite{gwpy}, each bin instead plots the ratio of the standard deviation of the PSD to the mean PSD value for that bin normalized such that stationary Gaussian noise will have a Rayleigh value of unity. Coherent processes such as sinusoids would have a value less than 1, and noise with non-Gaussian transients would have a value greater than 1.

In the spectrograms of Fig.~\ref{fig:rl_vs_lin_addl}, we see that the RL Policy substantially enhances the non-stationarity of the controller output above about 5\,Hz. These transient features are weak, and we have already shown in Fig.~\ref{fig:rl_vs_lin_strain_coords} that the temporal variation of the PSD of the controller output is modest (the 10th and 90th percentiles are close to the median). However, if the detector sensitivity improved, approaching its quantum limit, the transients might start to have an impact on the background evaluation of GW signal analyses. It is not possible at this time to estimate the impact on GW searches quantitatively for such a scenario, but clearly, given the additional complexity of incorporating transient noise features in our detector noise models, a goal of further RL Policy developments should be to reduce the non-stationarity of the controller output.

\subsection{IMC 40m results}
\label{s:40m}
\begin{figure}
    \centering
    \includegraphics[width=0.45\columnwidth]{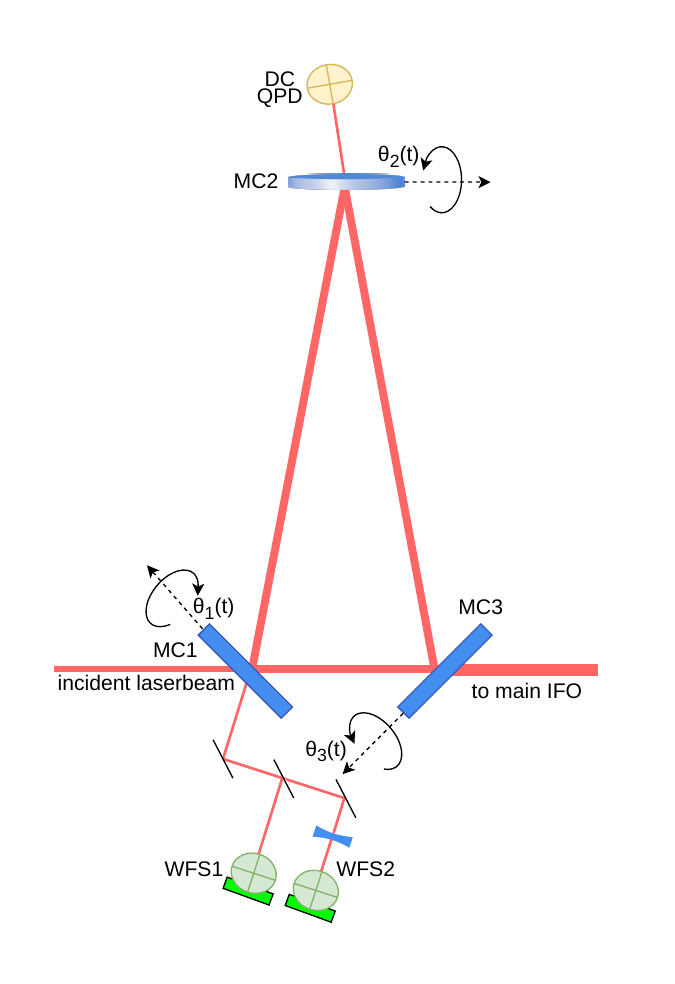}
    \caption{
    Simplified layout of the IMC at the 40m prototype at Caltech. It is the three-mirror ring cavity, where the two flat mirrors (MC1 and MC3) are partially transmitting with a transmissivity of 0.2\
    Errors in local or `mirror' coordinate system correspond to the physical angles of the mirrors, errors in the sensor coordinate systems are in counts of the sensor signal from the DAQ system (see details in text). }
    \label{fig:imc_layout}
\end{figure}
In our work, we have used the Caltech 40\,m prototype interferometer \cite{Wardeal2008} to test our approach in a low-risk, robust, high-availability environment.  
Here we are presenting results on a control policy deployed on the input mode cleaner (IMC) of the interferometer. The IMC is a 3-mirror ring cavity with suspended optics which simplified layout is shown in Fig.~\ref{fig:imc_layout}.
Note that we do not claim that this approach and result is directly operationally relevant or meaningful for the actual day-to-day operation of an IMC in an observation campaign, but rather serves to illustrate additional capabilities of the presented method.
We have also used the results as a means to verify our sim2real and real-time control capabilities before deploying policies on the actual LIGO Observatory. 
LLO is a big science instrument with the aim to maximize observation time and serving a large scientific community. 
It is a high-stakes environment, engineering testing time is limited, and downtime needs to be minimized.

The IMC subsystem allows to demonstrate some additionally interesting aspects and benefits of our approach: 
(1) Control of a multi-input-multi-output (MIMO) system 
(2) Ability to score flexibly in a variety and mix of relevant coordinate systems 
(3) Ability and potential to use model information to score on non-directly measured quantities, e.g., the mirror angles in our case. 
This information does not need to be present to the control policy at run time.
Importantly, there is no conceptual difference to the way we set up the learning problem for the 40\,m IMC and the LLO arm cavities. Besides a change in reward (to express control objectives) and environment/simulation (to express the physics of the target system), there are no changes to the setup used for the LLO arm cavity results.

Figure~\ref{fig:imc_layout} illustrates the layout of the 40\,m IMC. The 40\,m IMC is a triangular cavity composed of two flat mirrors (MC1 and MC3) separated by a distance of 17.5\,cm, and a spherical mirror (MC2) at the apex. The longer side of the isosceles triangle is 13.5\,m \cite{Drigg2006} long. 
For the ASC, two WFSs are placed at the reflection of MC1, while a DC QPD is located at the transmission of MC2. 
There are four degrees of freedom (DOFs), namely rotation and translation in both vertical and horizontal directions. The two WFSs generate error signals to control these four DOFs, whilst the MC2 QPD provides overall pointing control of the cavity itself.

Using a model of the geometry of the system, we can compute the mirror pitch angle errors from the sensor signal, and in fact, the Lightsaber simulation provides these quantities. As discussed elsewhere, one advantage of RL and thus our approach is that we can use such so-called privileged data to compute scores for the agent to learn from.

\begin{figure}
    \centering
    \includegraphics[width=0.32\columnwidth]{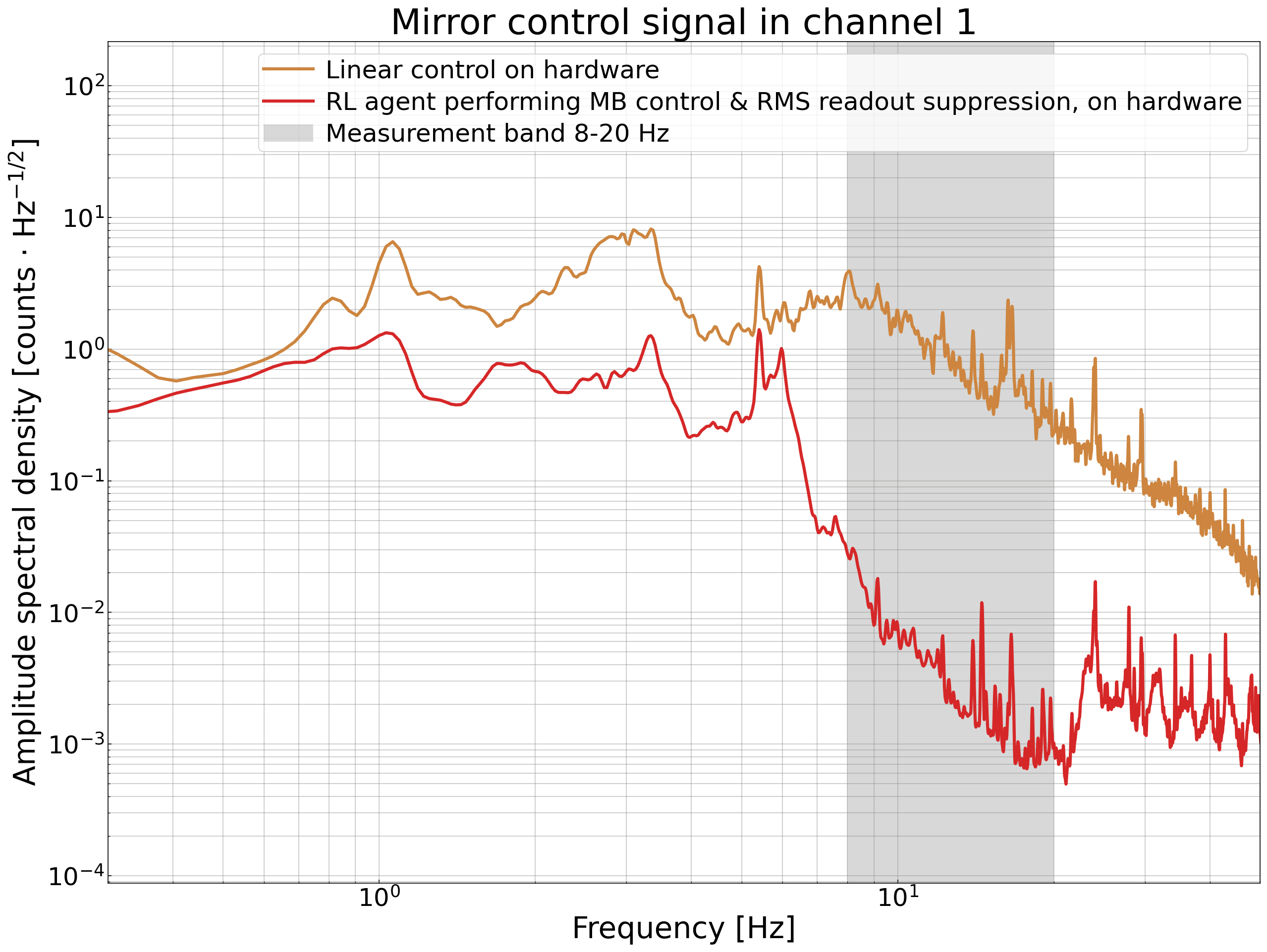}
    \includegraphics[width=0.32\columnwidth]{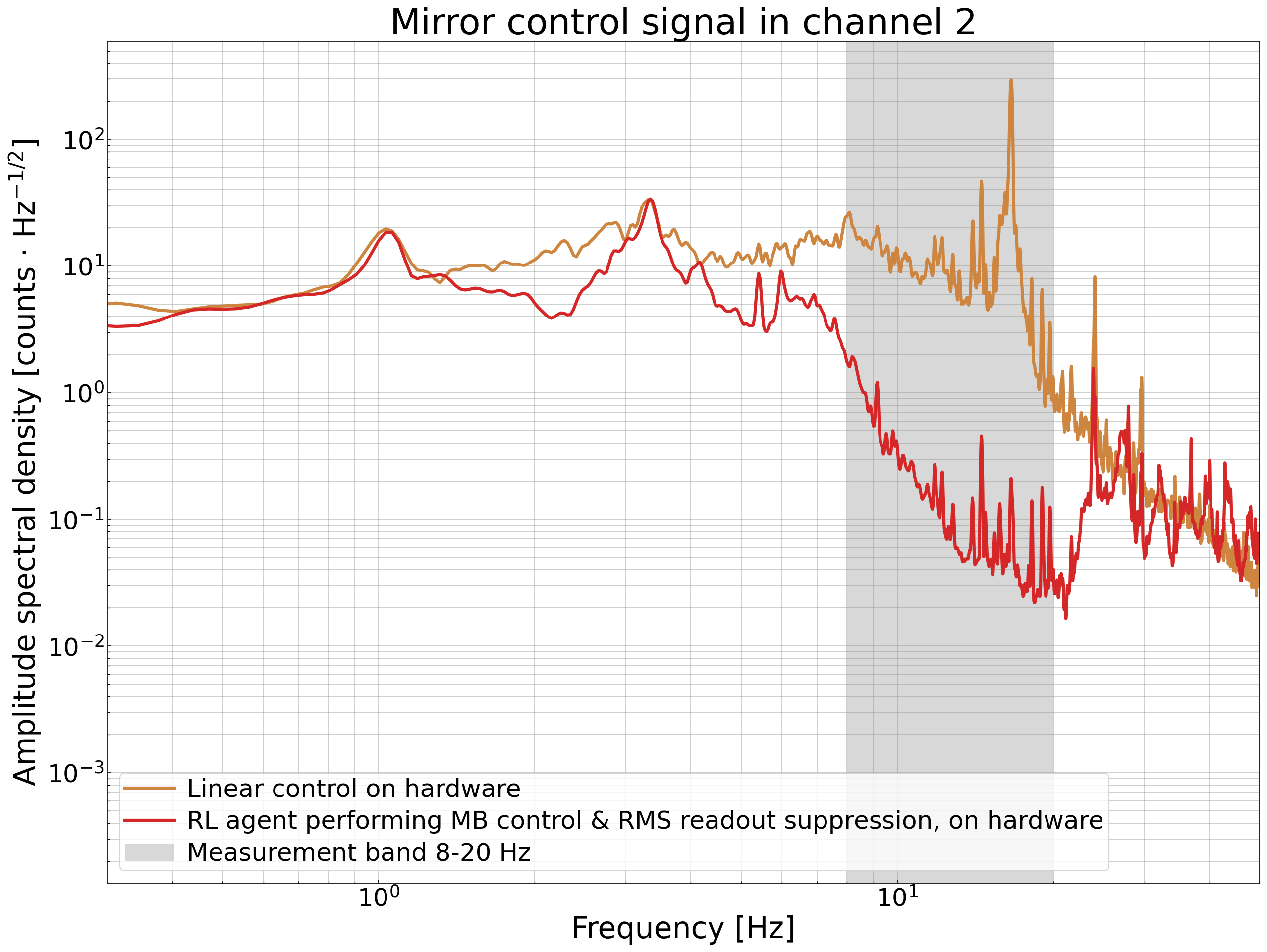}
    \includegraphics[width=0.32\columnwidth]{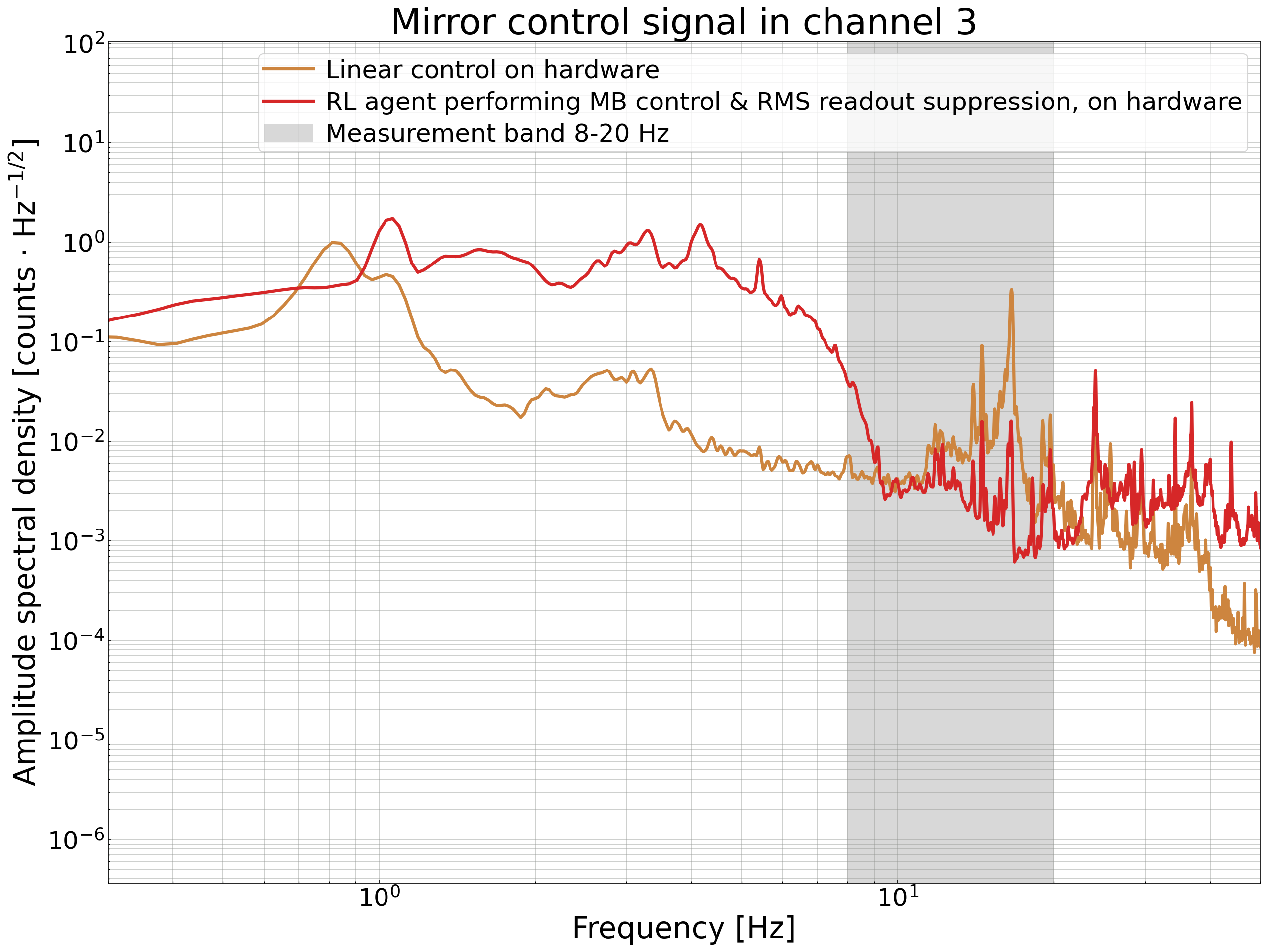}
    \includegraphics[width=0.32\columnwidth]{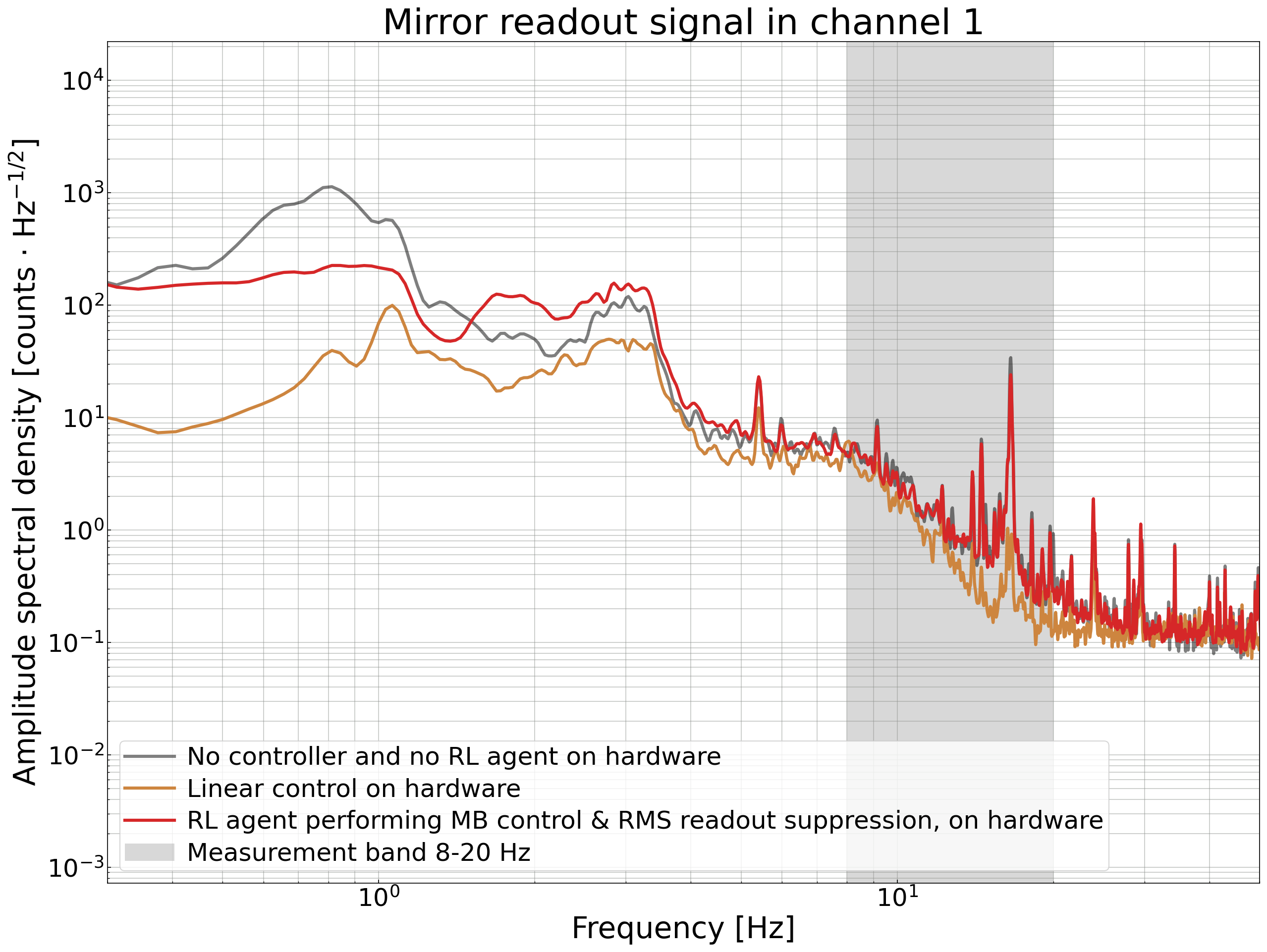}
    \includegraphics[width=0.32\columnwidth]{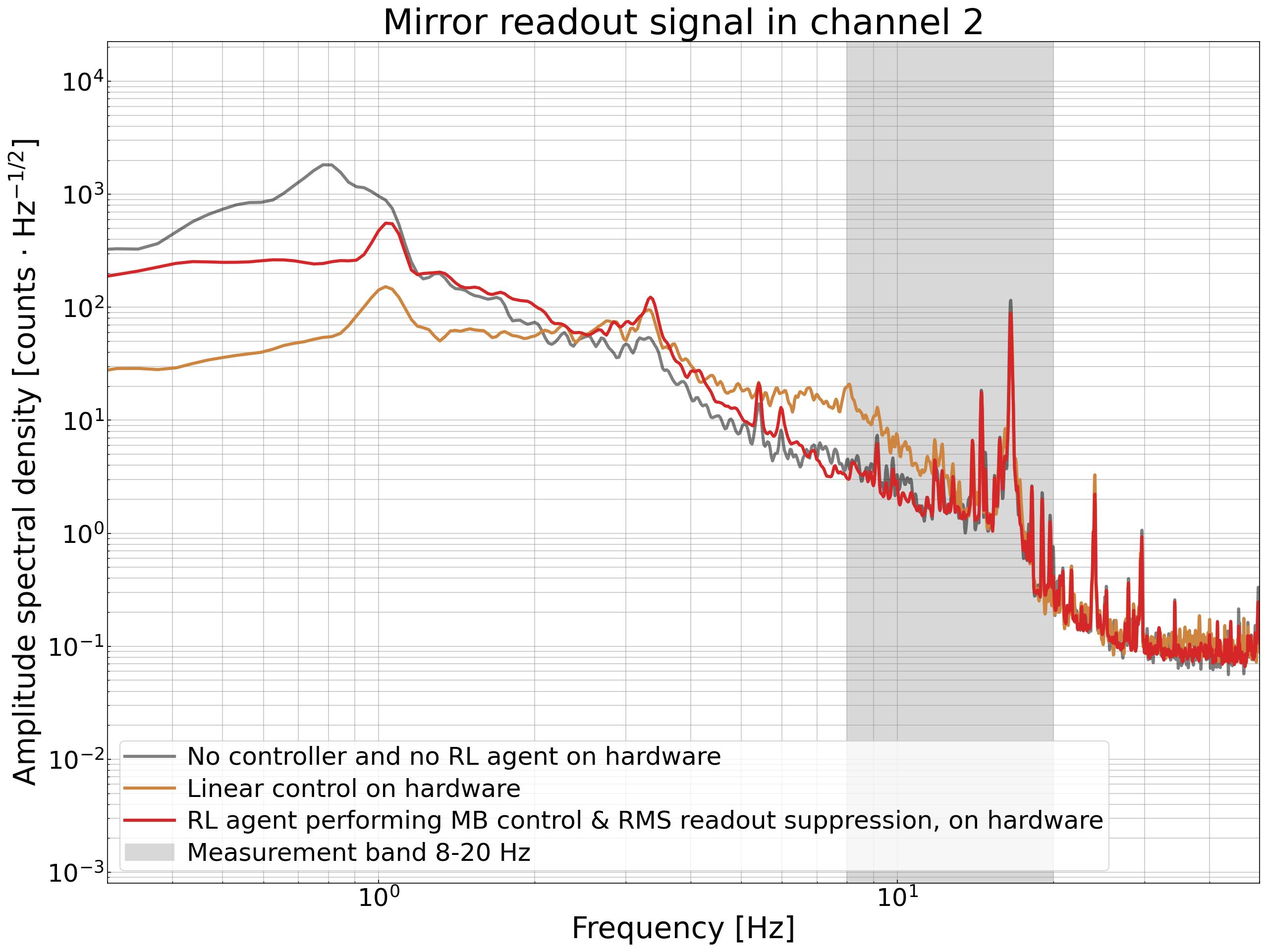}
    \includegraphics[width=0.32\columnwidth]{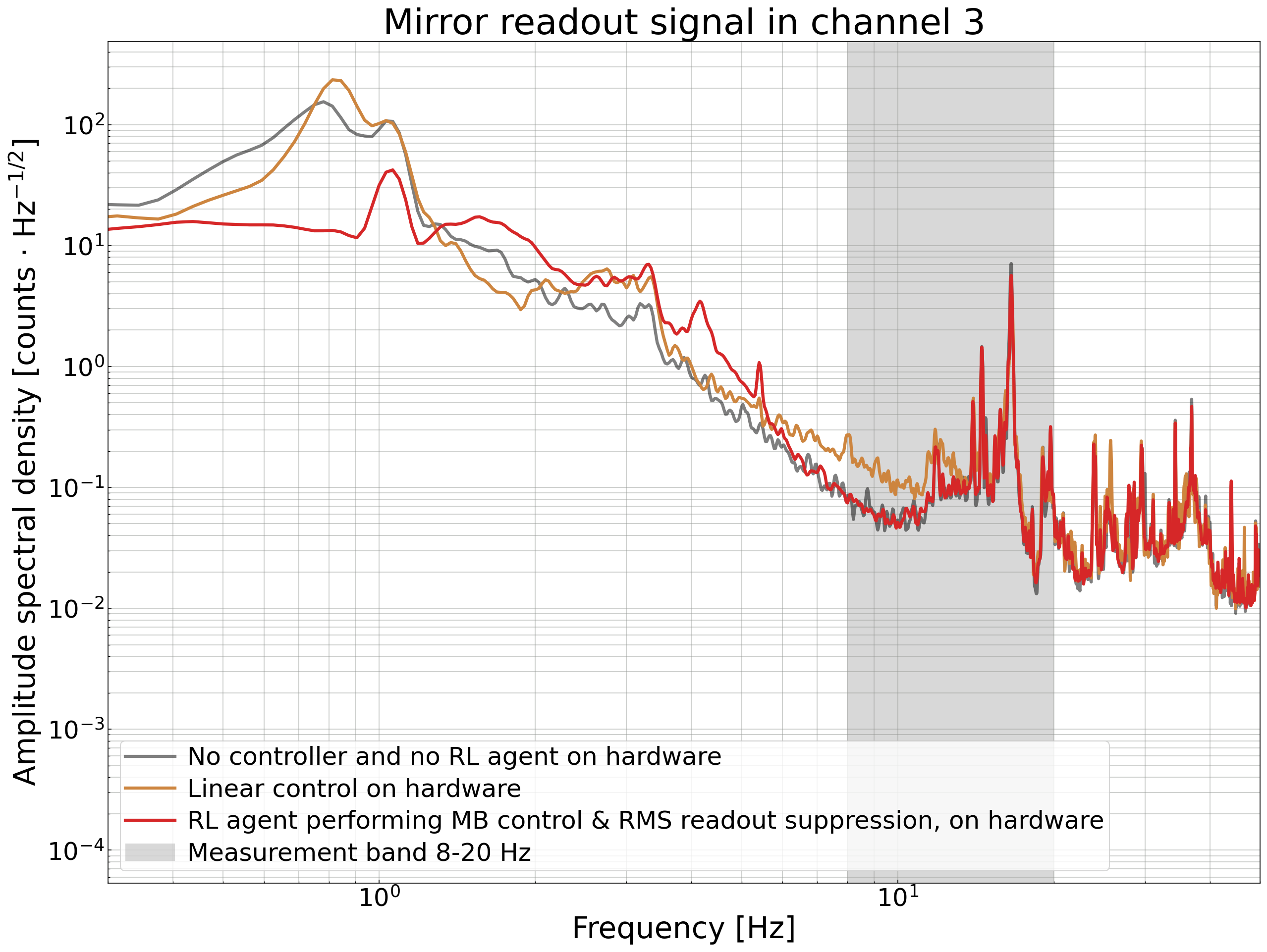}
    \includegraphics[width=0.32\columnwidth]{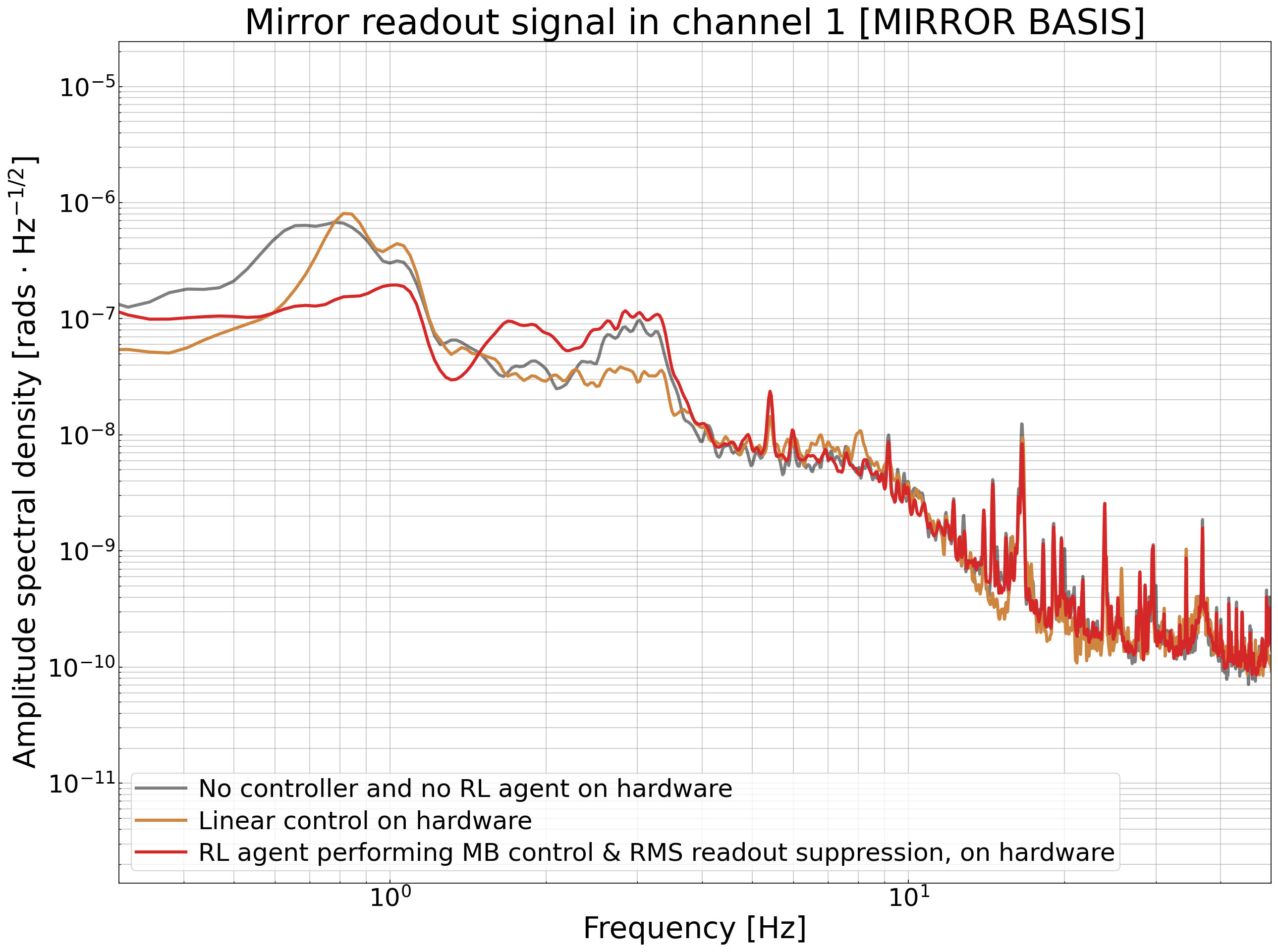}
    \includegraphics[width=0.32\columnwidth]{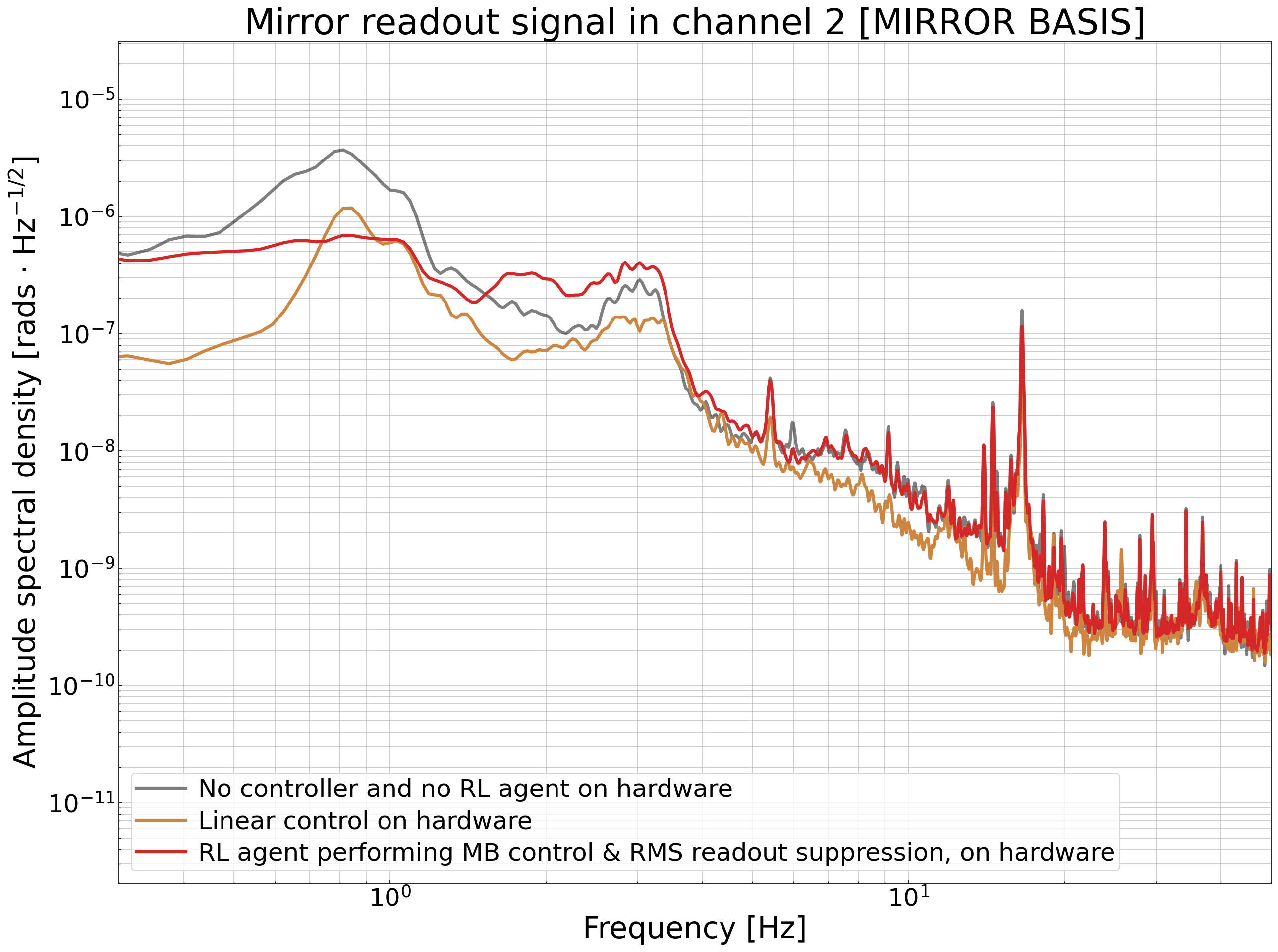}
    \includegraphics[width=0.32\columnwidth]{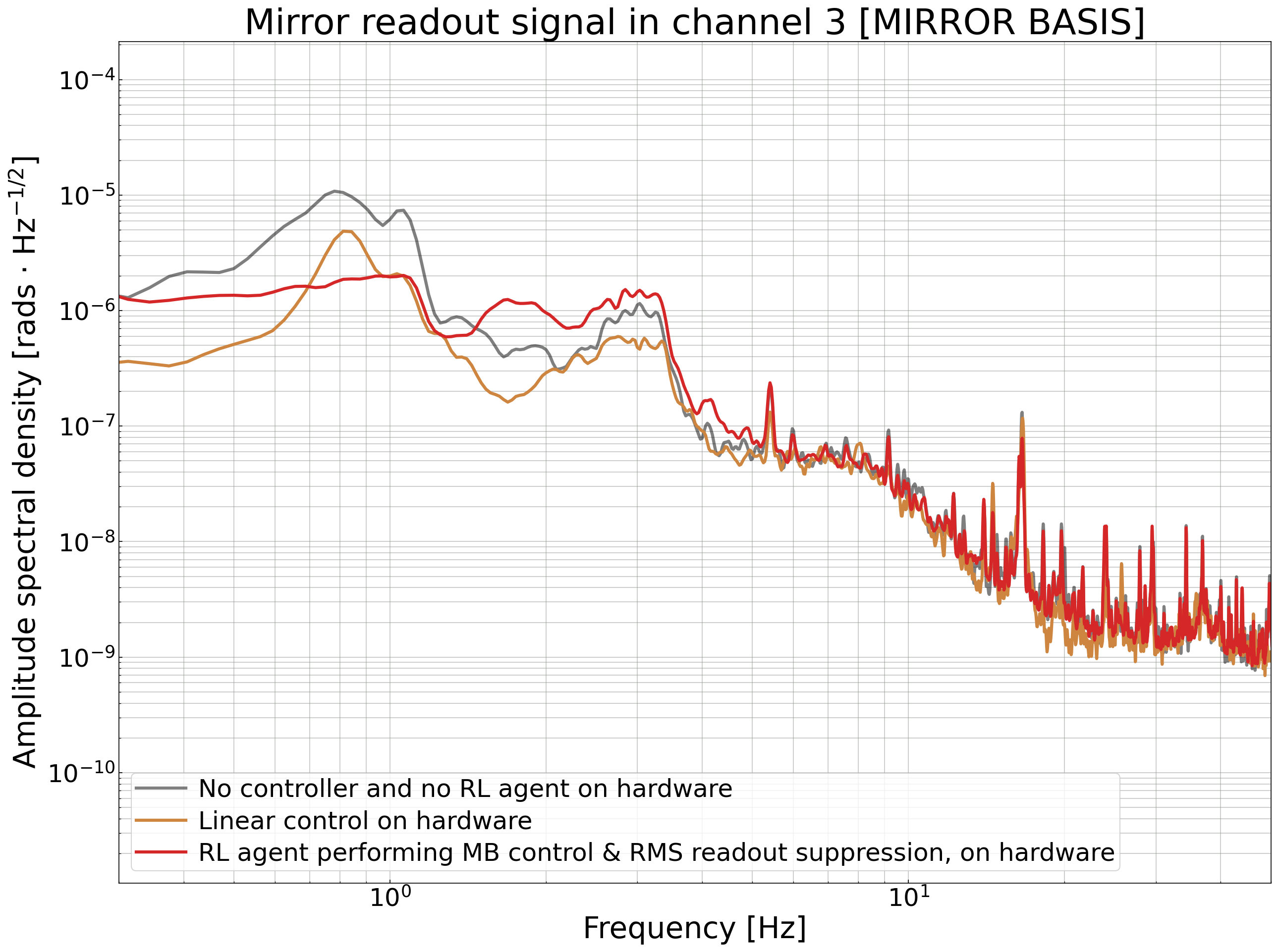}
    \caption{ RL policy on 40m IMC. (Top) Control signal in sensor basis (middle) error signal in sensor basis (bottom) error signal in local ('mirror') basis.  }
    \label{fig:gordian_40m}
\end{figure}

In the example, we have chosen two major control objectives. First, penalize the control effort in a chosen observation band (8-20Hz) in the `sensor basis' with the idea that it is in this coordinate system that the measurement-relevant noise is effective.
Second, penalize the overall RMS error in `local' (or mirror) basis, with the motivation that it is the actual physical excursion of each mirror that needs to be kept below a certain maximum level.\footnote{ We are aware that this does not necessarily correspond to the actual or only relevant space; i.e., maximum excursion of the laser beam on the sensor is practically important as well. It is straightforward to add additional penalties to account for such a requirement. In the spirit of the stated aim of a capability demonstration, we keep the example simple.} We show results of an exemplary run in Fig.~\ref{fig:gordian_40m}.

The rewards used for the IMC control are slightly more complex than the ones for the arm cavities due to the MIMO nature of the problem.
More concretely to achieve the above control objectives, we formulate the reward as follows:
\begin{description}
\item[Error RMS penalties:] In place of the filter, the absolute value of the error signal $|e(t)|$, this conceptually corresponds to using an all-pass in place of the filter. For each channel we apply the sigmoid squashing function individually with the parameters good=[0.0, 0.0, 0.0], bad=[$210^{-6}$, $210^{-6}$, $210^{-6}$] to compute individual scores in [0,1] for each channel. These scores are combined with a Smooth Max function with $\alpha=-3$ \cite{degrave2022} to yield a combined score.
\item[Observation band penalties:] We use an Elliptic IIR filter with settings for pass band
  $\omega_{p,low}=8.0\unit{Hz}$, $\omega_{p,low}=20.0\unit{Hz}$, stop band $\omega_{s,low}=3.0\unit{Hz}$, $\omega_{s,high}=35\unit{Hz}$, stop band gain $g_s = -60\unit{dB}$ and subsequent sigmoid scoring with
      good=[0.0, 0.0, 0.0], bad=[$210^{-6}$, $210^{-6}$, $210^{-6}$] to yield a combined observation band score using a Smooth Max function with $\alpha=-1$.
\item[Intermediate reward:] These two combined scores are further combined into the final, scalar intermediate reward $r(t)$ by multiplication, just as for the individual scores for the arm cavity.
\end{description}
This scoring approach can, in principle, be interpreted as a hierarchical soft-logic combination of the individual scores in [0,1] of the various channels.

The results demonstrate that the chosen control is effective in reducing the error in the observation band in sensor space, even though that does not necessarily mean being strictly improving in the mirror basis.
The same holds for the low frequency error suppressed by the RMS penalty, where we see the RL controller being strictly better in all channels in the basis the RMS penalty is applied in (local basis), whereas it is not strictly improving in all channels in sensor space.

\section{Baselines}

\subsection{Current operational linear control at LLO}
\label{s:llo-linear}
The 8x8 MIMO plant for the 4 coupled arm cavity mirrors is separated into sub-spaces as described in Section \ref{s:linearASC}.

The controllers currently used are linear and have been designed ``by-hand'' via pole placement mainly. The design criteria are for the loop to be stable in the presence of large, micro-seismic noise and also for the noise injected into the GW observation band to be less than that of the other limiting noise sources. Classical optimal control approaches have also been used over the years \cite{Hang2019,Mart2015,Tsang2024Hinf}, showing some modest improvement over the pole placement methods.

\subsection{Linear controllers derived with convex optimization}
\label{s:optimal_linear}
High-performance linear controllers can be designed by applying convex optimization techniques over finite impulse response filters \cite{barrattExampleExactTradeoffs1989, boydLinearControllerDesign1991, barrattInteractiveLoopshapingDesign1992, diamondCVXPYPythonEmbeddedModeling2016, agrawalRewritingSystemConvex2018}.  This method provides a systematic way to explore the fundamental trade-off between the performance metrics of pitch error RMS and observation-band control action RMS.  Realistic specifications on actuator effort, loop margin, and roll-off, and robustness to plant variation can be incorporated in these designs by imposing suitable constraints.

\begin{figure}
  \centering
  \includegraphics[width=0.75\textwidth]{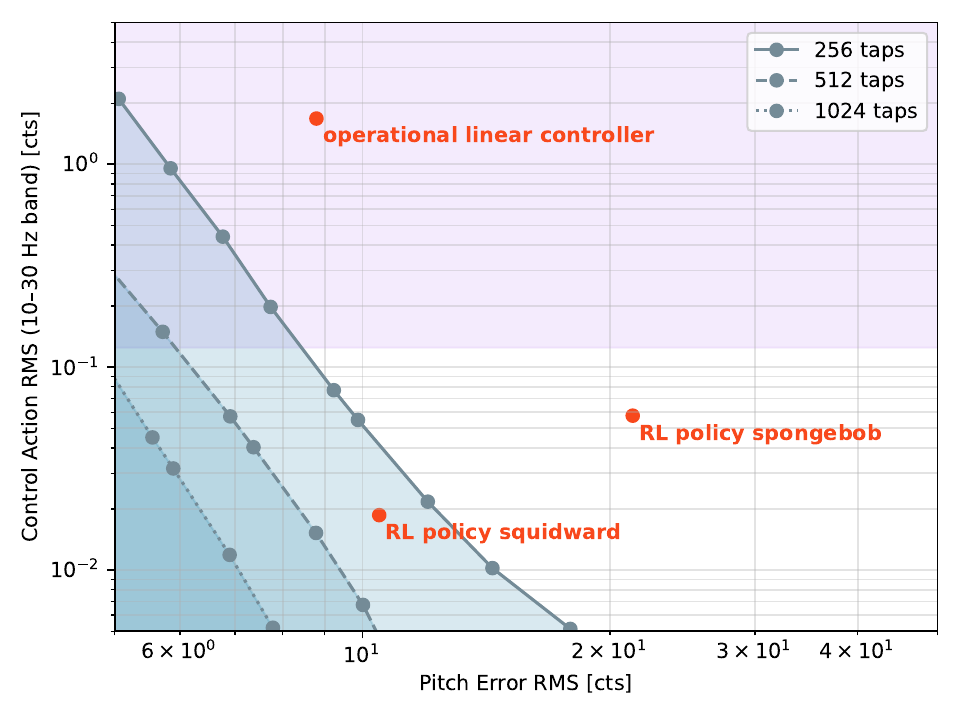}
  \caption{The plot shows the performance of the RL policies as well as the currently operational linear controller, as compared to that of a series of optimized linear controllers with varying filter tap line length.
  The purple region shows where the control action RMS in the observation band exceeds the design goal, motivated by the quantum limit.
  Blue regions indicate performance metrics inaccessible to linear control with convex-optimized finite impulse responses.
  }
  \label{fig:frontier}
\end{figure}

\begin{figure}
\centering
\includegraphics[width=0.75\textwidth]{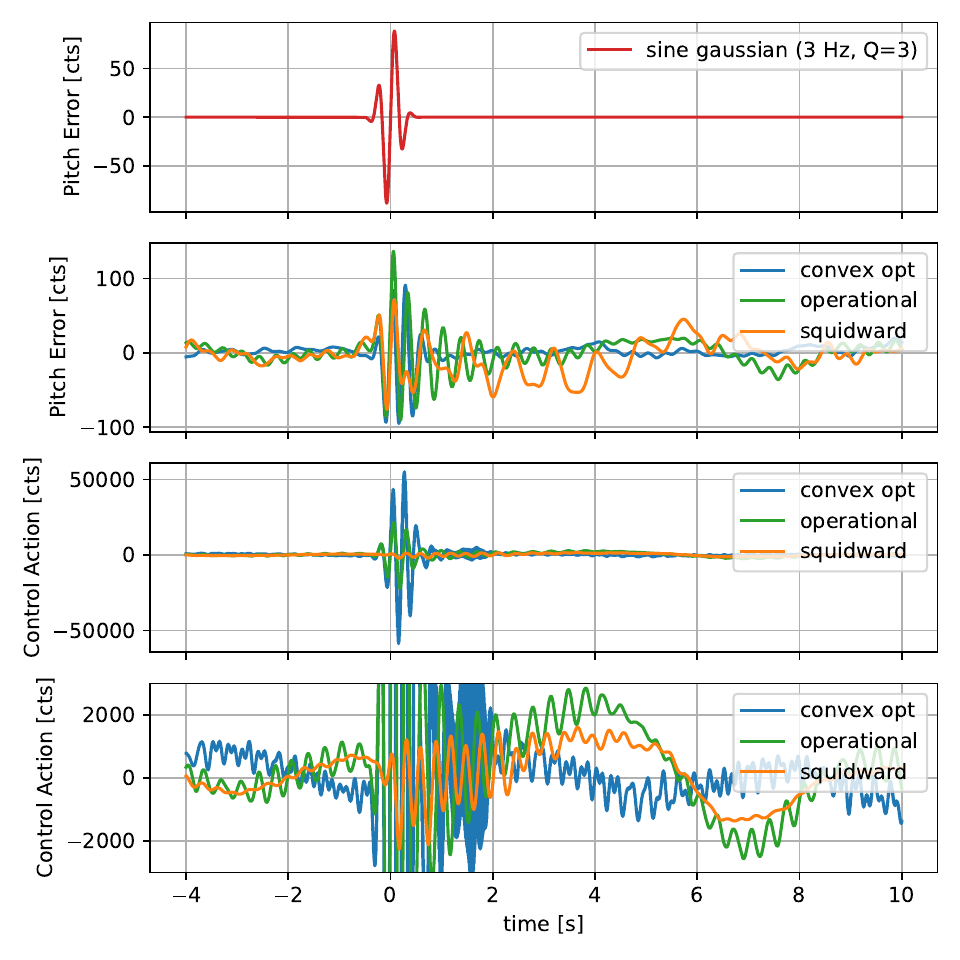}
\caption{Response to a sine-gaussian disturbance of the operational linear controller, a linear controller derived with convex optimization, and the deep loop shaping policy {\tt squidward}. The challenge of coping with this glitch is heightened by the proximity of the sine-gaussian peak frequency to that of the unstable mode of the plant. The resulting action of the optimized controller is highly aggressive, whereas the nonlinear neural-network-based policy has a notably sedate reaction to the disturbance. This finding indicates the advantage of the ability to (non-linearly) limit the RL policy by design (i.e., using a hyperbolic tangent in this case).}
\label{fig:sg}
\end{figure}

A comparison of the performance of convex-optimized linear controllers, the operational linear controller, and the RL policies is shown in Fig.~\ref{fig:frontier}.  An objective consisting of the weighted max of pitch error RMS and observation-band control action RMS has been used in the optimization, combined with the following constraints:
\begin{itemize}
\item Actuator effort: the total control action RMS was constrained to 3000\,ct, in the presence of the expected levels of seismic and sensing noise
\item Loop margin: a minimum disk margin \cite{seilerIntroductionDiskMargins2020} of $\frac{1}{2}$ was required
\item Loop roll-off: the loop gain was constrained to be below 0.1 at 8\,Hz, with a $1/f^4$ roll-off at higher frequencies
\item Robustness: applying the small-gain theorem, a constraint ensuring closed-loop stability when the plant's unstable resonance varied by up to $\pm 20$\
\end{itemize}
The first three constraints follow from the design goals embodied in the operational linear controller.  The robustness constraint is applied to guarantee stability when the frequency of the opto-mechanical resonance shifts, and it corresponds to the pole randomization condition used to train the RL policies.  However, since the small-gain theorem provides a sufficient but not necessary condition for stability, linear controllers designed with this constraint are not necessarily globally optimal.

The RL policy {\tt squidward} evidently surpasses the performance bound of the resulting convex-optimized controllers with comparable history length (256 taps). Moreover, it should be noted that due to the architecture of the convex optimization problem, linear controllers derived in this fashion benefit from working in tandem with an auxiliary stabilizing controller, while the RL policies were required to stabilize the plant on their own. The scaling of RL policy performance with history length, and the effects of training and operating RL policies in combination with a stabilizing controller, are topics to be explored in future work.

\subsection{Robustness}
\label{s:robustness}

Unlike with traditional linear robust control methods, with our method, we do not have theoretical mathematical predictors for robustness. When approximately solving the optimal control problem with Reinforcement learning, we encourage robustness through parameter variation and random noise added to the simulation (adding epistemic and aleatoric uncertainty, respectively). However, as we are deploying our policies in the high-stakes environment of the real LIGO observatory, we are seeking additional empirical assessment of robustness before we deploy the policy on the real system. We use a suite of simulation-based scenarios to test policies for robustness under perturbations and plant variations before deployment. We expose the candidate policy to additional robustness testing in simulation using a selected set of disturbances and non-nominal plant parameters. These conditions go beyond the variations that the policy has seen in training and the spectrum of specific disturbances that are modeled after populations of glitches observed on the real system (i.e., Sine-Gaussian disturbances).

The scenarios are:
\begin{itemize}
    \item Variation of seismic noise. We vary the overall level of the seismic noise using a scalar multiplier in the range [1-10].
    \item Variation of right half-plane (RHP) pole and other poles frequency. The pole frequencies are varied +/-20 \
    \item Input disturbances
    \begin{itemize}
      \item Step inputs in the range 0-100 counts. 
      \item Impulse inputs in the range 0-100 counts. 
      \item Sine-Gaussian inputs $d(t) = a \sin(2 \pi f_0 t) \exp(-t^2/\tau^2)  $ with $\tau = \frac{q}{\sqrt{2} \pi f_0}$. $a \in (0,100]$, $f_0 \in [3,30]$, $q \in [1,10]$.

    \end{itemize}
    \item Variation of time delays up to 50 control time steps. 
\end{itemize}

\subsection{Comparison of results against O4 stats}
\label{s:O4stats_section}
\begin{figure}
\centering
\includegraphics[width=0.35\columnwidth]{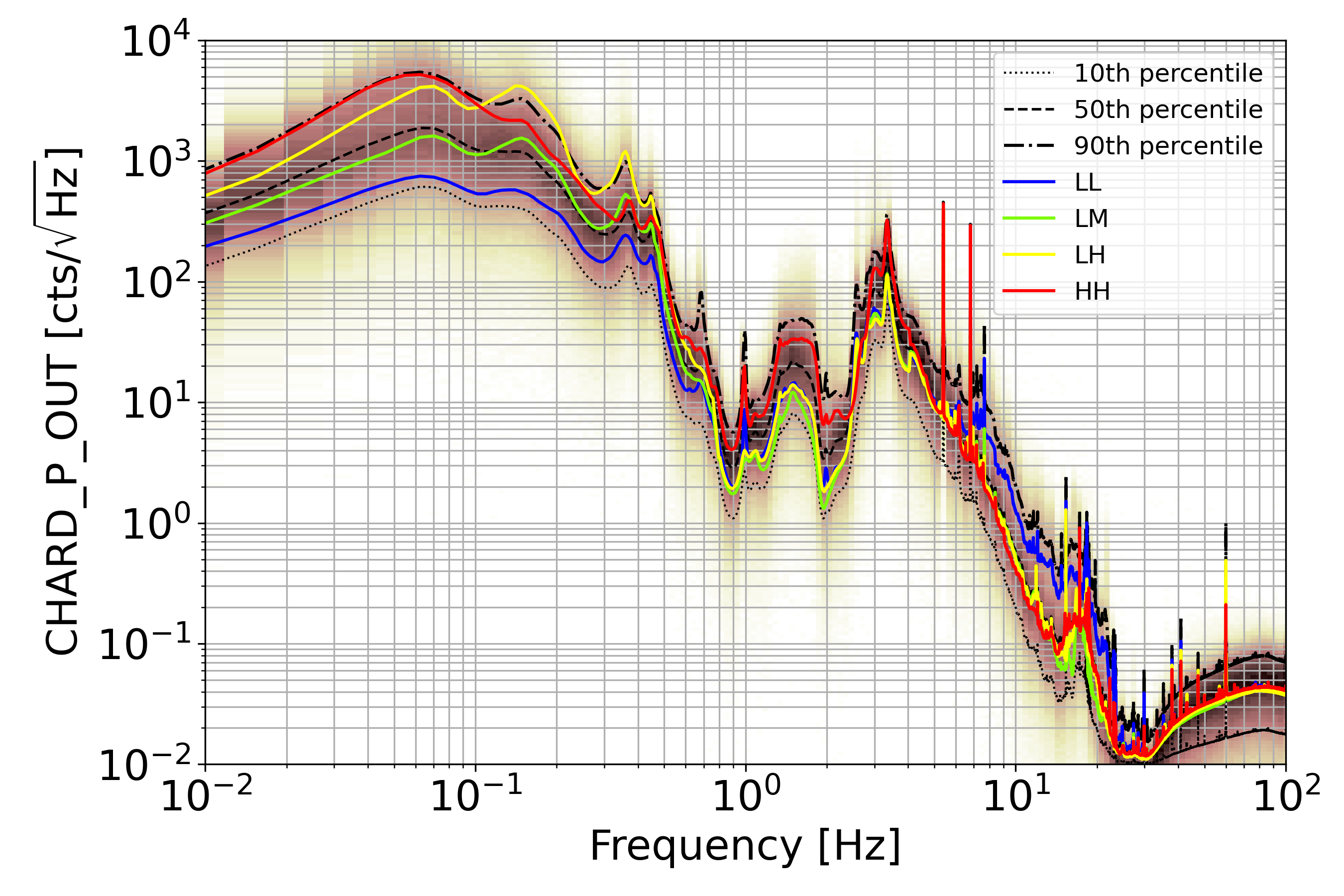}
\includegraphics[width=0.35\columnwidth]{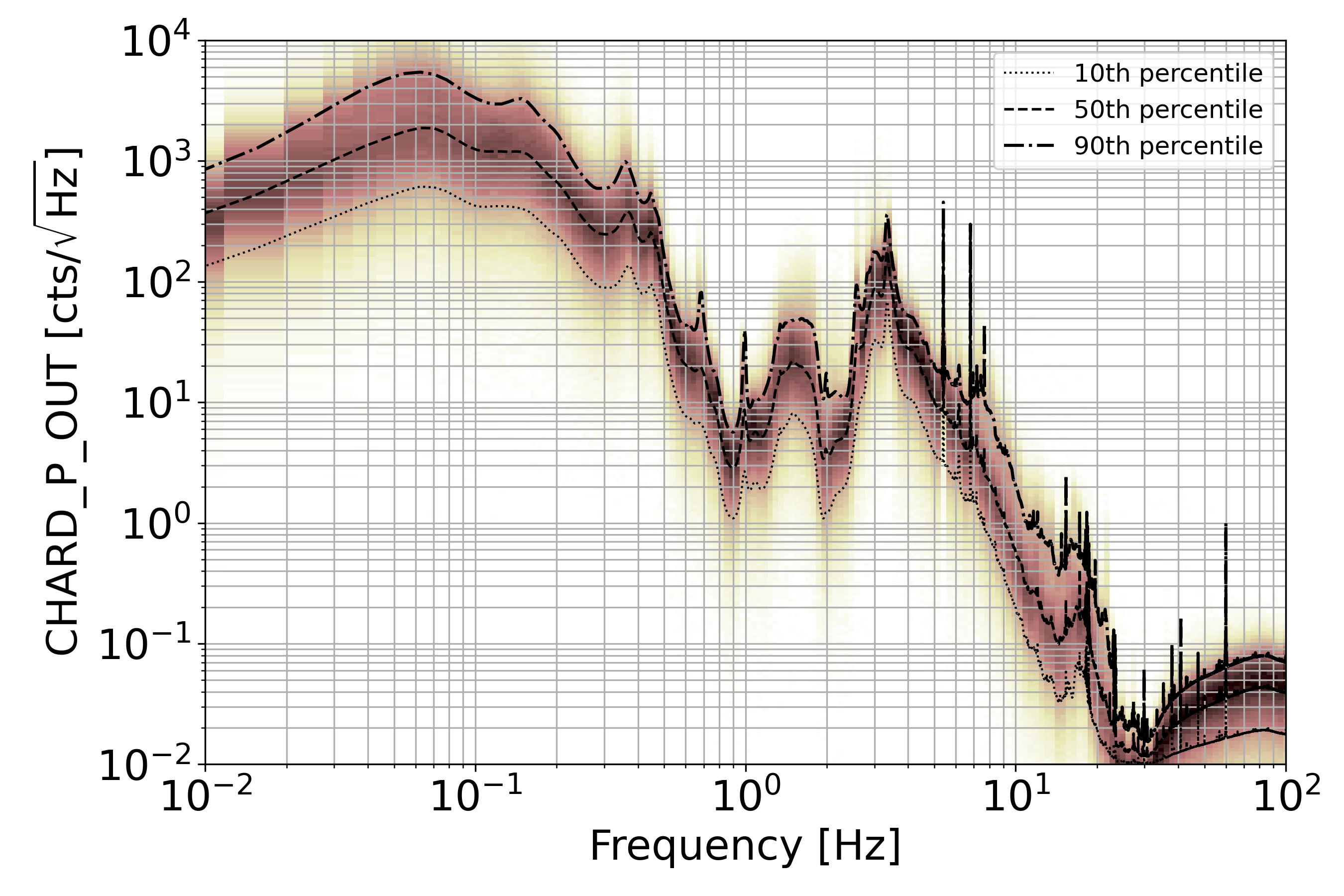}\\
\includegraphics[width=0.35\columnwidth]{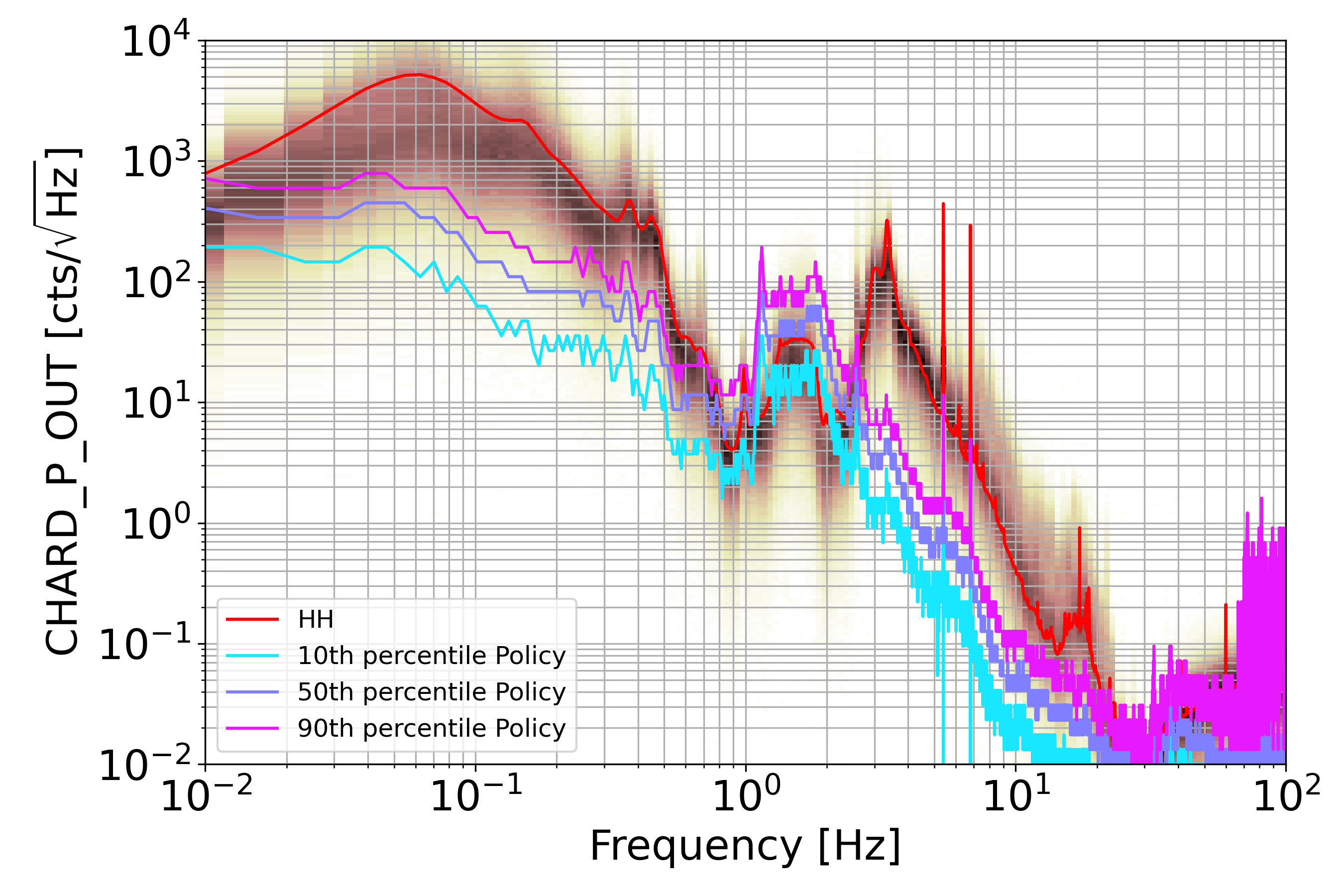}
\includegraphics[width=0.35\columnwidth]{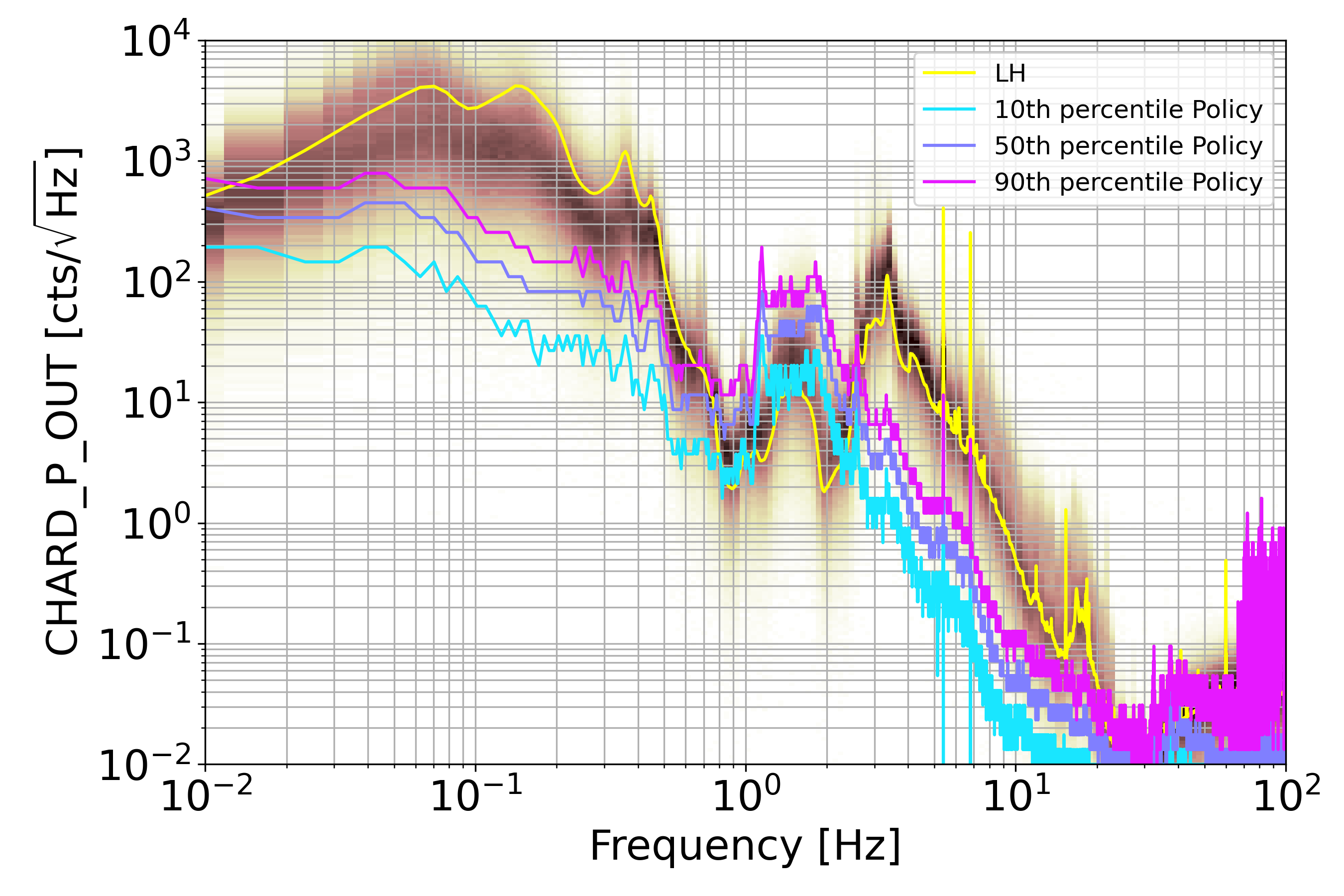}\\
\includegraphics[width=0.35\columnwidth]{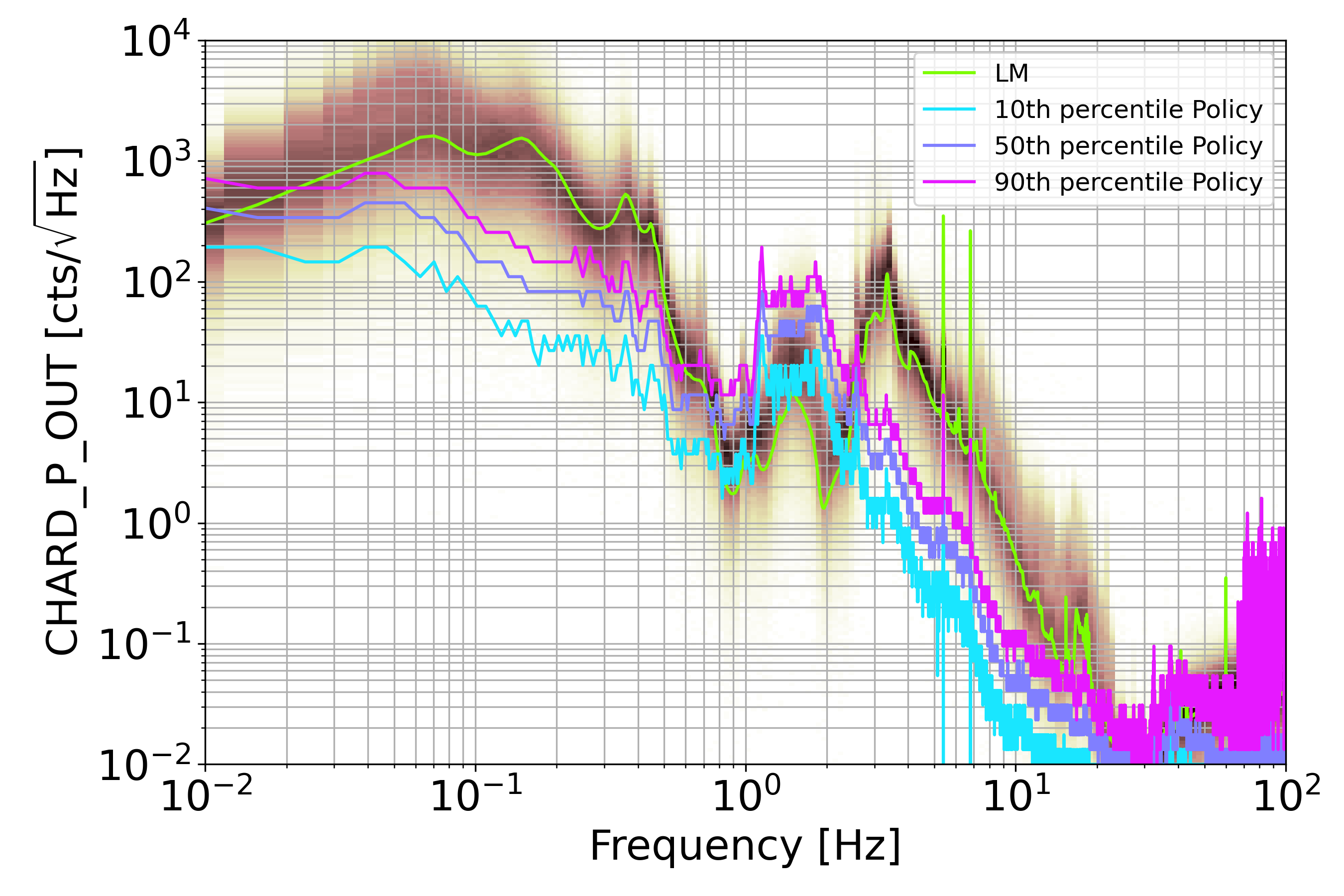}
\includegraphics[width=0.35\columnwidth]{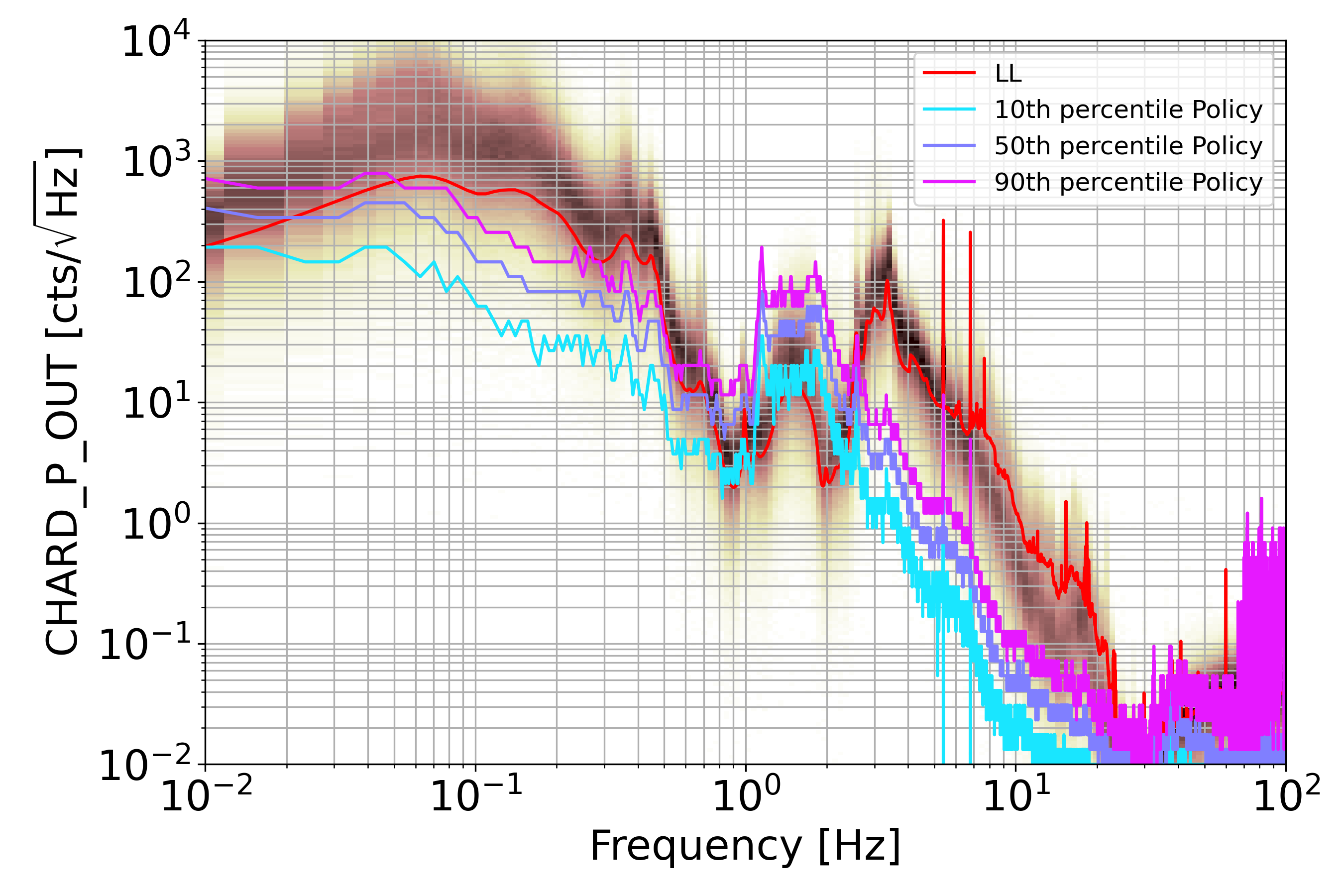}
\caption[Spectra Percentiles]{A spectral histogram of the \chardp{} feedback control loop for the entire O4 observing run is shown in the background of all six panels for reference. They are the data from when the detector was in observation mode,  excluding the first hour of the lock when the IFO is still thermalizing. 
It is compared with the spectra of a few typical ASDs used for training and with the RL policy spectra. 
The top left displays the 
10\textsuperscript{th}, 50\textsuperscript{th}, and 90\textsuperscript{th} 
percentiles of the O4 data, overlaid with cases representing four typical environmental conditions derived from seismic studies. Here abbreviations represent the following: LL (Low Baseline, Low Microseism), LM (Low Baseline, Med Microseism), LH (Low Baseline, High Microseism), HH (High Baseline, High Microseism). 
The top right shows the O4 data alone. 
The middle-left displays the HH case and the
10\textsuperscript{th}, 50\textsuperscript{th}, and 90\textsuperscript{th} 
percentiles of the RL policy spectra. 
The middle-right displays the LH case and RL policy percentiles. 
The bottom-left shows the LM case and RL policy percentiles, and the bottom-right shows the LL case with RL policy percentiles overlaid.
These panels highlight the interplay between environmental conditions and the RL policy performance, providing insight into the relationship between baseline noise, microseism levels, and control spectra.}
\label{fig:O4stats}
\end{figure}

In Fig.~\ref{fig:O4stats}, we compare our results against O4 run statistics, providing details in the figure's caption.

\end{document}